\documentclass{JHEP3} 

\usepackage{epsfig,multicol,bbm}
\usepackage{amssymb,amsmath}

\newcommand\fverb{\setbox\fverbbox=\hbox\bgroup\verb}
\newcommand\fverbdo{\egroup\medskip\noindent%
			\fbox{\unhbox\fverbbox}\ }
\newcommand\fverbit{\egroup\item[\fbox{\unhbox\fverbbox}]}
\newbox\fverbbox

\def \snu  { {\tilde{\nu}}}

\def \chii  {{\tilde{\chi}^0_i}}

\def \chai   { {\tilde{\chi}^{\pm}_i}}
\def \phot { \gamma}

\def \z {{Z}}

\def \w{{W}}
\def \h {{h}}
\def \H {{H}}
\def \A {{A}}

\newcommand{\lviol}{$\not\!{\rm L}~$}
\newcommand{\xisigma}{$\xi \sigma^{\rm (scalar)}_{\rm nucleon}$}
\newcommand{\mloop}{\Delta m^{\rm 1 loop}_{\rm neutrino}}

\title{Sneutrino cold dark matter, a new analysis: relic abundance and detection rates}

\author{Chiara Arina, Nicolao Fornengo\\ Dipartimento di Fisica Teorica, Universit\`a di Torino 
and Istituto Nazionale di Fisica Nucleare, Sezione di Torino, via P. Giuria 1, I--10125 Torino, Italy\\
E-mail: \email{arina@to.infn.it}, \email{fornengo@to.infn.it}}

\preprint{DFTT 17/2007}	

\abstract{We perform a new and updated analysis of sneutrinos as dark matter candidates, in different classes of supersymmetric models. We extend previous analyses by studying sneutrino phenomenology for
full variations of the supersymmetric parameters which define the various models. We first revisit 
the standard Minimal Supersymmetric Standard Model, concluding that sneutrinos are marginally compatible
with existing experimental bounds, including direct detection, provided they compose a 
subdominant component of dark matter. We then study supersymmetric models with the inclusion of right--handed fields and lepton--number violating terms. Simple versions of the lepton--number--violating
models do not lead to phenomenology different from the standard case when the neutrino mass bounds
are properly included. On the contrary, models with right--handed fields are perfectly viable: they
predict sneutrinos which are compatible with the current direct detection sensitivities, both
as subdominant and dominant dark matter components. We also study the indirect
detection signals for such successful models: predictions for antiproton, antideuteron and gamma--ray fluxes
are provided and compared with existing and future experimental sensitivities. The neutrino flux
from the center of the Earth is also analyzed.}


\begin{document} 


\section{Introduction}\label{sec:intro}

Sneutrino as a particle candidate to explain the Cold Dark Matter (CDM) present in the Universe has been investigated in the past in a number of interesting papers, where its relic abundance and its
scattering cross--section off nucleons, relevant for the direct detection searches of dark matter, have
been calculated and discussed. Sneutrinos in the Minimal Supersymmetric Standard Model (MSSM) have been studied in the past \cite{Ibanez:1983kw,Ellis:1983ew,Hagelin:1984wv,Goodman:1984dc,Freese:1985qw,Falk:1994es}, and from the direct detection searches they have been excluded as the 
major component of dark matter. The case of the inclusion of right--handed sneutrino fields 
\cite{Arkani-Hamed:2000bq,Grossman:1997is,Smith:2001hy,Smith:2002af,Tucker-Smith:2004jv} and of
lepton--number violating terms 
\cite{Arkani-Hamed:2000bq,Grossman:1997is,Smith:2001hy,Smith:2002af,Tucker-Smith:2004jv,Hall:1997ah,Hirsch:1997vz,KlapdorKleingrothaus:1999bd,Kolb:1999wx} 
have also been studied, in connection also to the problem of
neutrino masses. It has been shown that right--handed components may alter significantly neutrino
phenomenology, due to a reduction of the coupling with the $Z$ boson. Lepton--number violating 
terms, which can induce a mass splitting of the two mass eigenstates, also lead to a modified
phenomenology since they alter the coupling to the $Z$ boson, which is non--diagonal in the
CP basis \cite{Hall:1997ah}. These models offer a nice realization of inelastic dark matter, which has been
introduced in relation to direct detection searches \cite{Smith:2001hy,Smith:2002af,Tucker-Smith:2004jv,Hall:1997ah}. Sneutrinos in connection with the dark matter problem have also been discussed in different frameworks some of which may be found in Refs.~\cite{Han:1997wn,Han:1999jc,Hooper:2004dc,Asaka:2005cn,Lee:2007mt,Gopalakrishna:2006kr,Dimopoulos:1996gy}.

Typically, the relic abundance and the direct detection rate (which offers a very stringent
experimental bound to sneutrino dark matter) have been studied and some of the models have been
proposed in order to circumvent the character of exclusion of sneutrino dark matter in the minimal
version of the MSSM. However, a thorough analysis as is typically done for neutralino dark matter,
with a global study in the full parameter space of the supersymmetric models, has not been performed
for sneutrinos. Indirect detection signals, especially those coming from dark matter annihilation
in the Galaxy, have not been typically discussed in the literature.

In the present paper we wish to reconsider in a consistent way sneutrino as a cold relic from the early
Universe and study its phenomenology relevant both for Cosmology and for relic--particle detection. 
First of all, we explicitely consider both cosmologically dominant and sub--dominant sneutrino 
configurations: in fact, we are interested not only in those configurations which are able to solve the
CDM problem, but also those which provide a smaller amount of cosmological relic abundance but which
could be potentially detectable by means of various astrophysical signals. We then study
the sneutrino detection rates, both of direct and indirect type. Therefore, in addition to the quite
relevant direct--detection signal, coded into the scattering cross--section off nuclei and which typically provides the most stringent limits to sneutrino CDM once the relic abundance bound is imposed, we 
calculate also the indirect detection signals which come from sneutrino pair--annihilation in the galactic environment: predictions for antiproton, antideuteron and gamma--rays signals are provided and discussed. This is a novel analysis for sneutrino CDM, and we will show that relic sneutrinos could have a chance of
being detected by means of indirect searches. At the same time, for some particle physics models, indirect detection nicely complements direct searches in posing limits to the supersymmetric parameter space.
Predictions for the neutrino flux from the center of the Earth, relevant for neutrino telescope searches, are also discussed. 

The particle physics models we explicitely analyze in the paper are extensions of the Minimal Supersymmetric Standard Model (MSSM), where the sneutrino is the scalar super--partner of the left--handed neutrino. Since
neutrinos have masses, as is now clearly understood by a host of independent and very robust experimental
results and theoretical analyses 
\cite{GonzalezGarcia:2000sq,GonzalezGarcia:2002dz,Maltoni:2004ei,Fogli:2005cq,Fogli:2004as}, we will focus our attention on extensions of the MSSM which contain
terms in the supersymmetric lagrangian which can drive neutrino masses. Connections between neutrino physics and the phenomenology of sneutrino CDM will therefore arise, and we will explicitely consider them whenever relevant. The models which we will be considering are therefore natural and direct extensions of the MSSM which incorporate at the same time the new physics required to explain two basic problems of astro--particle physics: the origin of neutrino masses and the nature of dark matter. We do not attempt to be totally exhaustive on the type of supersymmetric models (something which would require an exceedingly large
analysis).
Instead we concentrate on a number of the most direct extensions of the MSSM and derive the phenomenology
of sneutrino CDM thoroughly.

We first review the case of the Minimal Supersymmetric extension of the Standard Model (MSSM), by enlarging the previous analyses to the case of sub--dominant sneutrino CDM. This is discussed in Sect. \ref{sec:std}. This model, not very appealing for sneutrino
CDM and actually already almost excluded by direct detection searches (we will explicitely show under what
circumstances a sneutrino CDM may still be viable in these model) sets the basis for the following extensions.
For the sake of clarity, and due to the complexity of the overall analysis of this paper, we will name the different
classes of models with specific labels. The standard MSSM will be called ``STD model'' in the discussion.

The first extension allows for the inclusion of a right handed field for the sneutrino
\cite{Arkani-Hamed:2000bq,Grossman:1997is,Smith:2001hy,Smith:2002af,Tucker-Smith:2004jv}
and is discussed in Sect. \ref{sec:lr}.
 We will 
label this class of models ``LR models''. Since we are enlarging the parameter space of the supersymmetric models by
adding terms (and therefore parameters) to the fundamental lagrangian, we will organize our discussion
in two steps. We first discuss the
sneutrino CDM phenomenology in terms of the new added parameters (by keeping a fixed configuration
for the rest of the parameter space, whenever relevant). This will be useful to disentangle specific 
features of the relic abundance and direct detection rate for this specific class of models. 
We then extend our analysis to include a full scan of the supersymmetric parameter space. 
All the relevant experimental bounds on searches of Supersymmetry at accelerators and supersymmetric contributions to rare processes are properly taken into account. Indirect detection signals are discussed for the full--scan case only. We will show that ``LR models'' contain viable sneutrino CDM candidates
for a large sector of its parameter space and possess a rich sneutrino CDM phenomenology with potentially
detectable direct and indirect detection rates. 

We then discuss a class of models with a lepton--number violating mass--term added to the lagrangian \cite{Hall:1997ah,Hirsch:1997is,KlapdorKleingrothaus:1999bd,Kolb:1999wx}.
In this case the two mass eigenstates are splitted and possess an off-diagonal coupling to the $Z$--boson
which may sizeably alter both the relic abundance and the direct detection rate
\cite{Hall:1997ah}. This class of models
is named here ``$\not\!{\rm L}$ models'' and is discussed in Sect. \ref{sec:cp}. In this case, the new term in the lagrangian can lead to 1--loop contributions to
the neutrino mass, which may be potentially large depending on the values of the parameters involved. We therefore bound our analysis to the neutrino mass limits, as an additional and relevant phenomenological
constraint. In this class of models, once all the experimental bounds are taken into account, including neutrino mass limits, we do not find a very interesting phenomenology for CDM: we therefore discuss only the general analysis, but we skip the 
full parameter--space scan and the discussion of indirect detection signals.

The last class of models we consider contains right--handed sneutrino fields and lepton--flavour violating
terms simultaneously \cite{Arkani-Hamed:2000bq,Grossman:1997is,Smith:2001hy,Smith:2002af,Tucker-Smith:2004jv,Hall:1997ah,Hirsch:1997vz,KlapdorKleingrothaus:1999bd,Kolb:1999wx} and is discussed in Sect. \ref{sec:maj}. These  models, which we
label as ``MAJ models'', are very interesting from the theoretical point of view since
they can successfully accommodate both Dirac and Majorana neutrino mass--terms with a renormalizable
supersymmetric lagrangian. Neutrino masses are then obtained by means of the see--saw mechanism \cite{Minkowski:1977sc,gellmann,yanagida,Mohapatra:1979ia,Mohapatra:1980yp,Schechter:1980gr}.  A nice exhaustive analysis of this topic has been recently performed in Ref \cite{Dedes:2007ef}.
Also for this class of models we first perform a study of the parameter space relevant for the sneutrino 
sector (and by taking into account also neutrino--mass theoretical and phenomenological consequences). We then perform a full scan of the supersymmetric parameter space for two specific relevant sectors of
the model, named ``MAJ[A]'' and ``MAJ[B]'', for which we discuss explicitely also the indirect detection rates. We will show that the class of models ``MAJ[A]'' (characterized by a TeV--scale Majorana mass--parameter in the neutrino sector) have again a rich phenomenology from the point of view of sneutrino CDM.

\section{Standard minimal MSSM}\label{sec:std}

In the minimal MSSM, sneutrinos are the scalar partners of the left--handed neutrinos. Superfields
$\hat L^{I}$ contains the fermionic SU(2)$_{L}$ doublets $L^{I} \equiv (\nu^{I}_{L}, \, l^{I}_{L})$ 
(where $I=e,\mu,\tau$ runs over the three families) and its corresponding scalar doublets 
$\tilde L^{I} \equiv (\tilde\nu^{I}_{L}, \, \tilde l^{I}_{L})$. The supersymmetric lagrangian is
constructed in the usual way (see, for instance, Refs. \cite{Haber:1984rc,Gunion:1984yn}).
We call this model ``STD model".
We do not explicitely detail here all the terms of the supersymmetric lagrangian which involve the
sneutrino fields. We instead quote only the terms which are relevant for our discussion. The
complete set of interaction lagrangians can be found in Ref. \cite{Haber:1984rc,Gunion:1984yn,arina2007}.

The part of the superpotential relevant for the leptonic sector is (we use notations as in Refs. \cite{Haber:1984rc} and \cite{Dedes:2007ef}):
\begin{equation}
W = \epsilon_{ij} (\mu \hat H^{1}_{i} \hat H^{2}_{j} - Y_{l}^{IJ} \hat H^{1}_{i} \hat L^{I}_{j} \hat R^{J} )
\end{equation}
where $\hat H^{1}$ and $\hat H^{2}$ are the two higgs--doublet superfields, $\hat R^{J}$ denotes the right--handed charged lepton superfields (which contain the right--handed sleptons), $\mu$ is the usual higgs--mixing parameters and 
$Y_{l}^{IJ}$ is a matrix which contains the Yukawa couplings. Repeated indices imply a sum--convention over them. In our analysis, which relies on a minimal version of the MSSM with a minimal set of
relevant parameters, $Y_{l}^{IJ}$ is real and diagonal in flavour space and the Yukawa couplings are linked to the charged--lepton masses by the usual relation $m_{I} = v_{1}Y_{l}^{II}$, where $v_{1}$ is the
vacuum expectation value of the neutral component of the $H^{1}$ Higgs field.

The soft--supersymmetry--breaking potential relevant for the sneutrino sector is:
\begin{equation}
V_{\rm soft} = (M_{L}^{2})^{IJ} \, \tilde L_{i}^{I \ast} \tilde L_{i}^{J} + 
[\epsilon_{ij}(\Lambda_{l}^{IJ} H^{1}_{i} \tilde L^{I}_{j} \tilde R^{J}) + \mbox{h.c.}]
\end{equation}
where $\Lambda_{l}^{IJ}$ is a matrix, which we take real and diagonal in flavor space, as we do for the Yukawa matrix 
$Y_{l}^{IJ}$. Also the mass matrix $M_{L}^{2}$, in the minimal version of MSSM, is taken to be diagonal and,
in order to reduce the MSSM to a minimal set of parameters, all the three entries equal to a common value which we denote as $m_{L}^{2}$. The scalar potential is determined as the sum $V = V_F + V_D + V_{\rm soft}$, 
where the D--term $V_D$ describes gauge--interactions \cite{Haber:1984rc,Gunion:1984yn} and the $F$--term $V_F$ is obtained by the superpotential 
by means of its derivatives over the scalar components $\phi_a$ as 
$V_F = \sum_a | \partial W /\partial \phi_a |^2$.

From all the above, the mass--term for the sneutrino field $\snu_{L}$ (for each family) is simply derived to be:
\begin{equation}
V_{\rm mass} = \left[ m_{L}^{2} + \frac{1}{2} m_{Z}^{2} \cos 2\beta \right] \snu_{L}^{\ast} \snu_{L}
\end{equation}
where $\beta$ is defined as usual from the relation $\tan\beta = v_{2}/v_{1}$ where $v_{2}$ is the
vacuum expectation value of the neutral component of the $H^{2}$ Higgs field and $m_{Z}$ is the $Z$--boson
mass. In this case, the three sneutrinos (one for each family) are also (degenerate) mass-- eigenstates 
with squared--mass $m_{1}^{2}=m_{L}^{2} + 0.5\, m_{Z}^{2} \cos(2\beta)$. 
Here and thereafter we will denote by $m_{1}$ the mass
of the lightest sneutrino mass--eigenstate, which in order to be a CDM candidate must also be the Lightest
Supersymmetric Particle (LSP).

The experimental bounds on MSSM sneutrinos come from searches for supersymmetry at accelerators and, for
sneutrinos lighter than $m_Z/2$, from their contribution to the invisible $Z$--width:
\begin{equation}
\Delta\Gamma_{Z} =  \frac{\Gamma_{\nu}}{2}\, \left[ 1- \left(\frac{2 m_1}{m_{Z}}\right)^{2}\right]^{3/2}
\,\;\theta(m_Z - 2 m_1)
\end{equation}
where $\Gamma_{\nu} = 167 $ MeV is the $Z$--boson invisible width into one neutrino species. The bound
we adopt is: $\Delta\Gamma_{Z} < 2$ MeV \cite{PDG}. This bound constrains MSSM sneutrinos to be heavier than
about 43.7 GeV for one sneutrino, and 44.7 for 3 degenerate sneutrino \cite{PDG}.
but will be evaded in non--minimal models, as we will discuss below.

The bound that comes from accelerator physics is induced by the non--observation of the corresponding charged
sleptons. Contrary to MSSM sneutrinos, which are purely left--handed fields, charged sleptons possess both
left-- and right--handed components. The mass bound on the charged slepton depends on the assumptions made
on the balance between left-- and right--handed components (and on some more assumptions on the MSSM parameter
space and mass splitting with neutralinos) and it is usually more conservative for the right--handed fields. For selectrons current limits are: 73 GeV for $\tilde e_R$ and 107 GeV for $\tilde e_L$ \cite{PDG,Heister:2002jc}. For smuons:
94 GeV for $\tilde \mu_R$. For staus: 81.9 GeV for a generic mixing of $\tilde \tau_L$ and $\tilde \tau_R$.

These limits refer to the mass eigenstates of charged sleptons, which depend on two mass parameters: $m_L$, the
soft--mass for the left--handed SU(2) doublet $\tilde L$ and the soft--mass parameter for the 
right--handed sleptons $m_R$.
These parameters are in general matrices in flavour space, but as we discussed above we assume them as diagonal
and common over the three families. The mass matrix for sleptons is, in this case:
\begin{equation}
 {\cal M}^2_{\tilde l}  = 
\begin{pmatrix}
m^2_L + m^2_Z \cos 2\beta(\sin^2\theta_{W}-\frac{1}{2})  + m^2_{l}
& m_{l} \left(A_{l}+\mu \tan\beta\right) \cr 
m_{l}\left(A_{l}+\mu \tan\beta\right) & 
m^2_R - m^2_Z \cos 2\beta \sin^2\theta_{W} +m^2_{l}
\end{pmatrix}
\end{equation}
where $m_l$ is the mass of the partner lepton (which is negligible, except when $l=\tau$) and $A_l$ is a trilinear coupling in the soft--breaking potential. The parameter $m_L$ is common with sneutrinos, while $m_R$ is an independent parameter. Therefore, in general, a bound to the mass of the lightest charged sleptons does not directly transfer to a mass limit to the corresponding sneutrino, but depends on the relative values of $m_L$ and $m_R$ 
(and, to a lesser extent, on $A_l$ and $\mu$ for the $\tau$ case). For instance, in a minimal Supergravity
scenario (mSUGRA), the values of the parameters at the electroweak--scale are induced by their renormalization--group
equation evolution from the GUT scale and they read:
\begin{eqnarray}
m_{R}^2 & = & m_0^2 + 0.15\,m_{1/2}^2 \\
m_{L}^2 & = & m_0^2 + 0.52\,m_{1/2}^2 \end{eqnarray}
where $m_0$ and $m_{1/2}$ are defined at the GUT scale and are the common value  for the soft susy--breaking mass parameters and the common gaugino mass parameter, respectively. In this framework the lower bound on the sneutrino 
mass is 84 GeV \cite{PDG,Heister:2002jc}

The version of the MSSM we adopt in this paper is a low--energy supersymmetric extension of the Standard Model,
which does not (necessarily) invoke mSUGRA relations. This model has been widely used to study neutralino dark matter and we extend it here to develop the sector where sneutrino is the LSP and the dark matter candidate. The basic features of the model and the parameter space not directly related to the sneutrino sector, together with the experimental constraints which we adopt, are reported in Appendix 
\ref{app:mssm}. In this minimal version of the MSSM,
we assume that all the soft--mass parameters of the charged sleptons are common at the electroweak--scale, and therefore we set $m_L=m_R$. In this case, the mass bound on the charged sleptons quoted above is transferred to a
lower limit on the mass of the three degenerate sneutrinos which can be as low as the bound coming from the invisible Z width, depending on the value of $\tan\beta$. This is therefore the lower bound on the sneutrino mass in this class of models.

 \begin{table}[ht!]
\begin{center}
\begin{tabular}{|c||c|c|}
\hline
Initial States & Annihilation Products & Available Channels\\
\hline
\hline
$\tilde{\nu}_{L}\tilde{\bar{\nu}}_{L}$ &$\nu\bar{\nu}$ & $\z (s)$,$\chii (t,u) \qquad  i=1,4$\\
 & $l\bar{l}$ & $\z (s),\h(s),\H(s)$,$\chai (t,u) \qquad  i=1,2$\\
 & $q\bar{q}$ & $\z (s),\h(s),\H(s)$\\
 & $\w^+\w^-$ & $\z(s),\h(s),\H(s),\tilde{e}_{L}(t),\mbox{ 4--point} $\\
 & $\z\z$ & $\h(s),\H(s),\tilde{\nu}_{L}(t,u),\mbox{ 4--point} $\\
 & $\h\h,\H\H,\h\H$ & $\h(s),\H(s),\tilde{\nu}_{L}(t),\mbox{ 4--point} $\\
 & $\A\A$ & $\h(s),\H(s),\mbox{ 4--point} $\\
 & $\A\h,\A\H$ & $\z(s)$\\
 & $\H^+\H^-$ & $\z(s),\h(s),\H(s),\tilde{e}_{L}(t),\mbox{ 4--point} $\\
 & $\w^+\H^-$ & $\h(s),\H(s),\tilde{e}_{L}(t,u) $\\
 & $\z\h,\z\H$ & $\z(s),\tilde{\nu}_{L}(t,u)$\\
 & $\z\A$ & $\h(s),\H(s)$\\
\hline
$\tilde{\nu}_{L}\tilde{\nu}_{L}$ & $\nu\nu$ &$\chii (t,u) \qquad  i=1,4$\\
\hline
$\tilde{e}_{L}\tilde{\bar{e}}_{L}$ & $\nu\bar{\nu}$ & $\z(s),\chai (t,u) \qquad  i=1,2$\\
 & $l\bar{l}$ & $\phot(s),\z (s),\h(s),\H(s)$,$\chii (t,u) \qquad  i=1,4$\\
 & $q\bar{q}$ & $\phot(s),\z (s),\h(s),\H(s)$\\
 & $\w^+\w^-$ & $\phot(s),\z(s),\h(s),\H(s),\tilde{\nu}_{L}(t),\mbox{ 4--point} $\\
 & $\z\z$ & $\h(s),\H(s),\tilde{e}_{L}(t,u),\mbox{ 4--point} $\\
 & $\phot\phot$ & $\tilde{e}_{L}(t,u),\mbox{ 4--point}$\\
 & $\z\phot$ & $\tilde{e}_{L}(t,u),\mbox{ 4--point}$\\
 & $\h\h,\H\H,\h\H$ & $\h(s),\H(s),\tilde{e}_{L}(t,u),\mbox{ 4--point} $\\
 & $\A\A$ & $\h(s),\H(s),\mbox{ 4--point} $\\
 & $\A\h,\A\H$ & $\z(s)$\\
 & $\H^+\H^-$ & $\phot(s),\z(s),\h(s),\H(s),\tilde{\nu}_{L}(t),\mbox{ 4--point} $\\
 & $\w^+\H^-$ & $\h(s),\H(s),\tilde{\nu}_{L}(t) $\\
 & $\z\h,\z\H$ & $\z(s),\tilde{e}_{L}(t,u)$\\
 & $\z\A$ & $\h(s),\H(s)$\\
\hline
$\tilde{e}_{L}\tilde{e}_{L}$ & $ll$ & $\chii (t,u) \qquad  i=1,4$\\
\hline
$\tilde{\nu}_{L}\tilde{\bar{e}}_{L}$ & $\nu\bar{e}$ & $\w^+(s),\chii (t,u) \qquad i=1,4$\\
 & $\w^+\z$ & $\w^+(s),\tilde{e}_{L}(t),\tilde{\nu}_{L}(t),\mbox{ 4--point} $\\
& $\w^+\phot$ & $\w^+(s),\tilde{e}_{L}(t),\mbox{ 4--point}$\\
& $\w^+\h,\w^+\H$ & $\w^+(s),\H^+(s),\tilde{e}_{L}(t),\tilde{\nu}_{L}(t)$\\
& $\w^+\A$ & $\H^+(s)$\\
& $\z\H^+$ & $\H^+(s),\tilde{e}_{L}(t),\tilde{\nu}_{L}(t)$\\
& $\phot\H^+$ & $\H^+(s),\tilde{e}_{L}(t)$\\
& $\A\H^+$ & $\w^+(s),\mbox{ 4--point}$\\
\hline
$\tilde{\nu}_{L}\tilde{e}_{L}$ & $\nu_{l}l$ &$\chii (t,u) \qquad  i=1,4$\\
\hline
\end{tabular}
\end{center}
\caption{Summary of the sneutrino annihilation and coannihilation channels. For definiteness, we report here the
case of the first family.}
\label{tab:ann}
\end{table}

\EPSFIGURE[t]{./Figures/1_Standard/STD_relic_m1_standard.eps,width=0.60\textwidth}
{STD model -- Sneutrino relic abundance $\Omega h^{2}$ as a function of the sneutrino mass $m_{1}$. 
The higgs
masses for the lightest CP--even higgs is 120 GeV, for the heaviest CP--even $H$ and the CP--odd $A$
is 400 GeV. The solid (dashed) curves  refer to models with (without) gaugino universality. The vertical line 
denotes the lower bound on the
sneutrino mass coming from the invisible $Z$--width. The horizontal solid and dotted lines delimit 
the WMAP interval for cold dark matter.
\label{fig:std-omega}}

\EPSFIGURE[t]{./Figures/1_Standard/STD_direct_dama_standard.eps,width=0.60\textwidth,}
{STD model -- Sneutrino--nucleon scattering cross section $\xi \sigma^{\rm (scalar)}_{\rm nucleon}$ 
as a function of the sneutrino mass $m_{1}$, for the same set of parameters of Fig. \ref{fig:std-omega}.
The solid (dashed) curves  refer to models with (without) gaugino universality. The vertical 
line denotes the lower bound on the sneutrino mass coming from the invisible $Z$--width.  
The dashed--dotted curve shows the DAMA/NaI region, compatible with the annual 
modulation effect observed by the experiment \cite{Bernabei:2003za,Bernabei:2005hj,Bernabei:2005ca,Bernabei:2006ya,Bernabei:2007jz}.
\label{fig:std-dama}}

\EPSFIGURE[t]{./Figures/1_Standard/STD_direct_cdms_standard.eps,width=0.60\textwidth}
{STD model -- Sneutrino--nucleon scattering cross section $\xi \sigma^{\rm (scalar)}_{\rm nucleon}$ 
as a function of the sneutrino mass $m_{1}$. 
Notations are as in Fig. \ref{fig:std-dama}, except for the experimental curves which
refer here to the upper limit from the CDMS experiment \cite{ArmelFunkhouser:2005zy,Akerib:2005kh,lightdirect}, as re--evaluated in 
Ref. \cite{lightdirect} for three different galactic halo models which delimit the uncertainty band. 
The dotted line refers to the 
standard isothermal sphere with $v_0 = 220$ km s$^{-1}$ and $\rho_0 = 0.3$ GeV cm$^{-3}$. The 
upper dashed line refers to a cored--isothermal sphere with a core radius of 5 Kpc (model B1 in 
Ref. \cite{lightdirect}) and with $v_0 = 170$ km s$^{-1}$ and $\rho_0 = 0.2$ GeV cm$^{-3}$.
The lower dashed--dotted line refers to
an axisymmetric density profile with a power--law potential (model C3 
in Ref. \cite{lightdirect}) with $v_0 = 270$ km s$^{-1}$ and $\rho_0 = 1.66$ GeV cm$^{-3}$. 
\label{fig:std-cdms}}

Let us turn now to the discussion of the sneutrino phenomenology relevant for dark matter. First of all, we have calculated the sneutrino relic abundance, by taking into account all the relevant annihilation channels and
co--annihilation processes which may arise when the sleptons are close in mass to the sneutrinos. Table \ref{tab:ann}
lists all possible channels for annihilation and coannihilation, referred to the first family for definiteness.
In this minimal MSSM models, the three neutrinos are degenerate in mass: they therefore must be considered jointly
in the calculation of the relevant processes.

For general values of the model parameter space, sneutrinos may occasionally also co--annihilate with neutralinos
and/or charginos. In the models we present in this paper we have explicitely neglected this case, by considering
configurations which possess neutralinos at least 30\% heaver than the lightest sneutrino. Co--annihilation with
neutralinos and charginos is more accidental than the one with sleptons, whose masses depend on some parameters 
(mostly $m_L$) which are common with the sneutrino mass sector. The relic abundance is calculated by numerically
solving the relevant Boltzmann equation, and the thermal--average of the (co)annihilation cross section is
performed numerically as detailed in Ref. \cite{Edsjo:1997bg}. We have developed a detailed numerical code for this pourpose \cite{arina2007}.

An example of sneutrino relic abundance $\Omega h^2$ for the minimal MSSM is plotted in Fig. \ref{fig:std-omega} 
as a function of the sneutrino mass (we report also the result for values of the mass which are below
the experimental bound discussed above for the sake of the discussion in the following Sections). In this plot
we have fixed the value of the higgs masses at 120 GeV for the lightest CP--even state $h$ and at
400 GeV for the heaviest CP--even $H$ and for the CP--odd state $A$. The lightest neutralino mass 
is $m_\chi = \min (294~\mbox{GeV},1.3\, m_1)$ for the solid curve, and $m_\chi = 1.3\, m_1$ for the dashed curve. 
This second case, which possesses light neutralinos, requires gaugino--non universality \cite{Ellis:2001msa,Roszkowski:2001sb,Belanger:2004ag} in order to evade the
neutralino mass lower bound of about 50 GeV which is instead derived for gaugino universal models. This is
nevertheless a perfectly viable possibility.

The sneutrino relic abundance is typically very small \cite{Ibanez:1983kw,Hagelin:1984wv,Falk:1994es}, much lower than the cosmological range for cold dark matter derived by the WMAP analysis \cite{wmap}:
\begin{equation}
0.092\le\Omega_{\rm CDM } h^2 \le 0.124
\label{eq:wamp}
\end{equation}
This means that sneutrinos in the minimal version of MSSM are not good dark matter candidates, except for masses in a  narrow range which we determine to be 600--700 GeV, consistent with previous analyses \cite{Falk:1994es}. For all the mass range from the experimental lower bound of about $m_{\rm Z}/2$ up to 600--700 GeV sneutrinos as the LSP are cosmologically acceptable ({\em i.e.} their relic abundance is below the WMAP bound) but they are typically underabundant. The three dips in the values
of the relic abundance refer (from left to right) to the $Z$, $h$ to the degenerate $H$ and $A$ poles in the (co)annihilation cross section, which occur when $m_1$ is close to half the mass of the exchanged particle. Since the mass of the Higgs particles are not fixed in the model, but they can span from their absolute lower bound of 90 GeV (which occurs for specific values of the Higgs mixing angle) to 114 GeV (for a SM--like light Higgs)\cite{:2001xwa,Affolder:2000rg} up to several hundreds of GeV (depending on naturalness assumptions), the dips shown in Fig. \ref{fig:std-omega}
may occur at different sneutrino masses. Since $h$ is upper bounded in the MSSM at a value around 140--150 GeV, a
dip below 75 GeV is always expected. The sharp drop in $\Omega h^2$ at $m_1 \simeq 80$ GeV is instead due to the opening of the $W^-W^+$ annihilation channel.

Dark matter direct search, which relies on the possibility to detect the recoil energy of a nucleus due to
the elastic scattering of the dark matter particle off the nucleus of a low--background detector, is known to
be a strong experimental constraint for sneutrino dark matter \cite{Goodman:1984dc}. The dependence of the direct detection rate on the DM particle rests into the particle mass and the scattering cross section. For sneutrinos, coherent scattering arises due to $Z$ and Higgs exchange diagrams in the $t$--channel. The
relevant cross section on nucleus is therefore:
\begin{equation}
\label{eq:direct}
 \sigma_{\cal N}=\sigma_{\cal N}^{Z}+\sigma_{\cal N}^{h,H}
\end{equation}
The $Z$--boson exchange cross section is:
\begin{equation}
\label{eq:directz}
 \sigma_{\cal N}^{Z}=\frac{G_F^2}{2\pi}\frac{m^2_{1}m^2_{\cal N}}{(m_1+m_{\cal N})^2}
 \left[ A_{\cal N}+2(2\sin^2\theta_W-1)Z_{\cal N}\right]^2
\end{equation}
where $m_{\cal N}$ denotes the nucleus mass, $A_{\cal N}$ and $Z_{\cal N}$ are the mass number and proton number of the nucleus and $G_{F}$ is the Fermi constant. The Higgs--bosons exchange cross section is:
\begin{equation}
\label{eq:directh}
 \sigma_{\cal N}^{h,H}= \frac{m^2_{\cal N}}{4\pi (m_{1}+m_{\cal N})^2} \left[ f_p Z_{\cal N}+f_n\left(A_{\cal N}-Z_{\cal N}\right) \right]^2
\end{equation}
where $f_{p}$ and $f_{n}$ denote the effective coupling of the sneutrino to the nucleus, whose determination 
(like in the case of neutralino--nucleus scattering) is rather involved and we do not reproduce it here. Details may be found for instance in Refs. \cite{Bottino:2001dj,Ellis:2005mb}. For the case of sneutrinos, the effective couplings may be written as:
\begin{displaymath}
 f_{i}= m_N\left(\sum_{q}^{u,d,s}k_q+\sum_{Q}^{c,b,t}k_Q\right) \qquad i=n,p
\end{displaymath}
where $k_q$ and $K_Q$ are defined as:
\begin{eqnarray}
 k_q & = & f_{T_q}\sum_{j=1,2} \frac{c^j_{\snu} c^j_q}{m^2_{\h_j}}\nonumber\\
 k_Q & =& \frac{2}{27} f_{T_G}\sum_{j=1,2} \frac{c^j_{\snu} c^j_Q}{m^2_{\h_j}}
\end{eqnarray}
with $c^j_{\snu}$ denotes the sneutrino-higgs couplings, $c^i_q,c^i_Q$ are the quark-higgs couplings and $f_{T_q}$ is the nucleon mass fraction due to light quark $q$ and $f_{T_G}= 1- \sum_{q}^{u,d,s}f_{T_q}$. 
From the analyses of Ref. \cite{Bottino:2001dj,Ellis:2005mb}, we derive the following values:
$f_{T_u}=0.023$, $f_{T_d}=0.034$ and $f_{T_s}=0.14$ for the proton and 
$f_{T_u}=0.019$, $f_{T_d}=0.041$ and $f_{T_s}=0.14$ for the neutron.
We remind that these quantities are affected by a sizeable uncertainty which can increase the direct
detection cross section up to a factor of a few \cite{Bottino:2001dj}. 

Comparisons with experimental results are more easily and typically performed by using the cross section
on a single nucleon $\xi \sigma^{\rm (scalar)}_{\rm nucleon}$. Two classes of experiments are currently running. The DAMA Collaboration uses a large--mass highly--radiopure and stable 100--Kg NaI detector, which is specifically designed to search for the annual modulation effect which is expected for direct detection, as a result of the yearly relative motion of the Earth around the Sun \cite{Drukier:1986tm,Freese:1987wu}. 
The DAMA/NaI Collaboration detects an annual modulation in its low--energy
rate over a period of 7 years (107731 kg $\times$ day of total exposure). The C.L. of this effect is 6.3 
$\sigma$ \cite{Bernabei:2003za,Bernabei:2005hj,Bernabei:2005ca,Bernabei:2006ya,Bernabei:2007jz}. When interpreted as due to a relic particle interaction, the allowed region shown in Fig. \ref{fig:std-dama}, whose contour is outlined by the dot--dashed curve, arises. The allowed region shown in Fig. 
\ref{fig:std-dama} has been calculated by taking into account the astrophysical uncertainties arising from
galactic halo modeling \cite{Belli:2002yt}. This is currently the only experiment which is designed to address the annual modulation effect. Further insight into the annual modulation effect are expected from the future results of the DAMA/LIBRA experiment, which is currently running with a mass of 250 Kg \cite{Bernabei:2006kc}.

A second class of experiments do not attempt to exploit the annual modulation signature, but instead
rely on the development of background--rejection techniques in order to reduce the background to the
sum of
neutrons plus dark matter recoils. These type of experiments provide upper bounds on the scattering cross
section. In Fig. \ref{fig:std-cdms} the upper bound from the CDMS experiment (72 kg $\times$ day with a
1.5 Kg Ge plus 0.6 Kg Si detector) \cite{ArmelFunkhouser:2005zy,Akerib:2005kh}. For three different galactic halo models, is shown as calculated in Ref. \cite{lightdirect}. The upper experimental bound has to be regarded as a conservative limit, and the span of the three curves reflects the uncertainty on the galactic halo modeling. Recently results from the XENON10 Collaboration (15 Kg active liquid Xe) \cite{Angle:2007uj} have been presented (for the standard isothermal model only), with an exposure of 136 kg $\times$ day: these limits appear to be more stringent than the CDMS ones by a factor which ranges from 2 to 10, depending on the mass. For a recent and detailed review on direct detection experiments, see Refs. \cite{belli,gerbier}.

Due to the different nature of the experimental results (positive indication vs. upper limits) and in
the absence of a solid criterion to consistently combine the various experimental results (which are
derived from different techniques and whose treatment would require to correctly quantify
uncertainties both of theoretical origin and of experimental nature, which are not under our
control) we will present our direct detection analysis by comparing our results separately with the DAMA/NaI region with the CDMS upper bounds (for which we consider the upper curve of Fig. \ref{fig:std-cdms} as the
conservative upper limit). In order not to exceed with the number of figures (already very large)
we will show our results by alternating them with either the DAMA/NaI region or the CDMS curves. The minimal MSSM is
the only case where we present both for the same theoretical predictions.

When compared with the DAMA/NaI annual modulation region in Fig. \ref{fig:std-dama} we see that direct detection is indeed a strong constraint on sneutrino dark matter in the minimal MSSM 
\cite{Falk:1994es}, 
but some very
specific configurations are still viable and could explain the annual modulation effect. We have to consider that, whenever the dark matter particle is subdominant in the Universe, also its local density $\rho_{0}$ 
in the
Galaxy is very likely reduced with respect to the total dark matter density. This means that the dominant component of dark matter is not the sneutrino, but still sneutrinos form a small amount of dark matter
and may be eventually detectable. In this case we rescale the local sneutrino abundance by means of the
usual factor:
\begin{equation}
\xi = \min (1, \Omega h^{2}/0.092)
\end{equation}
where the value 0.092 is the lower bound for CDM abundance from WMAP, given by Eq. (\ref{eq:wamp}). This situation is the same we encounter for neutralino dark matter, which also may be subdominant but nevertheless interesting to study as a relic from the early Universe. Since the direct detection rate is linearly dependent on the local dark matter density, and the experimental bounds are derived for a dominant dark matter candidate, the
theoretical scattering cross section has to be rescaled by the factor $\xi$. In the following, when we will
discuss indirect detection signals, $\xi$ will enter quadratically in the determination of the theoretical
fluxes, since in that case the signals depend on the square of the local density.

Comparison of Fig. \ref{fig:std-omega} and \ref{fig:std-dama} shows that the MSSM sneutrino is only viable
when strongly subdominant. The mass range 600-700 GeV where the sneutrino possesses a relic abundance in the
cosmologically relevant range is clearly excluded by direct detection. On the other hand, MSSM sneutrinos are still viable, but they require the special condition to stay on the pole of the annihilation cross section through Higgs exchange. This may appear a fine--tuned condition, but it is no different from the
case of relic neutralinos in mSUGRA models, where the relic abundance is acceptable mostly in very specific regions of parameter space, where strong coannihilation occurs or the pole of the $A$--boson exchange is met \cite{Ellis:2001msa,Roszkowski:2001sb,Belanger:2004ag}. We also notice that, since the masses of the Higgs bosons can vary, the dips in 
$\xi \sigma^{\rm (scalar)}_{\rm nucleon}$ (due to the rescaling factor) may occur at any value of the
sneutrino mass. This means that all the mass range from the lower bound on $m_{1}$ up to about 200 GeV
may be compatible with the DAMA/NaI results.

Fig. \ref{fig:std-cdms} compares our calculations with the CDMS upper limit  \cite{ArmelFunkhouser:2005zy,Akerib:2005kh}. For a standard isothermal halo
the bound completely excludes the possibility of sneutrino dark matter. Some very marginal room may still be present for the conservative upper limit, when the sneutrino mass matches the pole condition for annihilation through a light Higgs. Nevertheless, this option appears to be very marginal (we remind that
only the mass range above 70 GeV is allowed by accelerator searches).

The results of this analysis are in agreement with the previous analyses which excluded the MSSM sneutrino
as a dominant component of dark matter \cite{Falk:1994es,Beck:1993sb}. Differently from that studies, we explicitely consider subdominant sneutrinos as viable relic particles to be explored and we show that there is still an open
possibility for MSSM sneutrinos, although very marginal.

\section{Models with a right--handed sneutrino field}\label{sec:lr}

\EPSFIGURE[t]{./Figures/2_LeftRight_Analysis/LR_mn_m1_ml120.eps,width=0.60\textwidth}
{LR models -- Sneutrino mass $m_{1}$ as a function of the right--handed soft--mass $m_{N}$, for
different values of the off--diagonal parameter $F^2$: the [black] solid, [blue] dashed,
[green] dot--dashed and [red] dot--dot--dashed refer to $F^2 = 10, 10^{2}, 10^{3}, 10^{4}$ GeV$^{2}$, 
respectively. The left--handed soft--mass $m_{L}$ is fixed here at the value of 120 GeV.
\label{fig:lr-mnm1}}

\EPSFIGURE[t]{./Figures/2_LeftRight_Analysis/LR_mn_sinth_ml120.eps,width=0.60\textwidth}
{LR models -- Sneutrino left--right mixing angle $\theta$ as a function of the right--handed 
soft--mass $m_{N}$, for different values of the off--diagonal parameter $F^2$: the [black] solid, [blue] dashed,[green] dot--dashed and [red] dot--dot--dashed refer to $F^2 = 10, 10^{2}, 10^{3}, 10^{4}$ GeV$^{2}$, 
respectively. The left--handed soft--mass $m_{L}$ is fixed here at the value of 120 GeV.
\label{fig:lr-mnsinth1}}

\EPSFIGURE[t]{./Figures/2_LeftRight_Analysis/LR_mn_sinth_mltev.eps,width=0.60\textwidth}
{LR models -- Sneutrino left--right mixing angle $\theta$ as a function of the right--handed 
soft--mass $m_{N}$. Notations are as in Fig. \ref{fig:lr-mnsinth1}. 
The left--handed soft--mass $m_{L}$ is fixed here at the value of 1 TeV.
\label{fig:lr-mnsinth2}}

\EPSFIGURE[t]{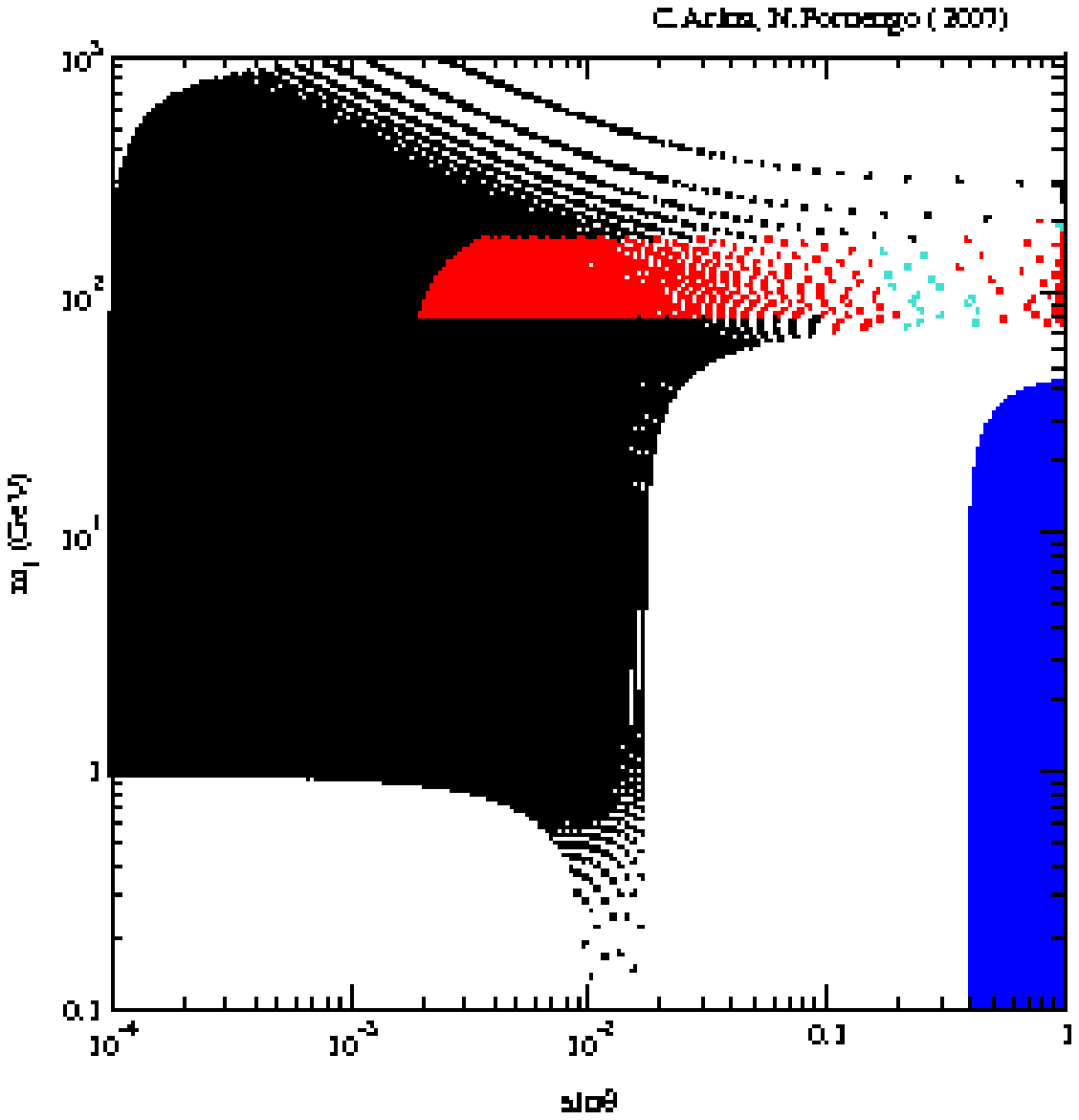,width=0.60\textwidth}
{LR models -- Sneutrino mass $m_{1}$ vs. the sneutrino left--right mixing angle $\theta$
for $F^{2} = 10^{2}$ GeV$^{2}$ and for a scan of the soft--mass parameters $m_{L}$ and $m_{N}$ in 
the ranges: $120~\mbox{GeV} \leq m_{L} \leq 1~\mbox{TeV}$ and 
$1~\mbox{GeV} \leq m_{N} \leq 1~\mbox{TeV}$.
The rightmost full [blue] area denotes the region which is excluded by the
invisible $Z$--width. The black darker area denotes the region where the sneutrino
relic abundance is in excess of the WMAP bound; the other areas 
cover the region where the relic abundance is acceptable; in addition, in the lighter [light blue] area, the
direct--detection scattering cross--section is inside the DAMA/NaI annual modulation region.
The white areas are not covered by models for the values of the parameters adopted here.
\label{fig:lr-m1sinth1}}

\EPSFIGURE[t]{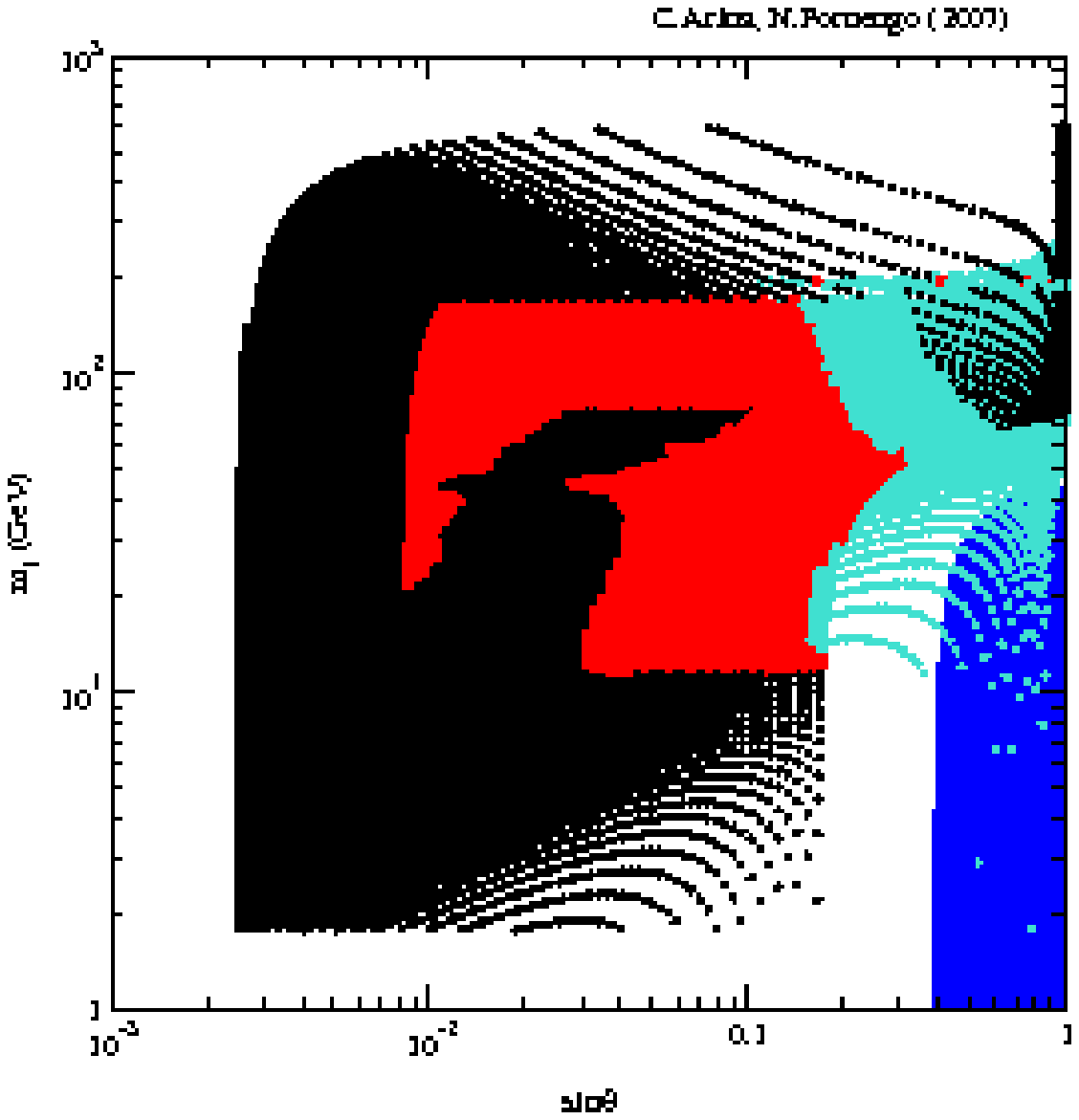,width=0.60\textwidth}
{LR models -- Sneutrino mass $m_{1}$ vs. the sneutrino left--right mixing angle $\theta$
for $F^{2} = 10^{3}$ GeV$^{2}$ and for a scan of the soft--mass parameters $m_{L}$ and $m_{N}$ in 
the ranges: $120~\mbox{GeV} \leq m_{L} \leq 1~\mbox{TeV}$ and 
$1~\mbox{GeV} \leq m_{L} \leq 1~\mbox{TeV}$. Notations are as in Fig. \ref{fig:lr-m1sinth1}, except
for the lighter [light blue] area where now the direct--detection scattering cross--section is 
below the CDSM conservative upper bound.
\label{fig:lr-m1sinth2}}

\EPSFIGURE[t]{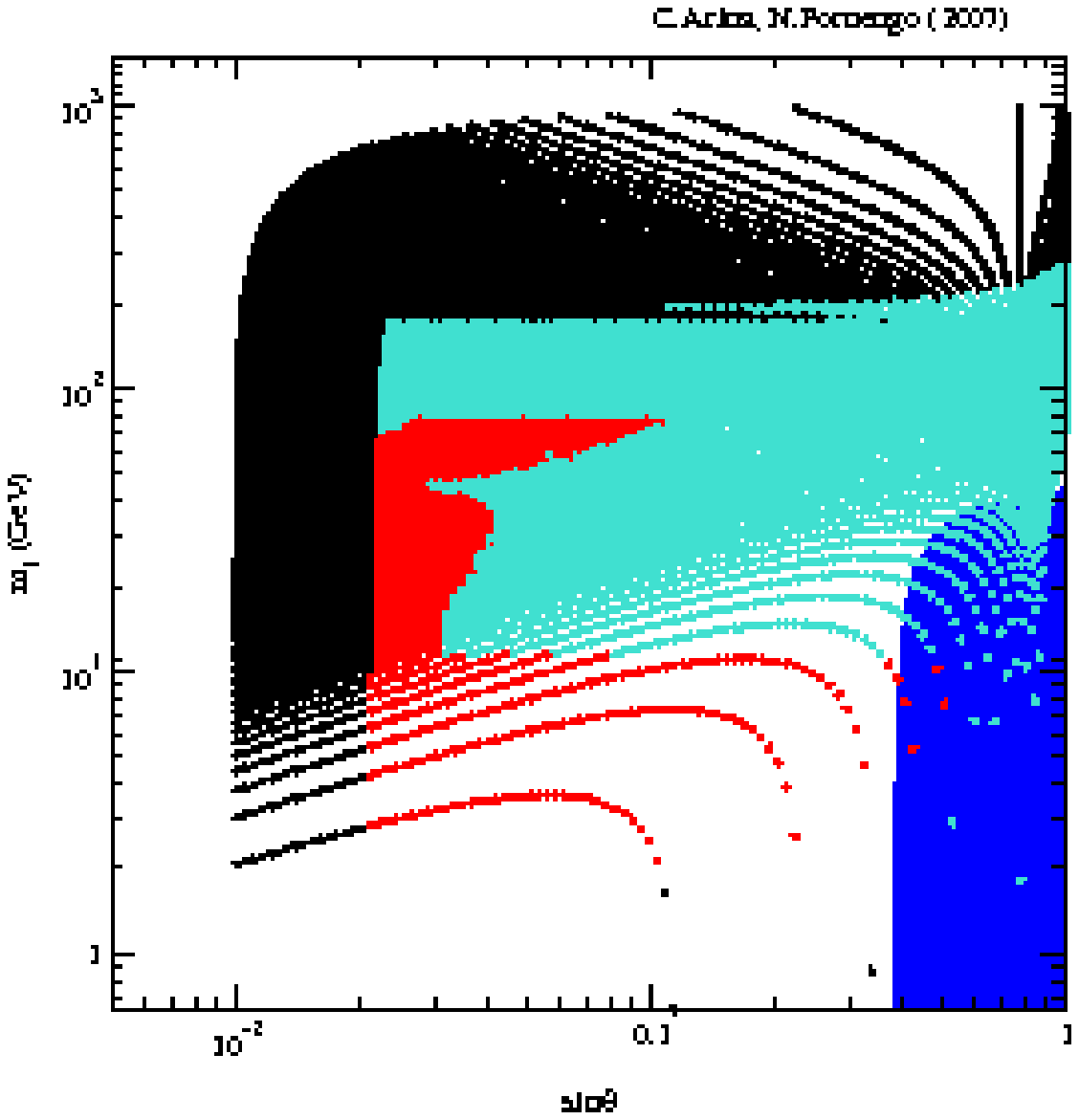,width=0.60\textwidth}
{LR models -- Sneutrino mass $m_{1}$ vs. the sneutrino left--right mixing angle $\theta$
for $F^{2} = 10^{4}$ GeV$^{2}$ and for a scan of the soft--mass parameters $m_{L}$ and $m_{N}$ in 
the ranges: $120~\mbox{GeV} \leq m_{L} \leq 1~\mbox{TeV}$ and 
$1~\mbox{GeV} \leq m_{L} \leq 1~\mbox{TeV}$. Notations are as in Fig. \ref{fig:lr-m1sinth1}, except
for the lighter [light blue] area where now the direct--detection scattering cross--section is 
below the CDSM conservative upper bound.
\label{fig:lr-m1sinth3}}

\EPSFIGURE[t]{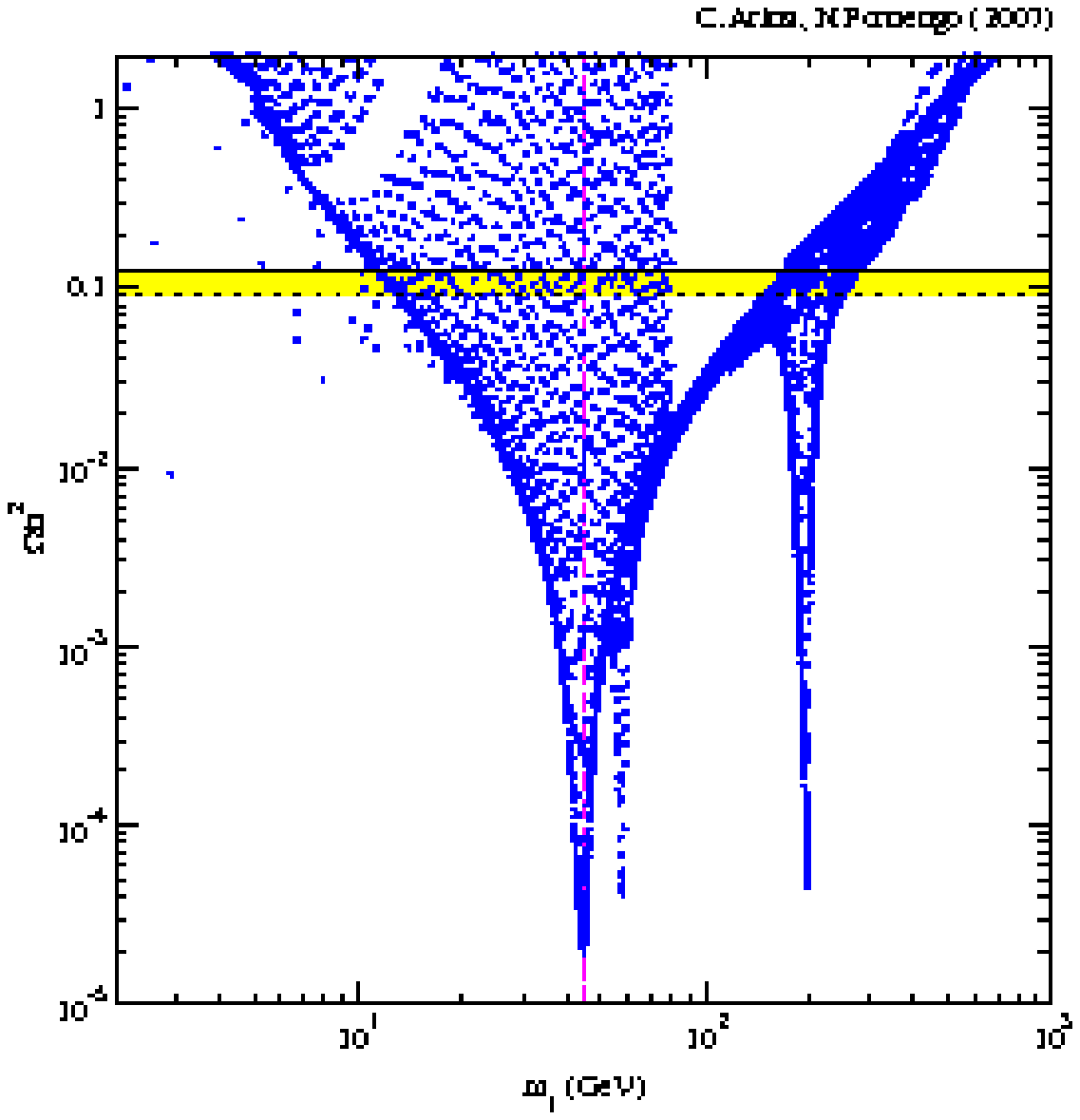,width=0.60\textwidth}
{LR models -- Scatter plot of the sneutrino relic abundance $\Omega h^{2}$ as a function of 
the sneutrino mass $m_{1}$, for supersymmetric models with the lightest neutralino 30\% heavier than the
sneutrino (which implies gaugino non--universality when the neutralino mass is light) and higgs
masses of 120 GeV for the lightest CP--even higgs $h$ and 400 GeV for the heaviest CP--even $H$
and the CP--odd $A$. The sneutrino parameters are varied as follows: 
$120~\mbox{GeV} \leq m_{L} \leq 1~\mbox{TeV}$, $1~\mbox{GeV} \leq m_{N} \leq 1~\mbox{TeV}$ and
$10~\mbox{GeV}^{2} \leq F^{2} \leq 10^{4}~\mbox{GeV}^{2}$. All the models shown in the plot 
are acceptable from the point of view of all experimental constraints. The horizontal solid and 
dotted lines delimit the WMAP interval for cold dark matter.
\label{fig:ls-omega}}

\EPSFIGURE[t]{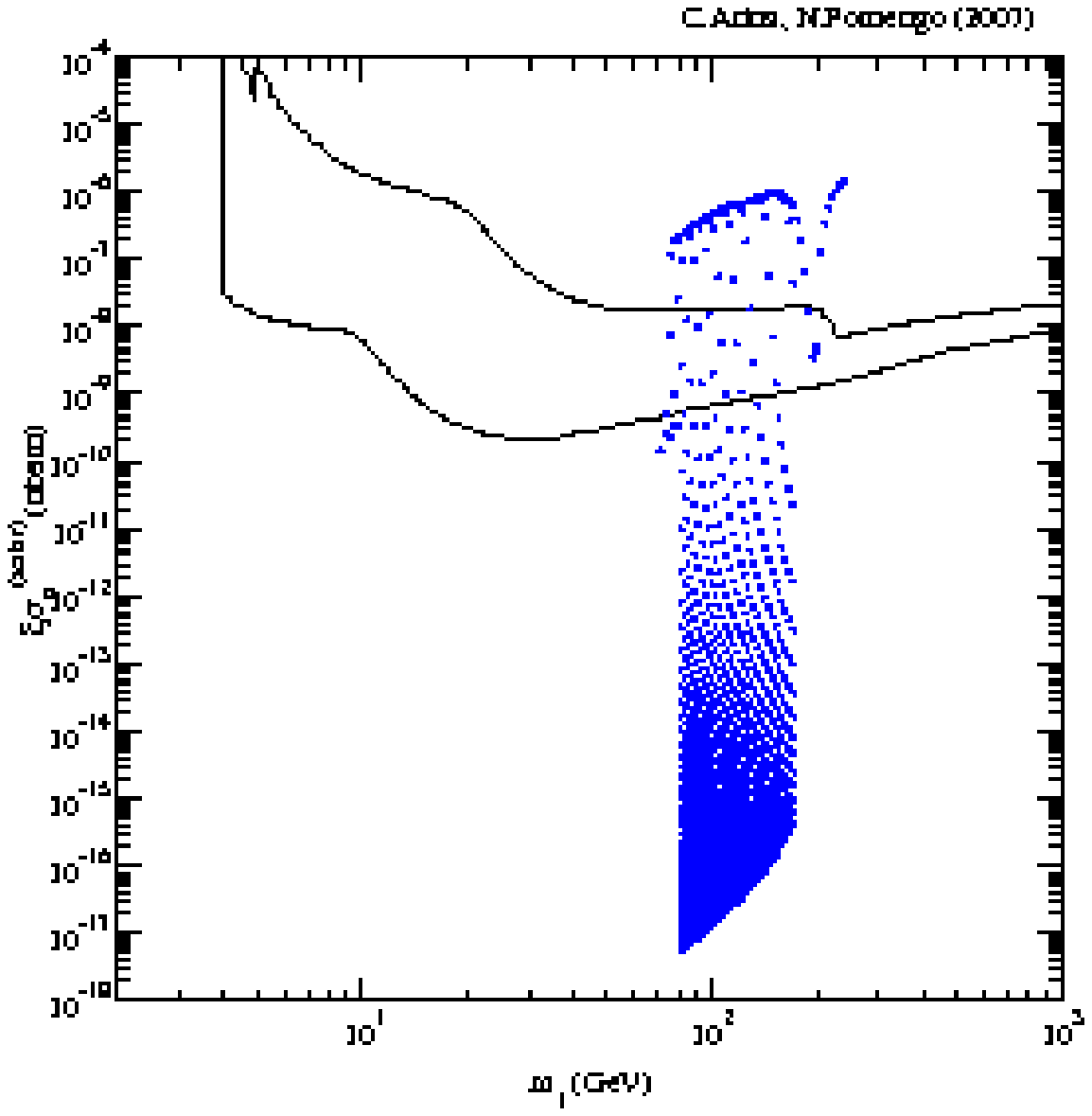,width=0.60\textwidth}
{LR models -- Sneutrino--nucleon scattering cross section $\xi \sigma^{\rm (scalar)}_{\rm nucleon}$ 
as a function of the sneutrino mass $m_{1}$, for $F^{2} = 10^{2}$ GeV$^{2}$ and for a scan of 
the soft--mass parameters $m_{L}$ and $m_{N}$ in the ranges: 
$120~\mbox{GeV} \leq m_{L} \leq 1~\mbox{TeV}$ and $1~\mbox{GeV} \leq m_{N} \leq 1~\mbox{TeV}$.
The other model parameters are as in Fig. \ref{fig:ls-omega}.
The solid curve shows the DAMA/NaI region, compatible with the annual 
modulation effect observed by the experiment \cite{Bernabei:2003za,Bernabei:2005hj,Bernabei:2005ca,Bernabei:2006ya,Bernabei:2007jz}. 
All the configurations shown possess relic abundance compatible with the WMAP bound.
\label{fig:lr-direct1}}

\EPSFIGURE[t]{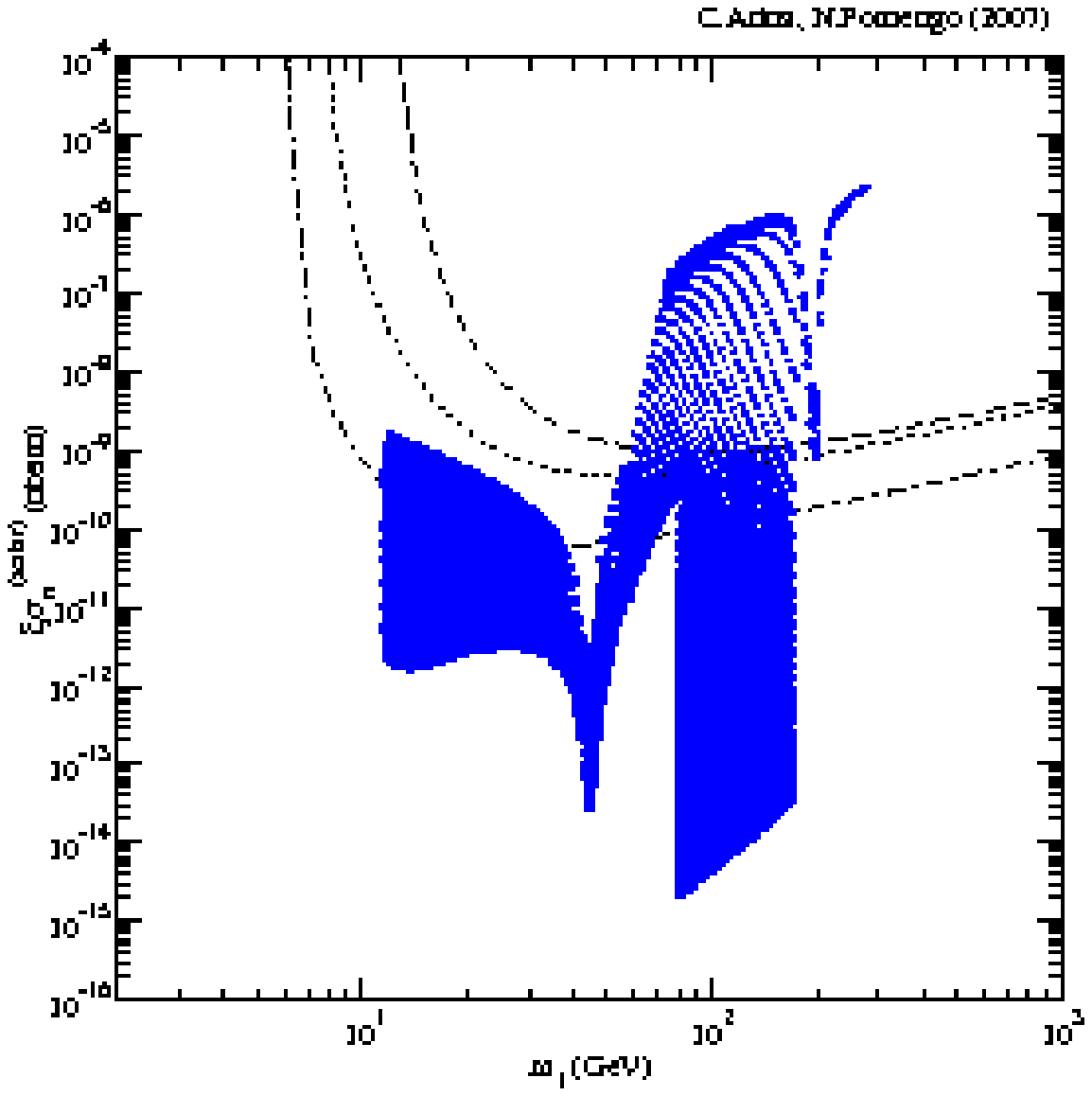,width=0.60\textwidth}
{LR models -- Sneutrino--nucleon scattering cross section $\xi \sigma^{\rm (scalar)}_{\rm nucleon}$ 
as a function of the sneutrino mass $m_{1}$, for $F^{2} = 10^{3}$ GeV$^{2}$,
$120~\mbox{GeV} \leq m_{L} \leq 1~\mbox{TeV}$ and $1~\mbox{GeV} \leq m_{N} \leq 1~\mbox{TeV}$.
The other model parameters are as in Fig. \ref{fig:ls-omega}.
The dashed, dotted and dot--dashed curves denote the CDMS upper bounds  \cite{ArmelFunkhouser:2005zy,Akerib:2005kh}, as in 
Fig. \ref{fig:std-cdms}.
\label{fig:lr-direct2}}

\EPSFIGURE[t]
{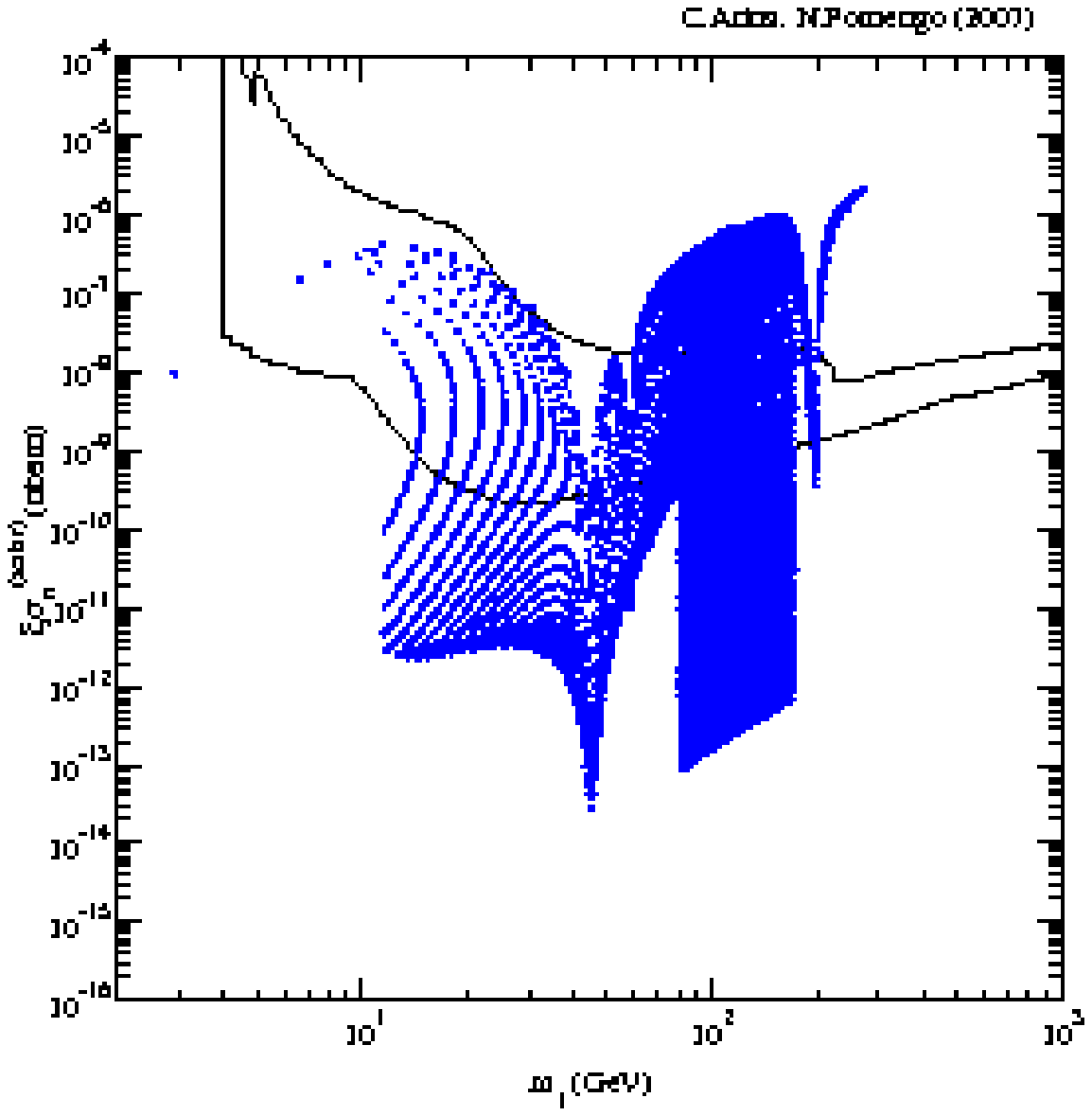,width=0.60\textwidth}
{LR models -- Sneutrino--nucleon scattering cross section $\xi \sigma^{\rm (scalar)}_{\rm nucleon}$ 
as a function of the sneutrino mass $m_{1}$, for $F^{2} = 10^{4}$ GeV$^{2}$,
$120~\mbox{GeV} \leq m_{L} \leq 1~\mbox{TeV}$ and $1~\mbox{GeV} \leq m_{N} \leq 1~\mbox{TeV}$.
The other model parameters are as in Fig. \ref{fig:ls-omega}.
The solid curve shows the DAMA/NaI region, compatible with the annual 
modulation effect observed by the experiment \cite{Bernabei:2003za,Bernabei:2005hj,Bernabei:2005ca,Bernabei:2006ya,Bernabei:2007jz}.
\label{fig:lr-direct3}}

\EPSFIGURE[t]{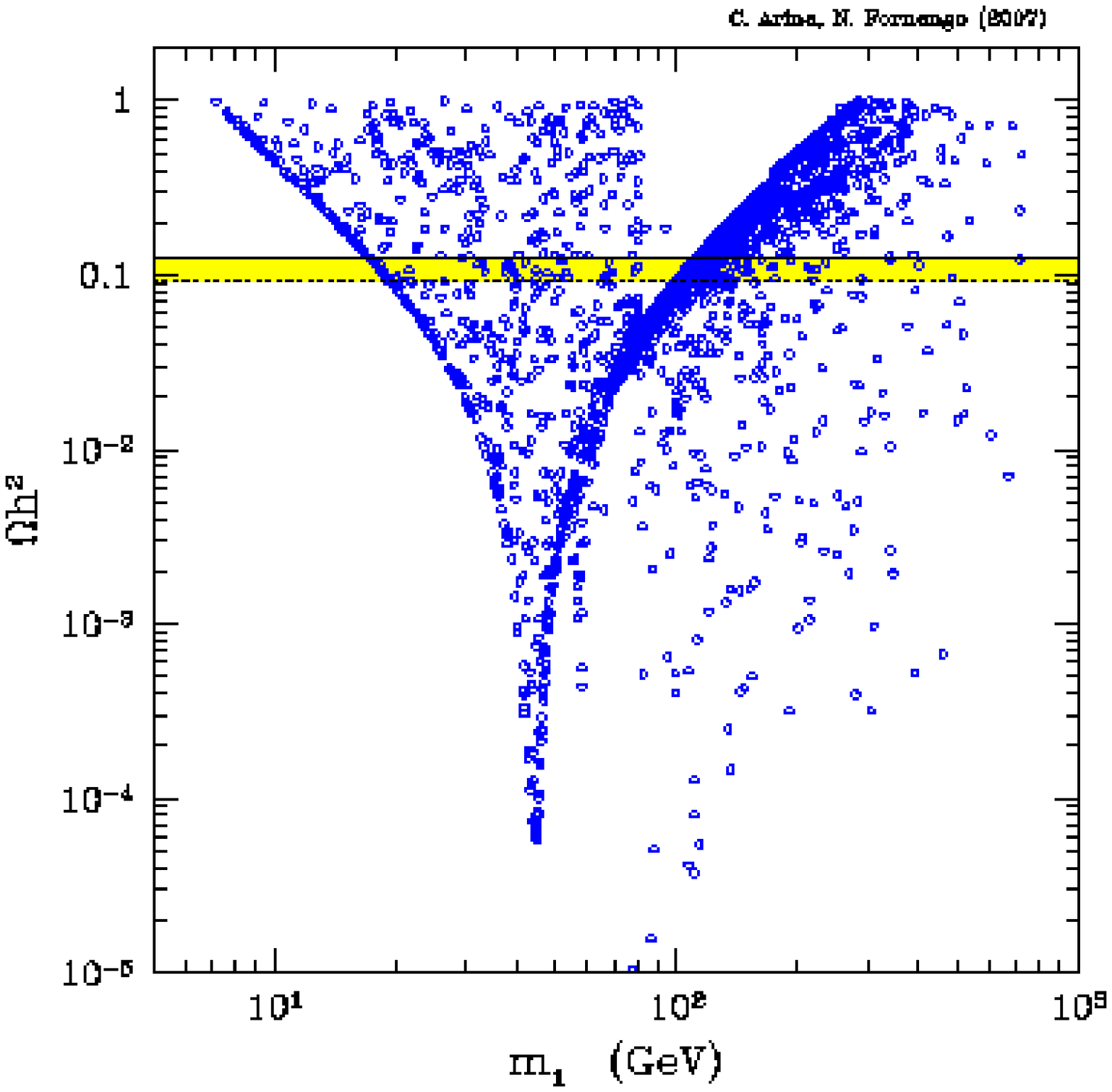,width=0.80\textwidth}
{LR models -- Scatter plot of the sneutrino relic abundance $\Omega h^{2}$ as a function of 
the sneutrino mass $m_{1}$, for a full scan of the supersymmetric parameter space, as
explained in the text.
The sneutrino parameters are varied as follows: $100~\mbox{GeV} \leq m_{L} \leq 3~\mbox{TeV}$, 
$1~\mbox{GeV} \leq m_{N} \leq 1~\mbox{TeV}$ and
$1~\mbox{GeV}^{2} \leq F^{2} \leq 10^{6}~\mbox{GeV}^{2}$.
The horizontal solid and 
dotted lines delimit the WMAP interval for cold dark matter.
\label{fig:lr-omegascan}}

\EPSFIGURE[t]
{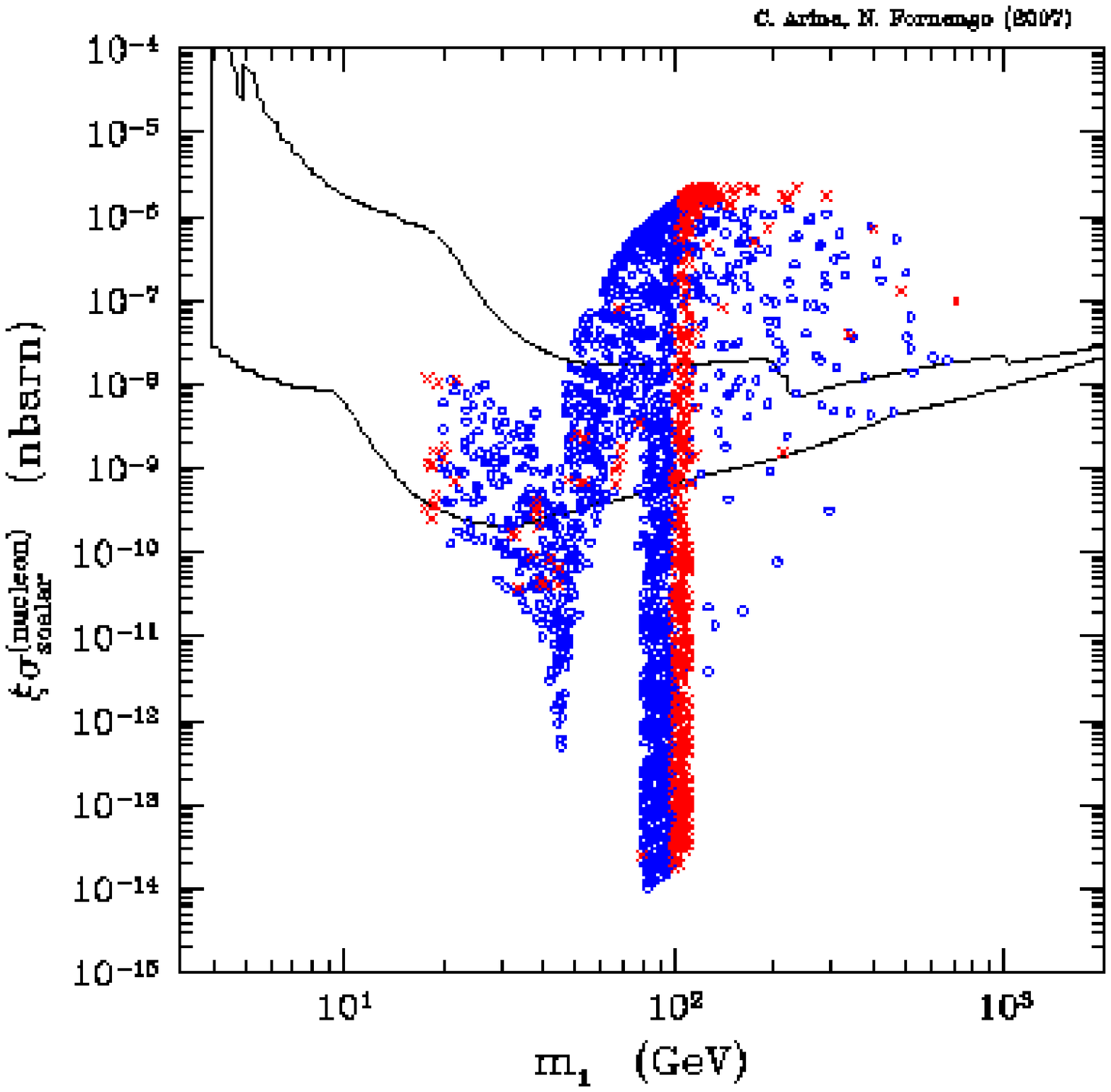,width=0.80\textwidth}
{LR models -- Sneutrino--nucleon scattering cross section $\xi \sigma^{\rm (scalar)}_{\rm nucleon}$ 
as a function of the sneutrino mass $m_{1}$, for a full scan of the supersymmetric parameter space.
Parameters are varied as in Fig. \ref{fig:lr-omegascan}. [Red] crosses refer to models with sneutrino relic abundance
in the cosmologically relevant range; [blue] open circles refer to cosmologically subdominant 
sneutrinos. The solid curve shows the DAMA/NaI region, compatible with the annual 
modulation effect observed by the experiment \cite{Bernabei:2003za,Bernabei:2005hj,Bernabei:2005ca,Bernabei:2006ya,Bernabei:2007jz}.
\label{fig:lr-directscan}}

\EPSFIGURE[t]{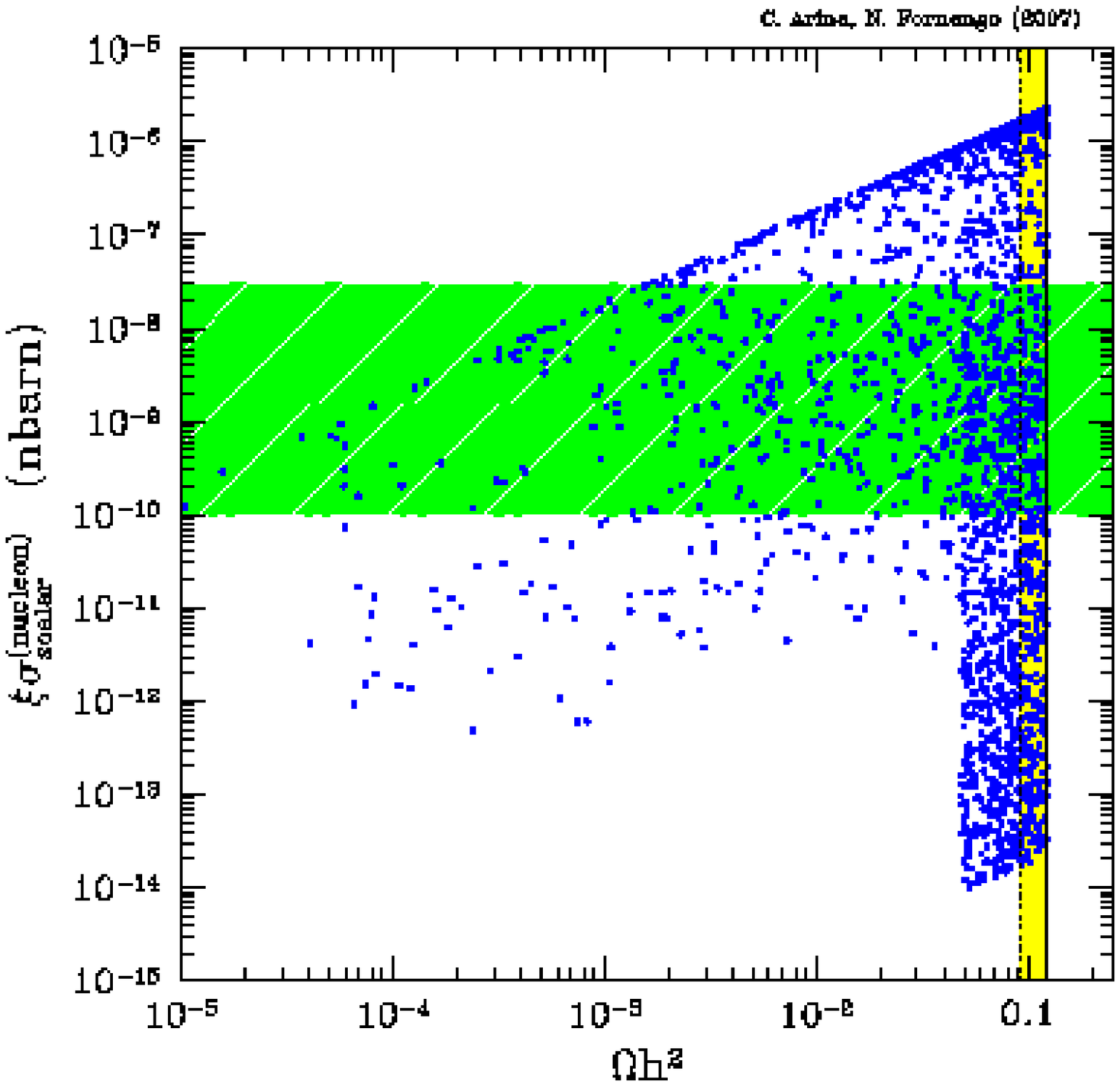,width=0.80\textwidth}
{LR models -- Sneutrino--nucleon scattering cross section $\xi \sigma^{\rm (scalar)}_{\rm nucleon}$ 
vs. the sneutrino relic abundance $\Omega h^{2}$, for a full scan of the supersymmetric 
parameter space. Parameters are varied as in Fig. \ref{fig:lr-omegascan}. The horizontal [green] band denotes the current sensitivity of direct detection experiments; the vertical [yellow] band
delimits the WMAP interval for cold dark matter.
\label{fig:lr-sigmaomega}}

\EPSFIGURE[t]{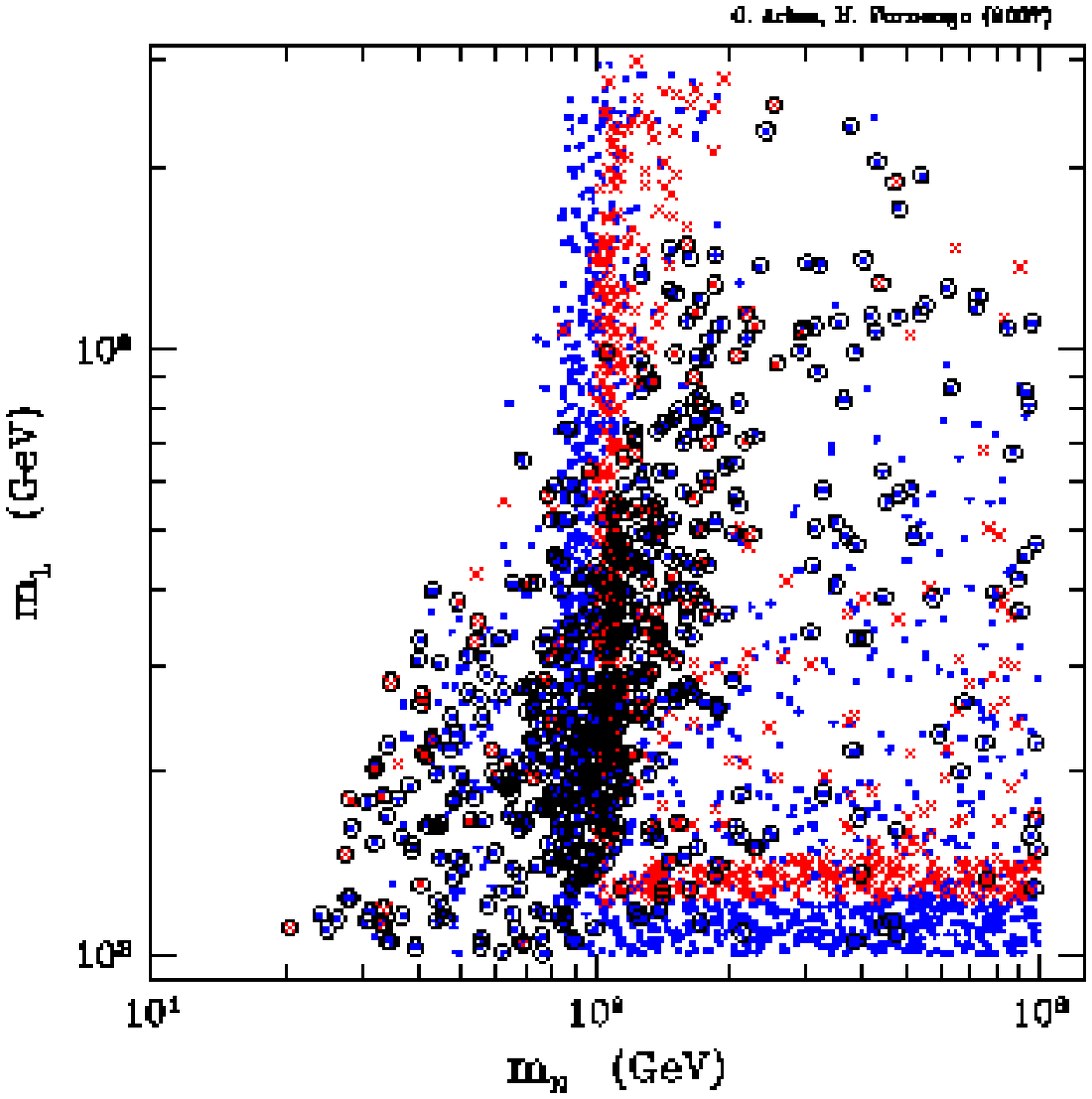,width=0.80\textwidth}
{LR models -- Distribution of cosmologically acceptable models in the  $m_{N}$ -- $m_{L}$ plane,
when a full scan of the supersymmetric parameter space is performed. Parameters are varied 
as in Fig. \ref{fig:lr-omegascan}. [Red] crosses refer to models with sneutrino relic abundance
in the cosmologically relevant range; [blue] dots refer to cosmologically subdominant 
sneutrinos; open dots mark the configurations which have a direct--detection cross section
in the current sensitivity range.
\label{fig:lr-mnml}}

\EPSFIGURE[t]{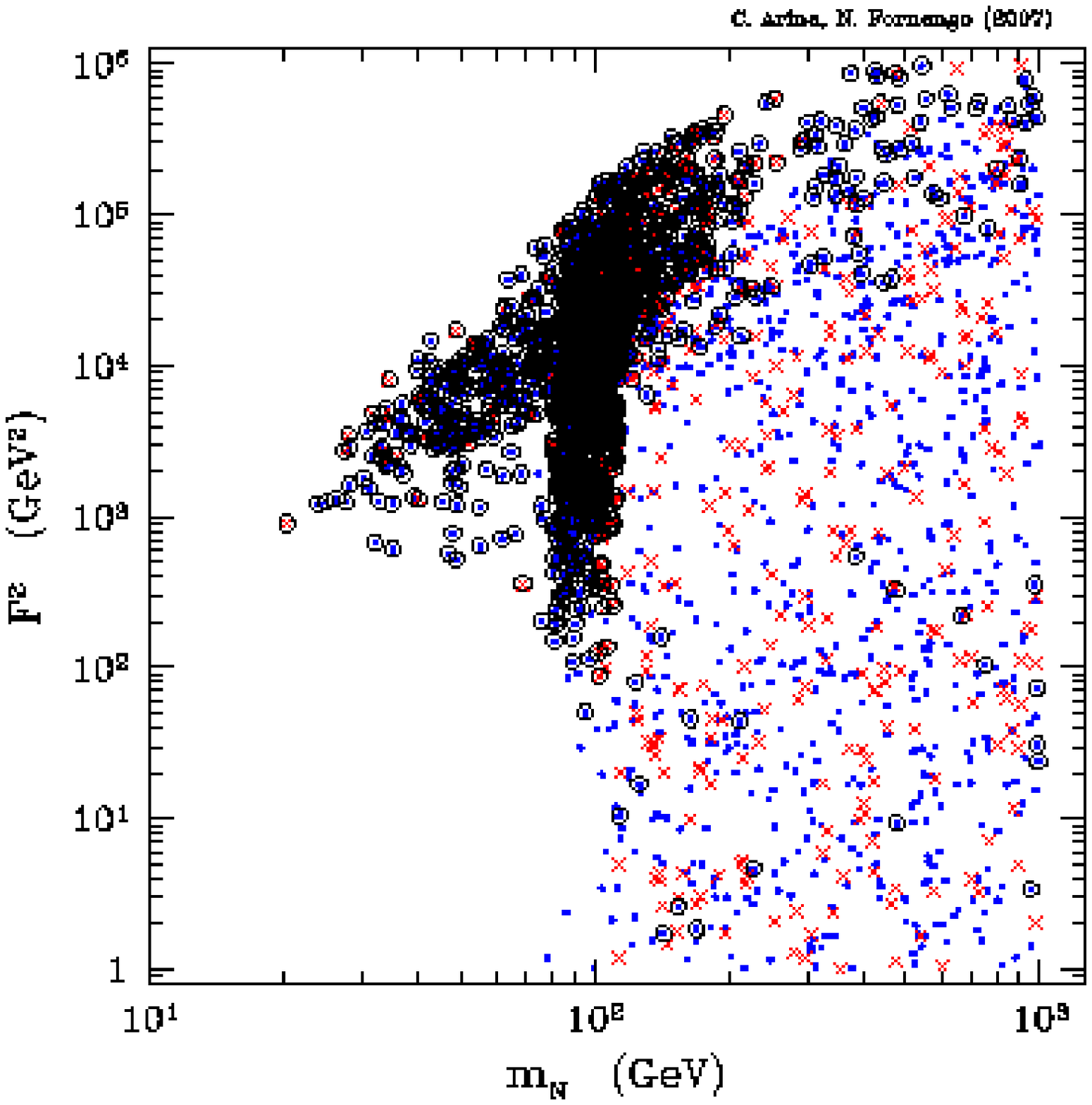,width=0.80\textwidth}
{LR models -- Distribution of cosmologically acceptable models in the  $m_{N}$ -- $F^{2}$ plane,
when a full scan of the supersymmetric parameter space is performed. Parameters are varied 
as in Fig. \ref{fig:lr-omegascan}. [Red] crosses refer to models with sneutrino relic abundance
in the cosmologically relevant range; [blue] dots refer to cosmologically subdominant 
sneutrinos; open dots mark the configurations which have a direct--detection cross section
in the current sensitivity range.
\label{fig:lr-mnf}}

\EPSFIGURE[t]{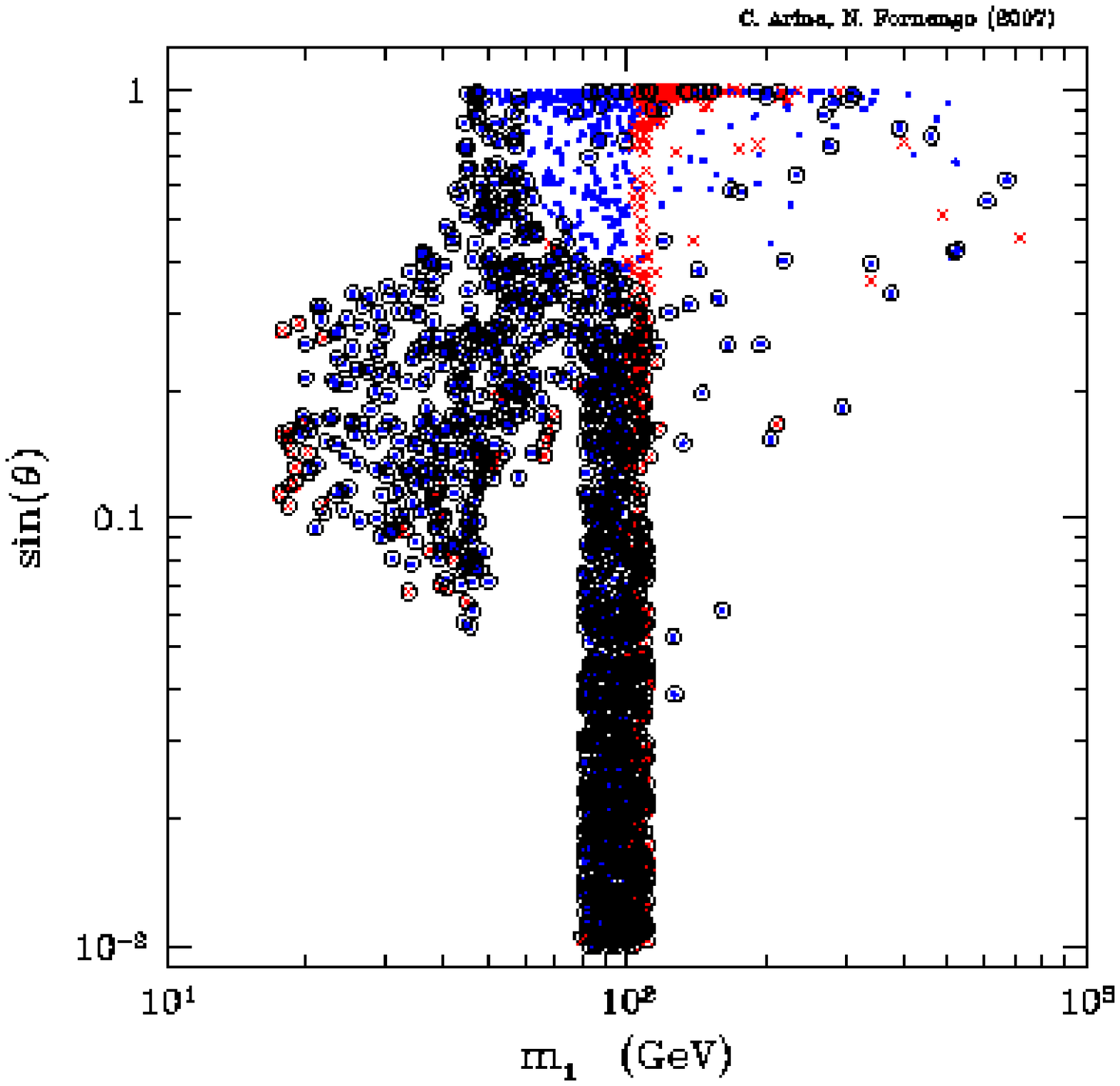,width=0.80\textwidth}
{LR models -- Distribution of cosmologically acceptable models in the  $m_{1}$ -- $\sin\theta$ plane,
when a full scan of the supersymmetric parameter space is performed. Parameters are varied 
as in Fig. \ref{fig:lr-omegascan}. [Red] crosses refer to models with sneutrino relic abundance
in the cosmologically relevant range; [blue] dots refer to cosmologically subdominant 
sneutrinos; open dots mark the configurations which have a direct--detection cross section
in the current sensitivity range.
\label{fig:lr-m1sinth}}

In this class of models, the neutrino/sneutrino sector is enlarged by the inclusion of right--handed superfields 
$\hat N^{I}$, whose scalar component are right--handed sneutrino fields $\tilde N^{I}$ \cite{Arkani-Hamed:2000bq,Grossman:1997is,Smith:2001hy,Smith:2002af,Tucker-Smith:2004jv}. We call this class
of models as ``LR models".
The right--handed fermionic
components lead to Dirac--type masses $(m_D)^I$ for the massive neutrinos. The relevant terms in the superpotential are now:
\begin{equation}
W = \epsilon_{ij} (\mu \hat H^{1}_{i} \hat H^{2}_{j} - Y_{l}^{IJ} \hat H^{1}_{i} \hat L^{I}_{j} \hat R^{J}
+ Y_{\nu}^{IJ} \hat H^{2}_{i} \hat L^{I}_{j} \hat N^{J} )
\end{equation}
where $Y_{\nu}^{IJ}$ is a matrix, which we again choose real and diagonal, from which the Dirac mass of neutrinos are obtained $m_D^{I} = v_{2}Y_{\nu}^{II}$. Also the soft--breaking potential gets modified:
\begin{eqnarray}
V_{\rm soft} &=& (M_{L}^{2})^{IJ} \, \tilde L_{i}^{I \ast} \tilde L_{i}^{J} + 
(M_{N}^{2})^{IJ} \, \tilde N^{I \ast} \tilde N^{J} - 
 [\epsilon_{ij}(\Lambda_{l}^{IJ} H^{1}_{i} \tilde L^{I}_{j} \tilde R^{J} + 
\Lambda_{\nu}^{IJ} H^{2}_{i} \tilde L^{I}_{j} \tilde N^{J})  + \mbox{h.c.}] \nonumber\\
&&
\end{eqnarray}
where we take both the matrices $M_{N}^{2}$ and $\Lambda_{\nu}^{IJ}$ real and diagonal, as we do 
for $M_{L}^{2}$ and $\Lambda_{l}^{IJ}$. The diagonal common entries for $M_{N}^{2}$ are denoted as 
$m_N^2$.

The sneutrino mass--term potential is now (we omit the family index for simplicity):
\begin{equation}
V_{\rm mass} = \left[ m_{L}^{2} + \frac{1}{2} m_{Z}^{2} \cos(2\beta) + m_D^2\right] \snu_{L}^{\ast} \snu_{L}
+ [m_N^2 + m_D^2] \tilde N^{\ast} \tilde N + F^2 (\snu_L^{\ast} \tilde N + \tilde N^{\ast} \snu_L)
\end{equation}
The parameter $F^2$, which mixes the left-- and right--handed sneutrino fields is:
\begin{equation}
F^2 = v \Lambda_\nu \sin\beta - \mu m_D\rm cotg\beta
\label{eq:f2}
\end{equation}
In the basis defined by the vector $\Phi^\dag=(\snu_L^\ast,\, \tilde N^\ast)$, we can define the sneutrino mass potential as: 
\begin{eqnarray}
V_{\rm mass} =\frac{1}{2}\, \Phi^{\dag}_{LR}\, \mathcal{M}^2_{LR}\, \Phi_{LR}
\end{eqnarray}
where the squared--mass matrix $\mathcal{M}^2_{LR}$ is:
\begin{eqnarray}
\mathcal{M}^2_{LR} & = & 
  \begin{pmatrix} m^2_L + \frac{1}{2} m_{Z}^{2} \cos(2\beta) + m_D^2  &  F^2 \cr
                  F^2 & m^2_N + m_D^2 
  \end{pmatrix}
  \label{eq:masslr}
\end{eqnarray}
The Dirac neutrino mass is small, and can be safely neglected. The parameter $m_N$ in general is 
independent of the other mass parameters, especially $m_{L}$, which instead is related to
the charged leptons masses, as discussed in the previous Section. We are therefore allowed to vary freely
$m_N$, and whenever $m_N$ is small enough, sneutrinos lighter than those encountered in the previous
Section are in principle viable. As long as the left--handed mass $m_{L}$ is compatible with the mass
lower bounds on the charged leptons (which occurs for $m_L\gtrsim 80-90$ GeV) $m_{1}$ can be small, provided that $m_{N} \ll m_{L}$, without
entering in conflict with accelerator bounds. In this case, light sneutrinos may arise, and the only relevant limit which remains is the one provided by the invisible $Z$--width, which we discuss in a moment.

The off--diagonal term $F^2$ is relevant for the mixing among the mass eigenstates, obtained by diagonalizing $\mathcal{M}^{2}_{LR}$. For $\Lambda_\nu$ aligned along the Yukawa matrix $Y_\nu$,
{\em i.e.} for $\Lambda_\nu = \eta Y_\nu$, $F^{2}$ is necessarily very small as compared to the diagonal
entries (especially the element $(\mathcal{M}^2_{LR})_{11}$ of the matrix), because in that case $F^2$ 
is proportional to the
neutrino mass $m_{D}$, and is therefore negligible since both $v$ and $\mu$ are electroweak--scale
parameters. However, $\Lambda_\nu$ is in general a free parameter. In this case $F^{2} \simeq v \Lambda_\nu \sin\beta$ may naturally be of the order of the other entries of the matrix, and induce a sizeable mixing
of the lightest sneutrino in terms of left--handed and right--handed fields. We define the mixing as follows:

\begin{eqnarray}
\left\{
\begin{array}{l}
 \snu_1 = -\sin\theta\;\snu_L + \cos\theta\;\tilde{N}\\
 \snu_2 = +\cos\theta\;\snu_L + \sin\theta\;\tilde{N}
\end{array}
\right.
\end{eqnarray}
where $\theta$ is the LR mixing angle. Sizeable mixings reduce the coupling to the $Z$--boson, which
couples only to left--handed fields, and therefore have relevant impact on all the sneutrino phenomenology,
as recognized in Refs. \cite{Arkani-Hamed:2000bq,Grossman:1997is,Smith:2001hy,Smith:2002af,Tucker-Smith:2004jv}. 

The first important consequence is that the lightest sneutrino may be lighter than $m_{Z}/2$ and,
due to the mixing angle, still pass the invisible $Z$--width constraint, which now is modified as:
\begin{equation}
\Delta\Gamma_{Z} = \sin^4\theta\,\frac{\Gamma_{\nu}}{2}
\left[ 1- \left(\frac{2 m_1}{m_{Z}}\right)^{2}\right]^{3/2} \,\;  \theta(m_Z-2\,m_1)
\label{eq:Zlr}
\end{equation}
Also $\snu_{1}$ annihilation and scattering cross sections which involve $Z$ exchange are reduced
because of the mixing. Also diagrams with higgs exchange are modified, but in a different way. For
details.

The free parameters in the sneutrino sector for the LR models are therefore: $m_{L}$, $m_{N}$ 
$F^{2}$. We will study the model by varying $m_{N}$ and $F^{2}$ freely, while for $m_L$ we will
assume a lower bound of 100 GeV, in order to assure that the mass limits on the charged sleptons
is satisfied.

In Figs. \ref{fig:lr-mnm1}, \ref{fig:lr-mnsinth1} and \ref{fig:lr-mnsinth2} we show some features of the
model relevant for our analysis. In Fig. \ref{fig:lr-mnm1} the mass $m_1$ of the lightest sneutrino is plotted versus the right--handed mass $m_N$ for various values of $F^2$ and for
$m_N=120$ GeV. Whenever $m_N \lesssim m_L$, sneutrinos are light, and largely mixed with the
right--handed component $\tilde N$, as is clear from Fig. \ref{fig:lr-mnsinth1} where the $\sin\theta$
is plotted against $m_N$, for the same set of configurations of Fig. \ref{fig:lr-mnm1}. Each line
starts from the value of $m_N$ below which a negative mass--squared eigenvalue $m_1^2$ occurs. If we increase 
the value of $m_L$, the tachionic bound is met for lower values of $m_N$ for any fixed $F^2$, and the
mixing angle diminishes ({\em i.e.} the mixing to the right--handed component increases). This is clear
from Fig. \ref{fig:lr-mnsinth2}, where $\sin\theta$ vs. $m_N$ is plotted for a larger value of
the left-handed mass parameter: $m_L = 1$ TeV. Figs. \ref{fig:lr-mnsinth1} and \ref{fig:lr-mnsinth2}
also show that, especially for large values of $m_L$ and $F^2$, small mixings may be obtained also 
for heavy sneutrinos.

Light sneutrinos and (very) small mixings may occur: the invisible $Z$--width may therefore be a relevant 
constraint. Figs. \ref{fig:lr-m1sinth1}, \ref{fig:lr-m1sinth2} and \ref{fig:lr-m1sinth3} show, 
in the plane $\sin\theta$--$m_1$ and for different values of $F^2$, the region which is excluded
by this constraints, as the rightmost full [blue] area. Clearly this bound applies only for $m_1<m_Z/2$,
and for light sneutrinos it is evaded for $\vert\sin\theta\vert \lesssim 0.4$, as is clear from Eq.~\ref{eq:Zlr}. The scatter plot shows the distribution
of points obtained by scanning the soft--mass parameters $m_{L}$ and $m_{N}$ in 
the ranges: $120~\mbox{GeV} \leq m_{L} \leq 1~\mbox{TeV}$ and 
$1~\mbox{GeV} \leq m_{N} \leq 1~\mbox{TeV}$. The off-diagonal parameter $F^2$ is fixed at different values
in each Figure. We notice that a large fraction
of points passes the $Z$--width bound and therefore represents viable models. We also see, in
Fig. \ref{fig:lr-m1sinth1}, that for
small values of $F^2$ the mixing angles are typically very small for light sneutrinos. Light eigenstates
are possible only for small values of $m_N$, since $m_L$ is lower bounded at 120 GeV in our scan, and therefore
the lightest sneutrino is already mostly right--handed, which implies almost vanishing mixing angle.
When $m_N > m_L$ the lightest sneutrino is
mostly left--handed, but when $m_N \simeq m_L$, even a small $F^2$ can allow for rotation
with small mixing angles. When $F^2$ increases, larger mixing angles are possible also for light
sneutrinos, and in this case the $Z$--width bound becomes important. This is clearly seen in Figs. 
\ref{fig:lr-m1sinth2} and \ref{fig:lr-m1sinth3}. These features are quite relevant for the sneutrino
relic abundance and detection rates, since a small mixing into $\snu_L$ implies reduced interactions,
and therefore larger relic abundance and smaller direct detection rate.

The relic abundance for the LR models is shown in Fig. \ref{fig:ls-omega}, for a full variation
of the sneutrino parameter space, and for a fixed configuration of the other supersymmetric parameters.
The sneutrino parameters are varied in the ranges:
$120~\mbox{GeV} \leq m_{L} \leq 1~\mbox{TeV}$, $1~\mbox{GeV} \leq m_{N} \leq 1~\mbox{TeV}$ and
$10~\mbox{GeV}^{2} \leq F^{2} \leq 10^{4}~\mbox{GeV}^{2}$. The other parameters are fixed at the same
values used in the previous Section for Figs. \ref{fig:std-omega}, \ref{fig:std-dama} and \ref{fig:std-cdms}:
the lightest neutralino is 30\% heavier than the
sneutrino (which implies gaugino non--universality when the neutralino mass is light) and higgs
masses of 120 GeV for the lightest CP--even higgs $h$ and 400 GeV for the heaviest CP--even $H$
and the CP--odd $A$. All the models shown in the plot 
are acceptable from the point of view of all experimental constraints, including the invisible $Z$--width.
The horizontal solid and  dotted lines delimit the WMAP interval for cold dark matter.

Contrary to the previous case of the minimal MSSM model, in LR models sneutrinos may represent the dominant
dark matter component for a wide mass range, which extends (for the specific supersymmetric configurations
discussed here) from a few GeV to about 300 GeV. This is in fact due to the
reduced $Z$ coupling which occurs as a consequence of the mixing to the right--handed field $\tilde N$. The
dips in the scatter plot are again due to the $Z$, $h$ and ($H$,$A$) poles in the annihilation cross
section (from left to right). We remind that the location of the higgs poles is a consequence of the choice of the higgs masses we are using for this specific case. The sharp drop at $m_1 = m_W$ is due to the opening of the
annihilation channel into the $W^+W^-$ pair. The points with an acceptable relic abundance at masses below
10 GeV are obtained for light neutralinos, since in this case the annihilation cross section into neutrinos
through the exchange of neutralinos gets enhanced, and therefore reduces the relic abundance to acceptable levels.
This possibility occurs if the gaugino sectors possesses non--universality \cite{Bottino:2002ry} (otherwise the lightest neutralino
mass is lower bounded at the value of 50 GeV by LEP2 analyses \cite{Colaleo:2001tc,delphi,:2001xwa,lep2}): gaugino non--universality therefore appears
instrumental in allowing light sneutrinos to be cosmologically viable \cite{Bottino:2002ry, Bottino:2003iu,Belanger:2002nr,Hooper:2002nq}.

Figs. \ref{fig:lr-m1sinth1}, \ref{fig:lr-m1sinth2} and \ref{fig:lr-m1sinth3} show how the configurations with acceptable relic abundance are distributed in the $\sin\theta$--$m_1$ plane. When $F^{2}$ is
small, it typically induces very small mixings for light sneutrinos. This is clearly shown in Fig. 
\ref{fig:lr-m1sinth1}, which refers to $F^{2}=10^{2}$ GeV$^{2}$. Light sneutrinos are
possible in these models, and due to the small mixing with the $Z$ boson they easily evade the invisible
$Z$--width bound. However, the small-mixing angle, typically much smaller than $2\cdot 10^{-2}$,
leads to values of the relic abundance in excess of the cosmological bound. 
The strong suppression of the $Z$ coupling for sneutrinos lighter than
$m_{W}$ in these models with small $F^{2}$, require the opening of the $W^+W^-$ channel in order to
reduce the relic abundance to acceptable levels.
Only for sneutrinos
in the mass range 80--200 GeV cosmologically acceptable configurations are found. When $F^{2}$ 
increases, the cosmologically allowed parameter space opens up. For $F^{2}= 10^{3}$ GeV$^{2}$,
shown in Fig. \ref{fig:lr-m1sinth2},
mixing angles are typically larger than in the previous case also for light sneutrinos, and 
they may reach values of 0.2. Cosmologically acceptable configurations are obtained for masses 
which range from 10 GeV up to 200--300 GeV. Larger values of $F^{2}$ may allow for a
further enlargement of the allowed mass range: for $F^{2}=10^{4}$ GeV$^{2}$, shown in
Fig. \ref{fig:lr-m1sinth3}, sneutrinos are cosmologically allowed starting from a few GeV mass up to 
200--300 GeV.

The sneutrino--nucleon cross section \xisigma, for the same supersymmetric configurations used in Fig.
\ref{fig:ls-omega}, is shown in Figs. \ref{fig:lr-direct1}, \ref{fig:lr-direct2} and \ref{fig:lr-direct3} for progressively larger values of $F^{2}$. The direct detection bound, although relevant for many
configurations, now is easily evaded. Most of the configurations are allowed, and a large fraction of these
would be a candidate to explain the DAMA/NaI annual modulation effect. The compatibility with sneutrino
dark matter and the annual modulation effect increase when $F^{2}$ is increased, especially for light
sneutrinos. Fig. \ref{fig:lr-direct1}, which refers to $F^{2} = 10^{2}$ GeV$^{2}$ shows the lower bound
of 80 GeV in the sneutrino mass which is due to the relic abundance bound, and that was discussed
in relation to Fig. \ref{fig:lr-m1sinth1}. For cosmologically acceptable sneutrinos, most of the configurations
are compatible with direct detection searches. Larger values of $F^{2}$ open up the possibility of
lighter sneutrinos, as is shown in Fig. \ref{fig:lr-direct2}, where $F^{2} = 10^{3}$ GeV$^{2}$
has been adopted. All configurations for sneutrinos lighter than 60 GeV are currently compatible
with direct detection searches. Fig. \ref{fig:lr-direct3}, which refers to $F^{2} = 10^{4}$ GeV, {\em i.e.} to a value of the same order of magnitude of the diagonal mass terms in ${\cal M}^{2}_{LR}$ of Eq. (\ref{eq:masslr}), show that in the whole mass range from few GeV up to 200 GeV sneutrinos are compatible with the annual modulation effect, without invoking any fine--tuned condition on the parameters. This is
a quite remarkable feature of LR models.

The analysis presented so far, useful to discuss the features of the sneutrino relic abundance
and direct detection rate, was specific to fixed values of the higgs masses and peculiar values 
of the neutralino and chargino masses. We now extend our analysis to a full scan of the supersymmetric
parameter space, analogous to the studies for neutralino dark matter. The parameter space is defined
in Appendix \ref{app:mssm}, where also the experimental constraints on the supersymmetric model are discussed. Parameters specific to the sneutrino sector are varied in the following intervals: $100~\mbox{GeV} \leq m_{L} \leq 3~\mbox{TeV}$, 
$1~\mbox{GeV} \leq m_{N} \leq 1~\mbox{TeV}$ and $1~\mbox{GeV}^{2} \leq F^{2} \leq 10^{6}~\mbox{GeV}^{2}$.

The sneutrino relic abundance is shown in Fig. \ref{fig:lr-omegascan}. The most relevant new feature
is that for the full supersymmetric scan, the mass range allowed by the cosmological constraints is
enlarged up to 800 GeV, and all the mass interval above the $Z$--pole may lead to strongly subdominant
sneutrinos. This is due to either to the occurrence of the higgs poles in the annihilation cross section discussed above or to the mixing with the right--handed field. The occurrence of sizeable mixings with the $\tilde N$ field is specially important for allowing lighter sneutrinos and for enhancing the relic abundance
around the $Z$ pole. From Fig. \ref{fig:lr-omegascan}
we can conclude that, for a full scan of the supersymmetric parameter space in LR models, after all experimental
(and theoretical) constraints are imposed, sneutrino dark matter is perfectly viable, both as a dominant and
as a subdominant component, for the whole mass range $15~\mbox{GeV} \lesssim m_1 \lesssim 800~\mbox{GeV}$.
The lower limit of 15 GeV represents therefore a cosmological bound on the sneutrino mass in LR models,
under the assumption of $R$--parity conservation.
This result may be confronted with the one obtained for relic neutralinos in gaugino non--universal models
\cite{Bottino:2002ry, Bottino:2003iu,Belanger:2002nr,Hooper:2002nq}, which is about $6$ GeV.
 
The sneutrino--nucleon cross section is shown in Fig. \ref{fig:lr-directscan}. Only points which are
accepted by the cosmological constraint are shown. We see that the presence of the mixing with the
right--handed $\tilde N$ fields opens up the possibility to have viable sneutrino cold dark matter.
A fraction of the configurations are excluded by direct detection, but now, contrary to the minimal MSSM 
case, a large portion of the supersymmetric parameter space is compatible with the direct detection bound,
both for cosmologically dominant (denoted by [red] crosses) and subdominant ([blue] points) sneutrinos. 
The occurrence of sneutrinos which are not in conflict with direct detection limits and, at the same 
time, are the dominant dark matter component, is a very interesting feature of this class of models. 
Fig. \ref{fig:lr-directscan} compares our theoretical calculations with the DAMA/NaI region (comparison
with the CDMS upper bounds in straightforward): it is clearly seen that sneutrino dark matter could
explain the annual modulation effect (as well as neutralinos do in many realizations of 
supersymmetric theories \cite{Bottino:2000gc,Bottino:2003cz,Cerdeno:2004xw}).

The correlation between the direct--detection relevant cross section \xisigma and the relic abundance
is shown in Fig. \ref{fig:lr-sigmaomega}. Cosmologically dominant (or slightly subdominant) sneutrinos 
are compatible with the current level of sensitivity of direct--detection experiments, 
as discussed above. Nevertheless, a fraction of these configurations refer to direct--detection  
cross--sections which are up to 4 orders of magnitude below current sensitivities. At the same time,
a fraction of the configurations which are at the level of direct detection sensitivity refer to
quite small values of the sneutrino relic abundance: these subdominant sneutrinos correspond to a case
where the dark matter is mostly composed of a different candidate, but nevertheless represent a relic from the early Universe potentially detectable in the laboratory (a very interesting situation by itself!). Fig. \ref{fig:lr-sigmaomega} shows that, provided a suitable model like the LR one discussed here, sneutrino dark matter may exhibit a phenomenology which is not very different from the neutralino in a low--energy effective MSSM.

Figs. \ref{fig:lr-mnml}, \ref{fig:lr-mnf} and \ref{fig:lr-m1sinth} show the distribution of the
allowed configurations in three different sections of the sneutrino parameter space. In all three
figures, [red] crosses refer to models with sneutrino relic abundance
in the cosmologically relevant range; [blue] dots refer to cosmologically subdominant 
sneutrinos; open dots mark the configurations which have a direct--detection cross section
in the current sensitivity range. Fig. \ref{fig:lr-mnml} shows that, when $m_N$ is below 100 GeV, cosmologically allowed 
configurations require $m_L$ to be close to its lower limit: this is necessary in order to
have a large enough mixing angle $\theta$, which otherwise would produce an exceedingly large
relic abundance. The off-diagonal parameter $F^2$ needs also to be tuned accordingly, in
a range from $10^2$ GeV$^2$ to $10^4$ GeV$^4$, as displayed in Fig. \ref{fig:lr-mnf}. When $m_N$ crosses the
lower bound on $m_L$, more wide possibilities open up. Light sneutrinos, with masses below
40--50 GeV, need values of the mixing angle such that $\sin\theta$ is around 0.05--0.5, with
the specific correlation shown in Fig. \ref{fig:lr-m1sinth}. The correlation with values of \xisigma
in the current direct--detection sensitivity range are also displayed: the most relevant feature is
that they require $F^2$ to be large when $m_N$ is large, in order to procude enough mixing
to reduce the direct detection cross--section to acceptable levels at large sneutrino masses.

Now that we have assessed the possibility to have viable sneutrino dark matter candidates,
let us now move to the study of their indirect detection signals. Dark matter particles distributes
in the galactic halo may annihilate in pairs and produce a bunch of possible signals. Among these 
signals antimatter, namely antiprotons and antideuterons \cite{Donato:1999gy}, and gamma--rays may be produced. Exhaustive
and detailed studies have been performed for the case of neutralino dark matter, but not yet for sneutrino
relics.

We start by studying the antiproton signal. The antiproton source term $q_{\bar p}^{\rm DM}(r,z;T_{\bar p})$
at a given position in the Galaxy, defined by the radial component along the galactic plane $r$ and the vertical
component $z$, is defined as:
\begin{equation}
q_{\bar p}^{\rm DM}(r,z;T_{\bar p})= \frac{1}{2}{\langle \sigma_{\rm ann} v\rangle_0} ~{ g_{\bar p}(T_{\bar p})}~ {\left[ \frac{\rho_{\snu_1}}{m_1}\right]^2}
\end{equation}
where $T_{\bar p}$ is the antiproton kinetic energy, $\langle \sigma_{\rm ann} v\rangle_0$ denotes the sneutrino
annihilation cross section averaged over the velocity distribution of sneutrinos in the Galaxy (which are
strongly non--relativistic since their average velocity is of the order of $\beta \sim 10^{-3}$), 
$g_{\bar p}(T_{\bar p})$ is the antiproton energy--spectrum per annihilation event and $\rho_{\snu_1}(r,z)$ denotes
the sneutrino density distribution, which is assumed to be proportional to the total dark matter density distribution
as $\rho_{\snu_1}(r,z) = \xi \rho_{\rm DM}(r,z)$, in order to take into account both dominant and subdominant sneutrino
relics. The energy--spectrum $g_{\bar p}(T_{\bar p})$ is a sum over the different energy spectra produced by the various final states of the annihilation process:
\begin{equation}
g_{\bar p}(T_{\bar p}) = 
\sum_F {\rm BR}(\snu_1 \snu_1 \rightarrow F)\, \left(\frac{dN_{\bar p}^F}{dT_{\bar p}}\right)
\end{equation}
where $F$ labels the final states. We have modelled the spectra $dN_{\bar p}^F/dT_{\bar p}$ as discussed in
Refs. \cite{Donato:2003xg}, by means of a semi--analytic calculation which follows the production and decay chain of each final state
until a quark is produced. We use detailed fits and interpolations over the results of PYTHIA \cite{Sjostrand:2000wi} simulations 
for the treatment of the processes involved in the quark hadronization the subsequent hadron decays \cite{Donato:2003xg}.
The antiproton source spectra are then propagated in the galactic environment to determine the antiproton flux
at the local position in the Galaxy $\Phi(R_0,0,T_{\bar p})$:
\begin{equation}
q_{\bar p}^{\rm DM}(r,z;T_{\bar p}) \longrightarrow \Phi(R_0,0,T_{\bar p})
\end{equation}
We model the galactic environment as a two--zone diffusion model and for the solution of the propagation equation, which takes into account antiproton diffusion, scattering, annihilation, energy losses, propagation against the galactic wind and reacceleration, we use the results of Ref. \cite{Donato:2003xg}, where a detailed analysis which takes into account astrophysical uncertainties (relevant for the antiproton signal) are discussed. Theoretical uncertainties are large and we will comment on this point: in all our figures we will
show the result for the median estimate of the antiproton signal, as defined in Ref. \cite{Donato:2003xg}. The antiproton flux has then to be propagated against the solar wind in order to provide fluxes at the Earth position, {\em i.e.} "top--of--atmosphere'' (TOA) fluxes. We will show our results for a period of solar minimum activity.

Antiproton fluxes from sneutrino annihilation in the galactic halo are provided in Fig. \ref{fig:lr-pbar023} and
\ref{fig:lr-pbar375} which show scatter plots of fluxes calculated at fixed antiproton kinetic
energies: $T_{\bar p} = 0.23$ GeV in Fig. \ref{fig:lr-pbar023} and $T_{\bar p} = 37.5$ GeV in  
Fig. \ref{fig:lr-pbar375}. The scatter plots refer to cosmologically acceptable configurations: the [red] crosses
denote cosmologically dominant sneutrinos, the [blue] points refer to the case of subdominant sneutrinos. The small
grey points show those configurations which are excluded by direct detection. In Fig. \ref{fig:lr-pbar023}, which refers
to the flux in a low--energy bin where the antiproton signal may have better possibility to be disentangled by the
background which is due to cosmic--rays spallation over the galactic medium. The [yellow] shaded area denotes the amount of exotic antiprotons which can be accommodated in the BESS data \cite{Orito:1999re,Maeno:2000qx} in that energy bin. This has been established
on the basis of the theoretical calculation of the antiproton background \cite{Donato:2001ms} and of the BESS measurements \cite{Orito:1999re,Maeno:2000qx}, by
determining the maximal amount of exotic antiproton flux which can be accommodated on the top of the background, without entering in conflict with the BESS data and its experimental error, at 90\% C.L. We see that the
theoretical predictions are not currently excluded by  BESS: the maximal predictions, which occur for 
sneutrino masses in the range 50--200 GeV, are at least one order of magnitude below the current BESS bound.
However, we have to remind that the theoretical estimates of the antiproton signal are largely affected by
astrophysical uncertainties related to the knowledge of the parameters which enter the diffusion equation. It has been shown in Ref. \cite{Donato:2003xg} that this uncertainty can lead to an increase of about a factor of 8 or a decrease of up to
a factor of 10. Therefore, our prediction in Fig. \ref{fig:lr-pbar023} may be altered by
this factor for different choices of the propagation parameters in their allowed ranges 
\cite{Maurin:2001sj}. In the case of the choice
of astrophysical parameters which produce the maximal antiproton signal, the scatter plot in Fig. \ref{fig:lr-pbar023} 
would be enhanced by a factor of 8: in this case, still BESS data would not exclude any sneutrino configuration,
but all the mass range from 50 to 200 GeV would have configurations potentially detectable with just a small 
increase in the experimental sensitivity.

Prospects for the future are shown by the dashed and dotted horizontal lines, which denote our
estimated sensitivity of the PAMELA \cite{pam} (dashed line) and AMS \cite{ams} (dotted line) detectors  to exotic antiprotons after a run of 3 years: the sensitivities are determined as admissible excess within the
statistical experimental uncertainty if the measured antiproton flux consists
only in the background (secondary) component. The estimate has been performed by
using the background calculation of Ref. \cite{Donato:2001ms}, and refers to a 1--$\sigma$
statistical uncertainty. All the supersymmetric configurations in Fig. \ref{fig:lr-pbar023} above
the dashed or dotted lines can be potentially identified by PAMELA or AMS as a signal over the
secondaries, while those which are below the dashed or dotted lines will not contribute
enough to the total flux in order to be disentangled from the background. We therefore see that, in the case
of the median antiproton flux shown in the Figure, only AMS will have the possibility to detect a signal
from sneutrino dark matter, for masses around 60 GeV or for masses in the range 100--200 GeV. We notice that
most of these configurations refer to subdominant sneutrinos: this justifies the approach to consider
relic particle candidates also when they are not dominant dark matter components, since a signal from relic particles
in the galaxy may well be discovered. 

Fig. \ref{fig:lr-pbar023} also shows that, in the case of astrophysical propagation parameters close to the
values which provide the maximal antiproton signal, both PAMELA and AMS will have good chances of detection, for
sneutrinos in the mass range 50--200 GeV. We also notice that in this case a large fraction of configurations
with masses in the range 65--130 GeV, and which could be potentially detectable by AMS, are actually already excluded by direct detection (grey points). This shows a very nice interplay between different dark matter searches techniques. This is further elucidated in Fig. \ref{fig:lr-pbardirect}, where the antiproton signal is plotted against
the direct detection cross section \xisigma. The current bound from both BESS data and from direct detection
experiments is shown, as well as the foreseen capabilities of PAMELA and AMS, together with the current direct--detection sensitivity region. From this figure we see that a fraction of the configurations which are currently
excluded by direct detection would have been in reach of PAMELA and AMS.
More interestingly, Fig. \ref{fig:lr-pbardirect} shows that direct detection and antiproton searches offer a good
deal of complementarity. Direct detection is sensitive to configurations which will be hardly probed by antiproton
searches (those points inside the vertical [green] band and which refer to low antiproton fluxes). On the contrary,
some configurations which refer to a large antiproton signal but are below current direct detection sensitivity are 
also present. Nevertheless, from this figure direct detection appears to be a stronger probe to sneutrino dark matter
(in the sense that it can explore a wider region of the parameter space) but good chances of antiproton detection
are also present.

We come back now to Fig. \ref{fig:lr-pbar375}, where the antiproton signal is calculated for a higher antiproton
kinetic energy: $T_{\bar p} = 37.5$ GeV. At this energy, the CAPRICE experiments reports the detection of a flux which is potentially in excess of the theoretical background \cite{caprice}. Although this excess has a small statistically significance, nevertheless is an intriguing possibility to be studied. In Fig. \ref{fig:lr-pbar375} we show the band
which refers to a signal which would fill the CAPRICE excess. The scatter plot of the theoretical prediction for
sneutrino dark matter is not able to reach the level of the CAPRICE excess, for the median choice of astrophysical
parameters. We nevertheless comment that for the maximal choice, a marginal compatibility would arise, for sneutrino
masses in the range 200--500 GeV. In Fig. \ref{fig:lr-pbar375} we also show our estimated sensitivities for PAMELA and AMS, determined as we discussed above for the lower energy bin. For the median choice of astrophysical parameters, AMS will have a marginal detection potential, while for the maximal case, both AMS and PAMELA will probe configurations in the mass range 200--500 GeV. Light sneutrinos are not probed at this antiproton energies: they cannot produce
antiprotons at energies above their mass, since annihilation occurs almost at rest.

\EPSFIGURE[t]{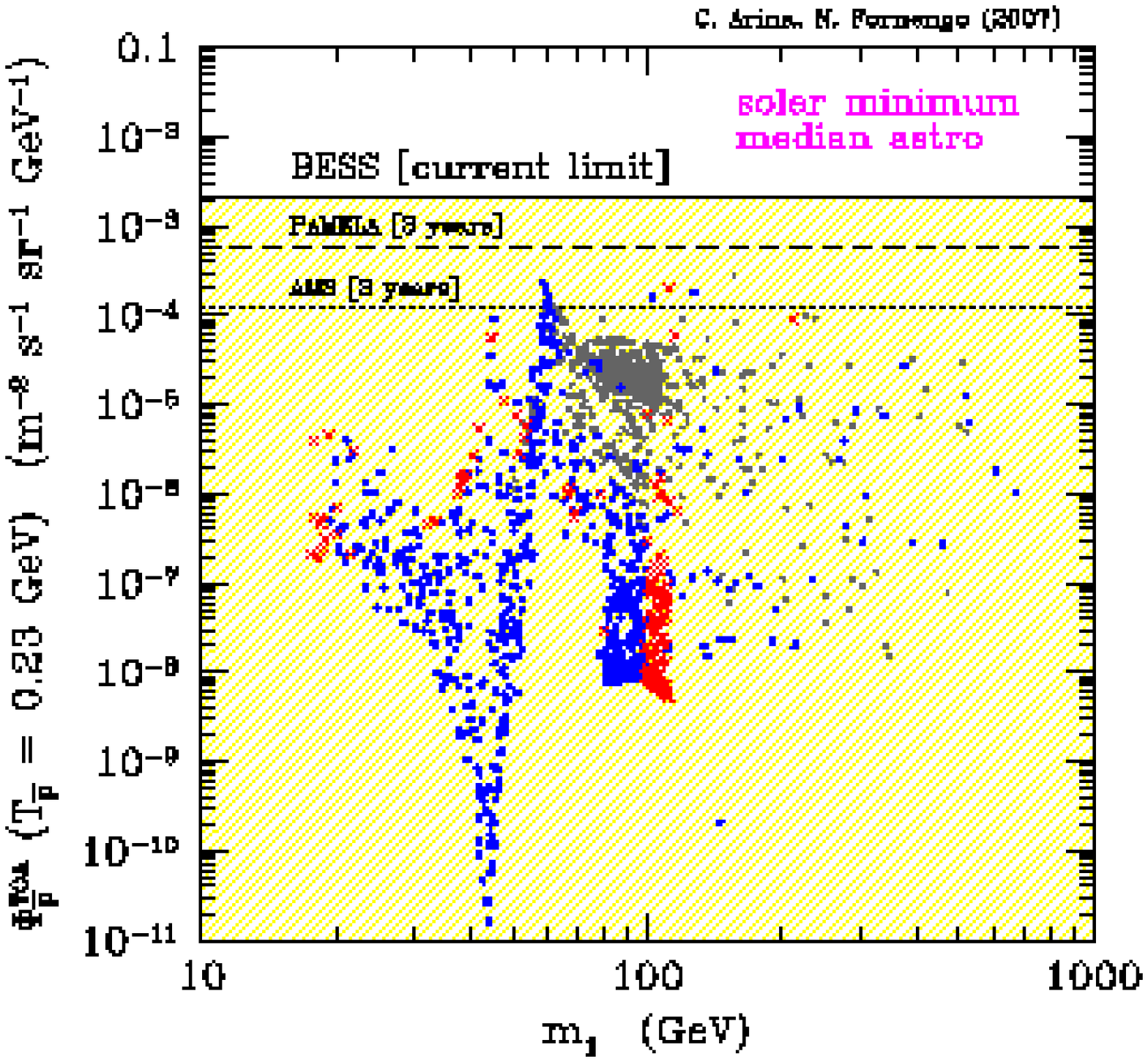,width=0.80\textwidth}
{LR models -- Antiproton flux at the antiproton kinetic energy $T_{\bar p} = 0.23$ GeV as a 
function of the sneutrino mass $m_{1}$, for the galactic propagation parameters which provide the
median value of antiproton flux and for a solar activity at its minimum. [Red] crosses refer to models with sneutrino relic  abundance in the cosmologically relevant range; [blue] dots refer to cosmologically 
subdominant sneutrinos; light gray points denote configurations which are excluded by direct
detection searches. The [yellow] shaded area denotes the amount of exotic antiprotons which can be accommodated in the BESS data \cite{Orito:1999re,Maeno:2000qx}. The dashed and dotted lines show the
PAMELA \cite{pam} and AMS \cite{ams} sensitivities to exotic antiprotons for 3 years missions, respectively. 
\label{fig:lr-pbar023}}

\EPSFIGURE[t]{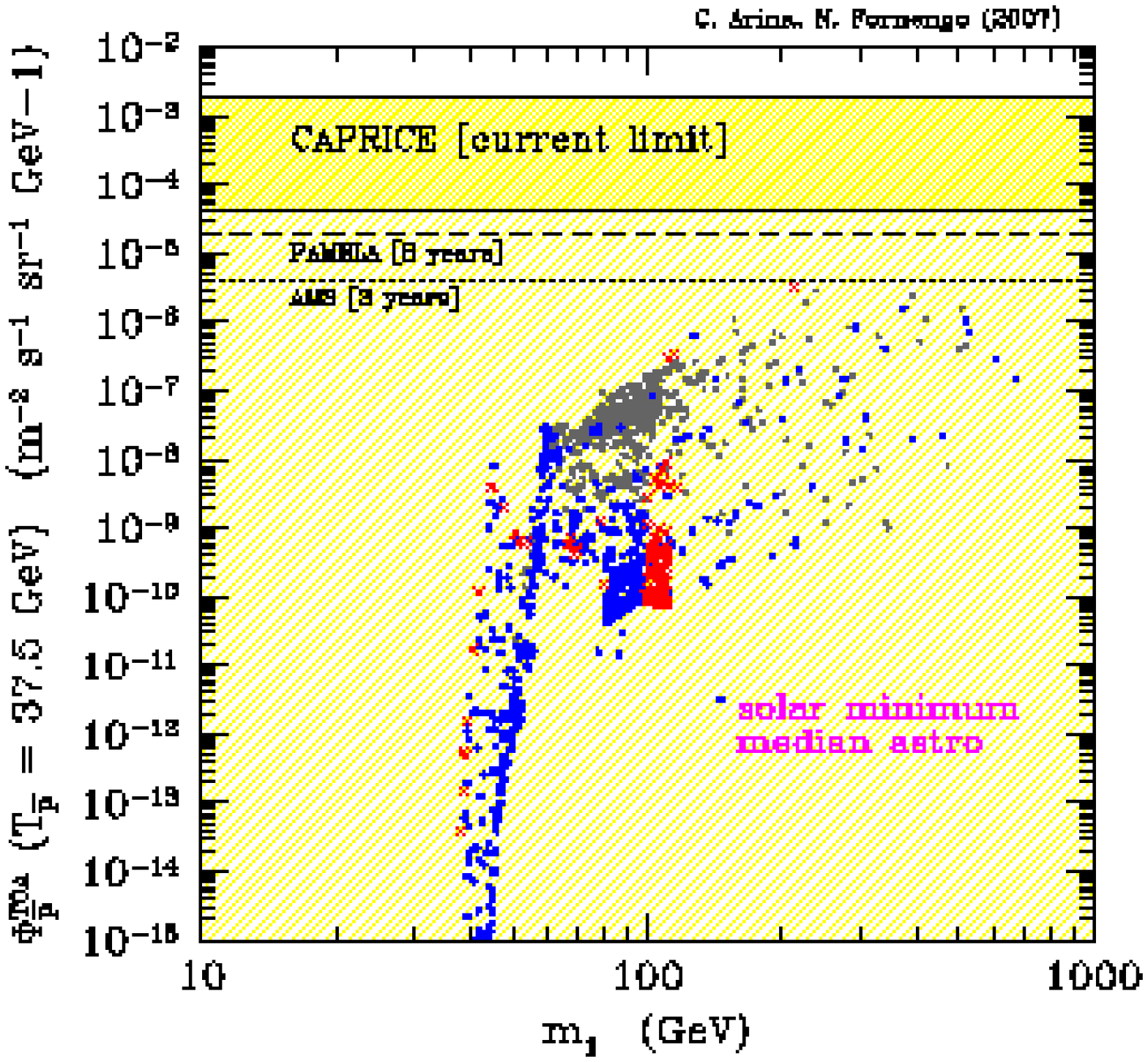,width=0.80\textwidth}
{LR models -- Antiproton flux at the antiproton kinetic energy $T_{\bar p} = 37.5$ GeV as a 
function of the sneutrino mass $m_{1}$. Notations are as in Fig. \ref{fig:lr-pbar023}, except
for the upper [yellow] shaded band, which delimits the possible excess over the background 
in the CAPRICE data \cite{caprice}. The lower [yellow] area refers to fluxes compatible with
CAPRICE.
\label{fig:lr-pbar375}}

\EPSFIGURE[t]{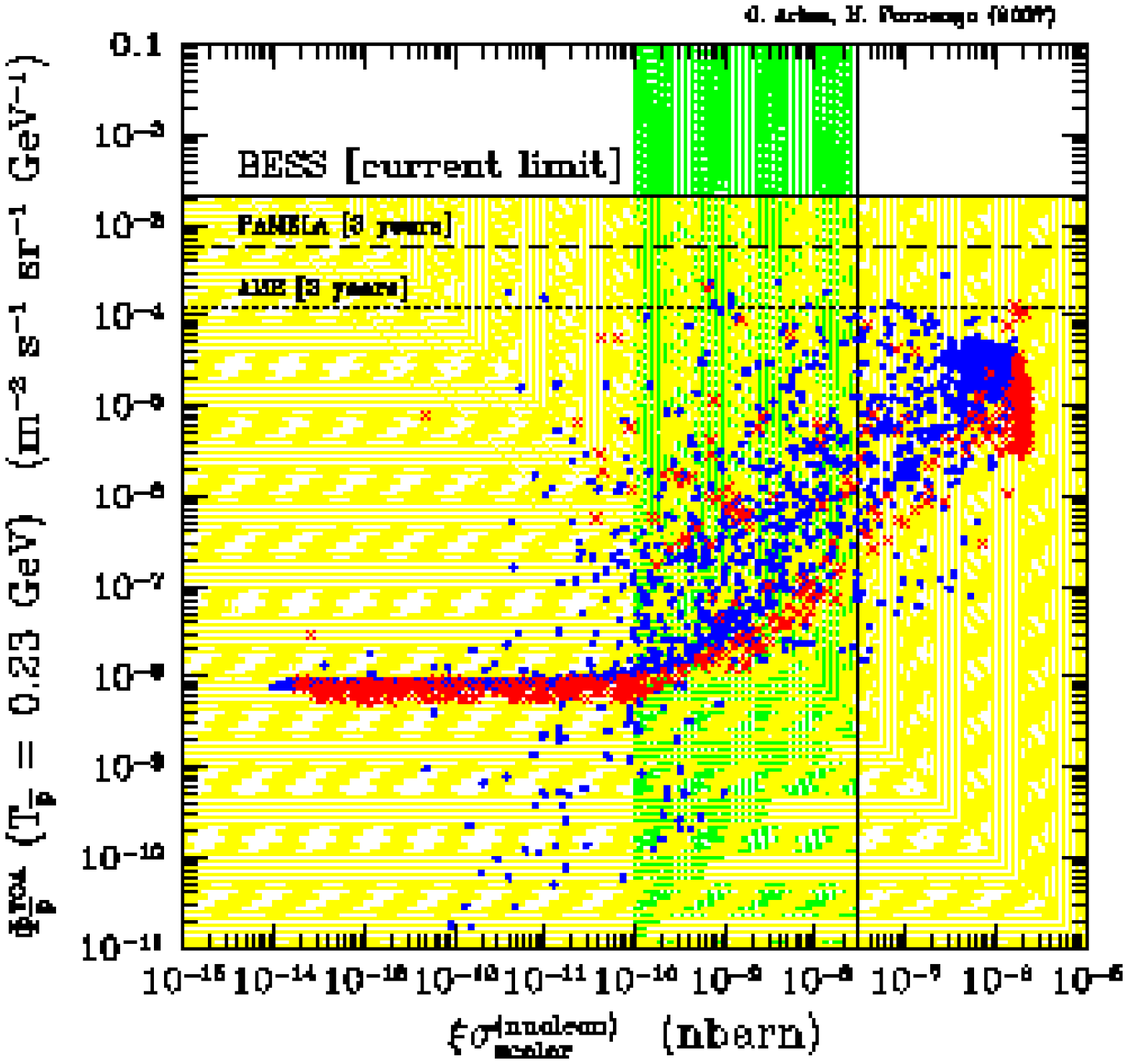,width=0.80\textwidth}
{LR models -- Antiproton flux at the antiproton kinetic energy $T_{\bar p} = 0.23$ GeV vs. the
sneutrino--nucleon scattering cross section $\xi \sigma^{\rm (scalar)}_{\rm nucleon}$.
[Red] crosses refer to models with sneutrino relic abundance in the cosmologically relevant range; [blue] dots refer to cosmologically subdominant sneutrinos. The horizontal solid line denotes the upper 
limit from BESS \cite{Orito:1999re,Maeno:2000qx} and the [yellow] shaded area shows the amount of exotic antiprotons which 
can be accommodated in the BESS data. The dashed and dotted horizontal lines show the
PAMELA \cite{pam} and AMS \cite{ams} sensitivities to exotic antiprotons for 3 years missions, respectively. The vertical solid line denotes a conservative upper limit from direct detection searches
and the [green] vertical shaded area refers to the current sensitivity in direct detection searches.
\label{fig:lr-pbardirect}}

\EPSFIGURE[t]{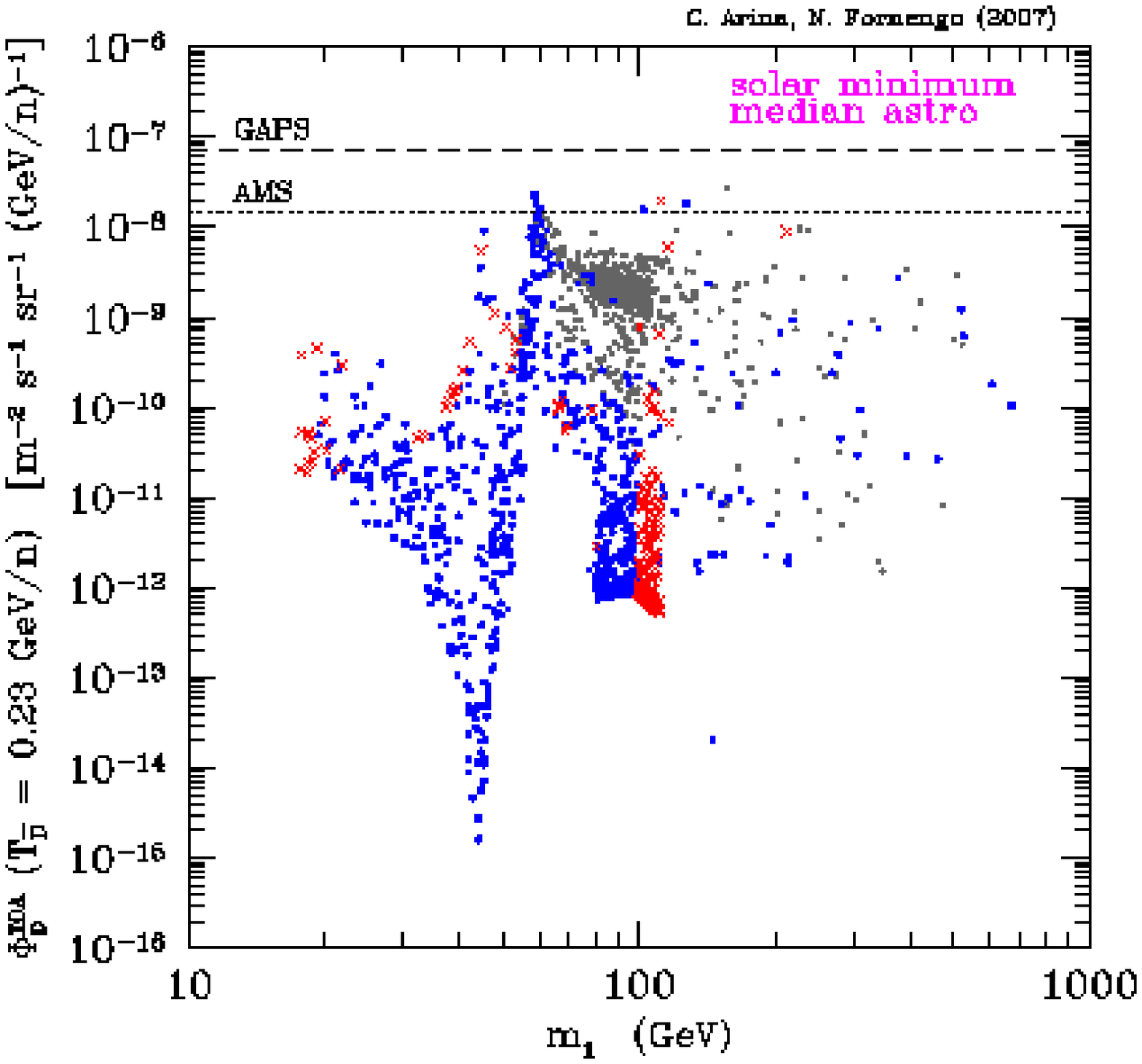,width=0.80\textwidth}
{LR models -- Antideuteron flux at the antideuteron kinetic energy (per nucleon) $T_{\bar p} = 0.23$ GeV/n,
as a function of the sneutrino mass $m_{1}$. Notations are as in Fig. \ref{fig:lr-pbar023}.
The dashed and dotted lines show the GAPS \cite{Mori:2001dv,koglin} and AMS \cite{ams} sensitivities.
\label{fig:lr-dbar023}}

\EPSFIGURE[t]{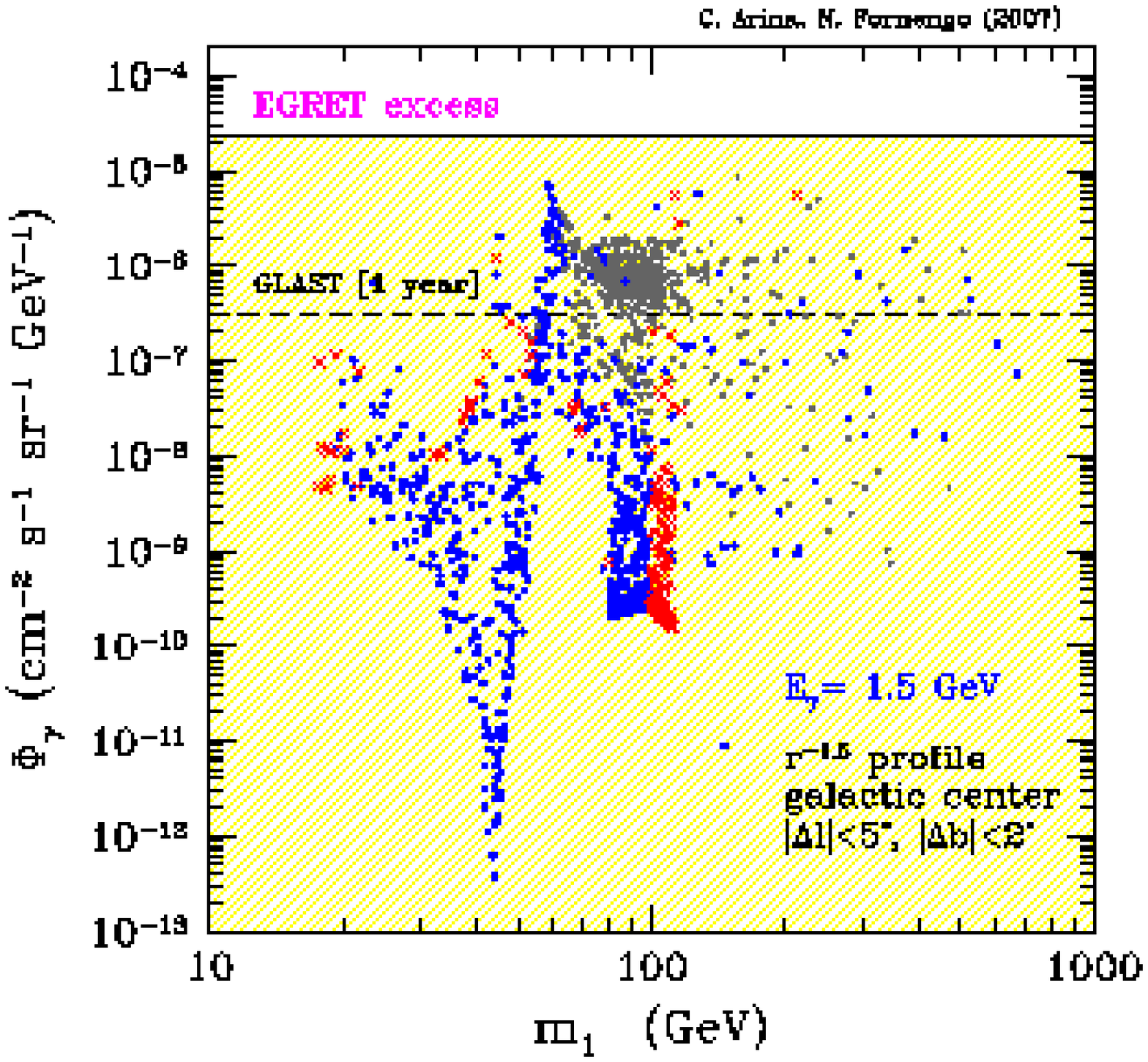,width=0.80\textwidth}
{LR models -- Gamma--ray flux from the galactic center at the photon energy $E_{\gamma} = 1.5$ GeV,
as a function of the sneutrino mass $m_{1}$, for a galactic profile of Moore's type \cite{Moore:1999nt,Moore:1999gc} 
and for the angular resolution of EGRET \cite{Egret,MayerHasselwander:1998hg}. [Red] crosses refer to models with sneutrino relic  abundance in the cosmologically relevant range; [blue] dots refer to cosmologically 
subdominant sneutrinos; light gray points denote configurations which are excluded by direct
detection searches. The [yellow] shaded area denotes the amount of exotic gamma--rays compatible
with the EGRET excess \cite{Egret,MayerHasselwander:1998hg}. The dashed line shows the GLAST \cite{glast} sensitivity for a 1 year data-taking and for the same EGRET angular bin.
\label{fig:lr-gamma}}

\EPSFIGURE[t]{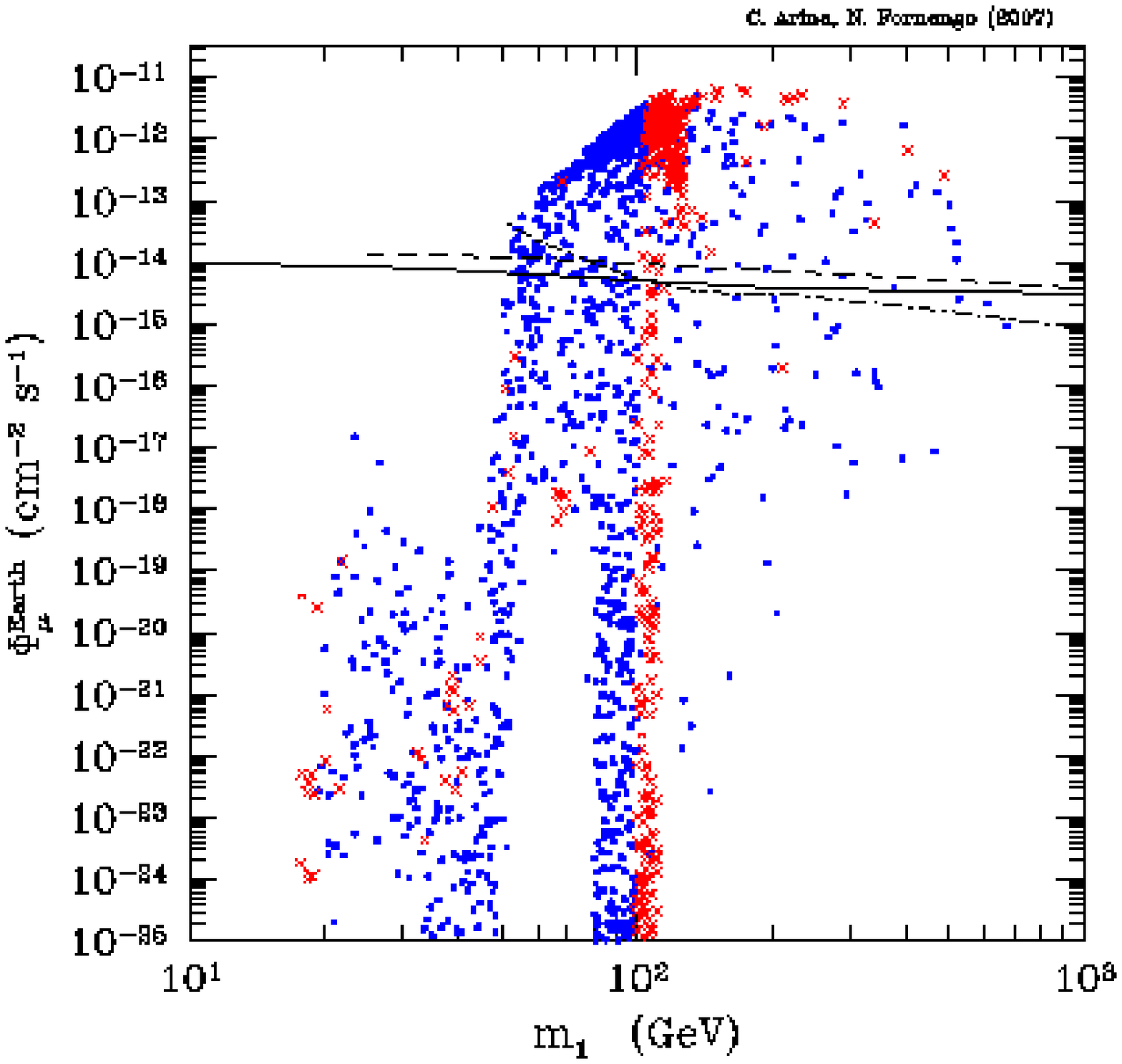,width=0.80\textwidth}
{LR models -- Upgoing muon flux from sneutrino pair annihilation in the center of the Earth
$\Phi_{\mu}^{\rm Earth}$, as a function of the sneutrino mass $m_{1}$. [Red] crosses refer to models with sneutrino relic abundance in the cosmologically relevant range; [blue] dots refer to cosmologically 
subdominant sneutrinos. The solid, dashed and dot--dashed lines denote the upper limits from
the SuperKamiokande \cite{Superk}, MACRO \cite{macro} and AMANDA \cite{amanda} detectors, 
respectively.
\label{fig:lr-fluxe}}

\EPSFIGURE[t]{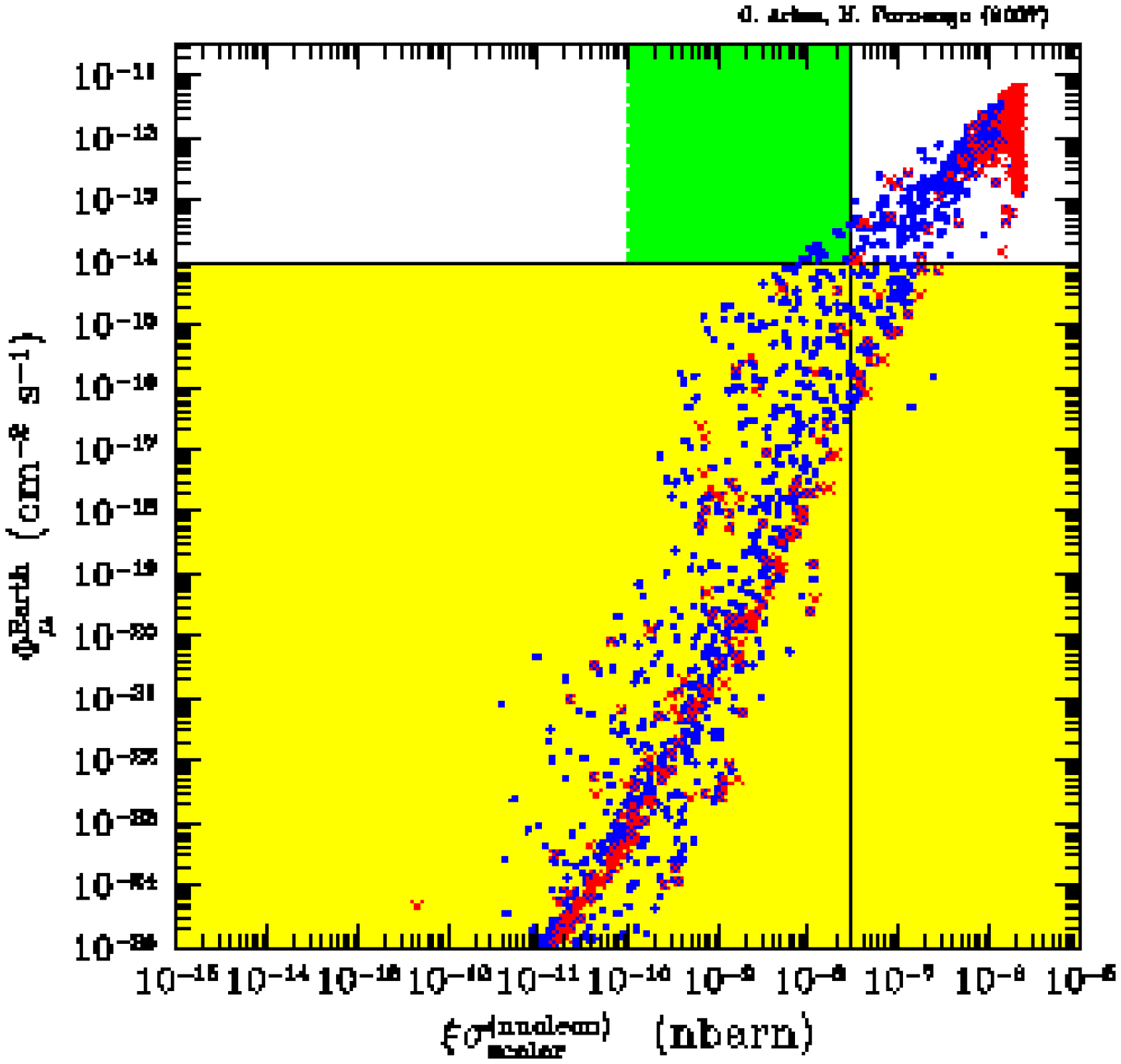,width=0.80\textwidth}
{LR models -- Upgoing muon flux from sneutrino pair annihilation in the center of the Earth
$\Phi_{\mu}^{\rm Earth}$ vs. the sneutrino--nucleon scattering cross section 
$\xi \sigma^{\rm (scalar)}_{\rm nucleon}$. [Red] crosses refer to models with sneutrino 
relic abundance in the cosmologically relevant range; [blue] dots refer to cosmologically 
subdominant sneutrinos. The horizontal solid line denotes the current upper limit from
neutrino telescopes. The vertical solid line denotes a conservative upper limit from direct detection 
searches and the [green] vertical shaded area refers to the current sensitivity in direct detection searches.
\label{fig:lr-fluxedirect}}

Another indirect detection signal which is very promising is the production of antideuterons \cite{Donato:1999gy}. In this paper especially it has been shown that the low energy tail of the antideuteron signal offers a good possibility to disentangle the signal
from the background, since kinematical conditions allow low--energy antideuterons to be produced in the
dark matter pair annihilation easily, while the spallation process is suppressed below a kinetic energy per--nucleon of 
$T_{\bar D} =$ 1--3 GeV/n. The predictions for sneutrino dark matter are shown in Fig. \ref{fig:lr-dbar023} for $T_{\bar D} = 0.23$ GeV/n. No experimental limit is currently available, but detectors, which will be able to reach good sensitivities, are under development: Fig.  \ref{fig:lr-dbar023} shows the expected sensitivity of the GAPS \cite{Mori:2001dv,koglin} and AMS \cite{ams} detectors. We see that 
antideuteron searches will offer a sensitivity to sneutrino dark matter similar to antiproton searches. A signal
detectable in one channel by two different detectors, will be detectable also in the other channels, again
in two different experiments. Moreover, one of the detector, AMS, has the capabilities to detect a signal in both
channels. The possibility to cross--correlate different signals and to complement their information
would be an extraordinary opportunity for dark matter searches. This is further complemented by direct detection, as discussed above.

We now move to discuss the signal which consists in the production of a diffuse gamma-rays flux. Sneutrino annihilations may produce, through the decay chain of their annihilation products, also gamma--rays, which
mostly come from the decay of neutral pions and other mesons produced in the annihilation process. The gamma ray flux arriving at the Earth from
a given angular position $\psi$ in the sky is the integral along the line of sight of all the differential
contributions of sneutrino annihilation:
\begin{equation}
\Phi_\gamma^{\rm DM}(E_\gamma,\psi) = \frac{1}{4\pi}
\frac{\langle \sigma_{\rm ann} v \rangle_0}{2 m_\chi^2}~{ g_\gamma(E_\gamma)}~{I(\psi)}
\end{equation}
where $g_\gamma(E_\gamma)$ is the gamma--ray spectrum, defined analogously as we defined above the antiproton spectrum $g_{\bar p}(T_{\bar p})$ and
$I(\psi) $ denotes the line--of--sight integral of the squared of the dark matter density in the direction $\psi$:
\begin{equation}
I(\psi) = \int_{\rm l.o.s.} \rho_{\snu_1}^2[r(\lambda,\psi)]d\lambda
\end{equation}
For the production of gamma--rays arising from the hadronization of quarks and the subsequent decay of hadrons,
we use the detailed fit of PHYTIA simulations of Ref. \cite{Bottino:2004qi,Fornengo:2004kj}. 

Fig. \ref{fig:lr-gamma} shows our predictions for the gamma--ray flux at $E_\gamma = 1.5$ GeV coming from the center of the Galaxy, in an angular bin which corresponds to the EGRET \cite{Egret,MayerHasselwander:1998hg} field of view. The choice of the energy bin refers to
the case where EGRET detects an excess of gamma--rays over the background from the galactic center. The [yellow] shaded
area in facts refers to this excess: exotic gamma--ray fluxes inside this band are compatible with the EGRET
measurement and those close to the solid horizontal line, which delimits the area, are able to explain the excess.

The gamma--ray signal is strongly sensitive to the behavior of the dark matter density profile toward
the inner regions of the Galaxy. In Fig. \ref{fig:lr-gamma} we have used a strongly peaked profile: the
radial behavior is $r^{-1.5}$, like the Moore et al. profile \cite{Moore:1999nt,Moore:1999gc} or those obtained from milder distributions
by effects due to the growth of a black hole \cite{Gondolo:1999ef,Merritt:2002vj,Merritt:2003qk} or of baryon dissipation \cite{Prada:2004pi,Gnedin:2004cx}. For a NFW profile with a $r^{-1}$
behavior, our theoretical estimates would decrease by a factor of 60  and an additional factor of 10
for cored profiles \cite{Bottino:2004qi}.

Fig. \ref{fig:lr-gamma} shows that, for an $r^{-1.5}$ profile, sneutrino dark matter is at the level of explaining
the EGRET excess, for masses in the range 60--250 GeV. Most of these configurations refer to subdominant sneutrinos.
Clearly for an NFW profile the predicted fluxes are too low to fill the EGRET excess.

Fig. \ref{fig:lr-gamma} also shows our estimate for the capabilities of GLAST, for a 1--year data--taking. We have taken into account GLAST effective area as shown in Ref. \cite{glast}, and we have derived our predictions for the same
angular energy bin of EGRET. We see that GLAST will be sensitive to configurations of masses between 30 GeV and
600 GeV, and will be close to access also very light sneutrinos with further live--time of data. Again we have to 
warn that in the case of less steep profiles, the sensitivity of GLAST will be restricted to a smaller mass range
(60--300 GeV), but nevertheless a good deal of configurations will be probed. We comment also that the angular resolution of GLAST will be much better than the EGRET one and therefore the capabilities of GLAST would be even more promising than those shown in Fig. \ref{fig:lr-gamma}.

Finally, we mention that indirect detection of dark matter could also be performed at neutrino telescopes, since
dark matter may accumulate in the central regions of the Earth and the Sun by gravitational capture and there
annihilate. The only annihilation products which can escape are neutrinos, and these can be searched for as upgoing
muons in a neutrino telescope. In Fig. \ref{fig:lr-fluxe} we show the predictions for upgoing muons from the Earth \cite{Cirelli:2005gh}
and compare them with the current experimental limits from SuperKamionande, MACRO and AMANDA. We see that
neutrino telescopes are sensitive to a large fraction of sneutrino configurations, although many of the
configuration which are excluded by this technique are also excluded by direct detection. This is manifest
in Fig. \ref{fig:lr-fluxedirect}, which shows the correlation of the upgoing muon flux from the Earth with
the direct detection cross section \xisigma.

\section{Models with a lepton--number violating term}\label{sec:cp}

Models with lepton--number violating terms can allow for Majorana neutrino masses. The most direct way
to include a Majorana mass term is to introduce a gauge--invariant dimension--5 operator of the type 
\cite{Arkani-Hamed:2000bq,Hirsch:1997vz,Dedes:2007ef,Hirsch:1997is}:

\begin{equation}
{\cal L} = \frac{g_{IJ}}{M_{\Lambda}} (\epsilon_{ij} L^I_i H_j)(\epsilon_{kl} L_k^J H_l) + \mbox{h.c.}
\end{equation}
In this case, a Majorana mass term for the neutrino is generated when the neutral component of the
Higgs field acquires a vacuum expectation value and the neutrino mass which arises is of the order 
of $m_M \sim g v^2/M_{\Lambda}$. This can be made compatible with neutrino mass bounds for $M_{\Lambda}$
close to the GUT scale. The dimension--5 operator is clearly not fundamental
as an extension, and makes the supersymmetric lagrangian non--renormalizable. Nevertheless, this operator
may arise as an effective term from new physics at the high energy scale $M_{\Lambda}$. 

If we apply this extension at the MSSM lagrangian, we are allowed for a $L$--number violating term
also in the sneutrino lagrangian, which modifies the mass--term potential as:
\begin{equation}
V_{\rm mass} = \left[ m_{L}^{2} + \frac{1}{2} m_{Z}^{2} \cos(2\beta) \right] \snu_{L}^{\ast} \snu_{L} +
\frac{1}{2}m_B ^2 (\snu_L\snu_L + \snu_L^\ast \snu_L^\ast)
\end{equation}
where $m_B$ is a mass parameter that makes the mass lagrangian no longer diagonal in
the $(\snu_L \, \snu_L^\ast)$ basis. In this basis the squared--mass matrix reads:
\begin{eqnarray}
\mathcal{M}^2_{\not{\rm L}} & = &
\begin{pmatrix}
m^2_L+\frac{1}{2} m_{Z}^{2} \cos(2\beta)  & m^2_B \cr
m^2_B & m^2_L+\frac{1}{2} m_{Z}^{2} \cos(2\beta) 
\end{pmatrix}
\end{eqnarray}
and it may be conveniently diagonalized by a rotation into a basis defined by the 
CP--even $\snu_{+}$ and CP--odd $\snu_{-}$ sneutrino eigenstates \cite{Arkani-Hamed:2000bq}:
\begin{eqnarray}
\left\{
\begin{array}{c}
\snu_+ = \left. \frac{1}{\sqrt{2}} \right. (\snu + \snu^{\ast})\\
\snu_- = \left. \frac{-i}{\sqrt{2}} \right. (\snu - \snu^{\ast})
\end{array}
\right.
\label{eq:cpstates}
\end{eqnarray}
The states $\snu_+$ and $\snu_-$ are also mass eigenstates. The squared--mass eigenvalues are easily computed:
\begin{equation}
m^2_{1,2} =  m^2_L+\frac{1}{2} m_{Z}^{2} \cos(2\beta) \pm m_B^2
\label{eq:mb2}
\end{equation}
which implies $\Delta m^2 \equiv m_2^2-m_1^2 = 2 m_B^2$. The mixing angle is fixed at the value $\theta=\pi/4$.
We name this models ``\lviol models". They have the same field content as the minimal MSSM, but include the additional
\lviol terms. The parameters relevant for the sneutrino sector are $m_L$ and $m_B$.

The $Z$ coupling to sneutrinos is non--diagonal in the ($\snu_+$, $\snu_-$) basis, which are now
non--degenerate in mass. The first consequence is that the invisible $Z$--width decay occurs via the process
$Z \rightarrow \nu_1\nu_2$.
%
%
By keeping $m_L \gtrsim 100$ GeV (which assures that the charged lepton mass bounds are satisfied) and
using suitable values of the off-diagonal $m_B$ parameter, we may obtain lightest sneutrinos lighter than
in the minimal MSSM models. From Eq. (\ref{eq:mb2}) is clear that when $m_B$ is large (close to its allowed upper limit, which is due to preventing a tachionic sneutrino) $m_1$ can be small. At the same time $m_2$ is large and the sum of the two never gets smaller than $m_Z$ (for our parameter intervals). In this case, the $Z$ invisible width does not get contributions and light sneutrinos are possible. 

In \lviol models, however, we have an additional bound, which is related to neutrino physics. The \lviol terms may induce radiative contributions to the neutrino masses 
\cite{Grossman:1997is,Hirsch:1997vz,Dedes:2007ef,Hirsch:1997is}. At 1--loop, these corrections arise from diagrams involving sneutrinos and neutralinos. We include this constraint by calculating the 1--loop radiative contribution to the
neutrino mass $\mloop$ as detailed in Ref. \cite{Dedes:2007ef}, and then imposing that $|\mloop|$ does not exceed the experimental upper
bound on the neutrino mass. This constraint coming from the 1--loop contribution $\mloop$
introduces a direct connection of the sneutrino sector to neutrino physics. The 1--loop correction $|\mloop|$ is
basically proportional to the mass difference between the two mass eigenstates $\Delta m_{\rm sneutrino} = m_2 - m_1$
(which in turn depends on the parameter $m_B$, as well as $m_N$). Sneutrino dark matter phenomenology will be therefore
bounded by neutrino physics in a non trivial way. This fact will occur also for the class of models of Section \ref{sec:maj}, where a \lviol term will also be present.

\EPSFIGURE[t]{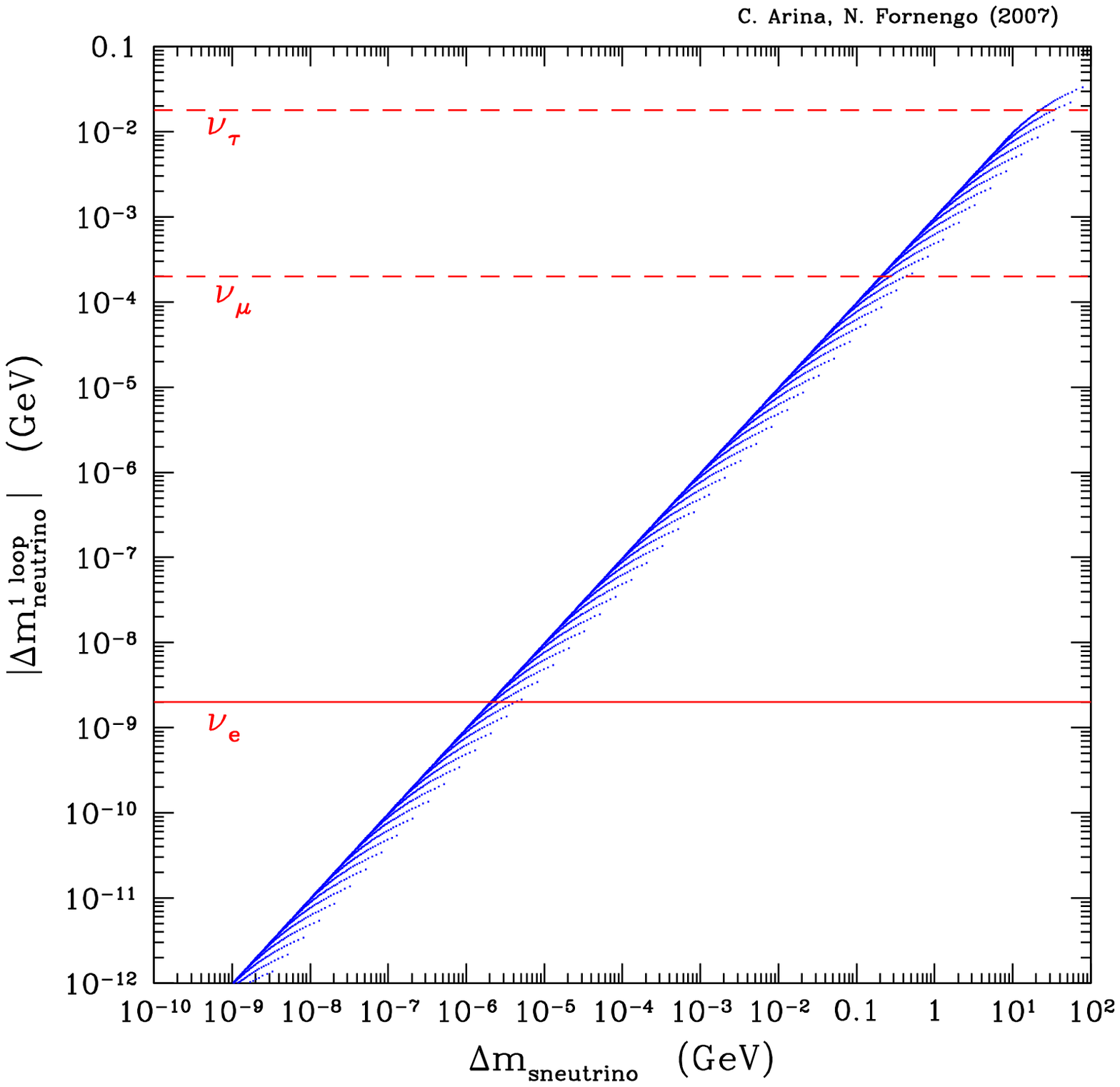,width=0.80\textwidth}
{\lviol models -- Absolute value of the 1--loop contribution to the neutrino mass as a function of
the mass difference of the two sneutrinos. The sneutrino mass parameters are
varied as: $80~\mbox{GeV} \leq m_{L} \leq 1000~\mbox{GeV}$ and 
$10^{-4}~\mbox{GeV} \leq m_{B} \leq 10^{2}~\mbox{GeV}$. The lightest neutralino is a pure bino of 1 TeV mass.
The horizontal lines denote the
upper limits on the neutrino mass, as labelled.
\label{fig:cp-mneutrino}}

\EPSFIGURE[t]{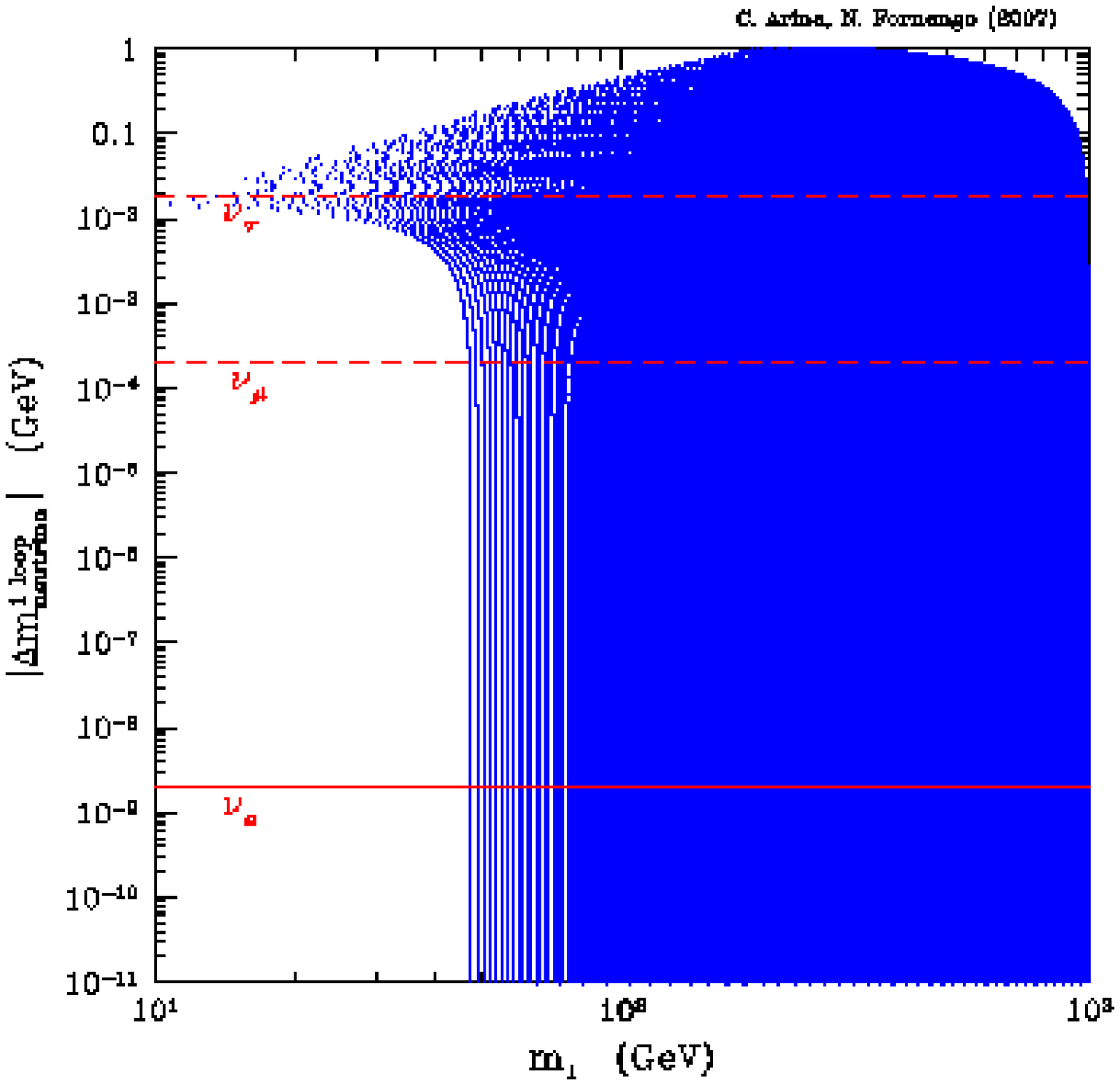,width=0.80\textwidth}
{\lviol models -- Absolute value of the 1--loop contribution to the neutrino mass as a function of
the mass of lightest sneutrino. The parameters as as in Fig. \ref{fig:cp-mneutrino}.
The horizontal lines denote the
upper limits on the neutrino mass, as labelled.
\label{fig:cp-light}}

\EPSFIGURE[t]{./Figures/4_CP_Analysis/CP_omega_m1.eps,width=0.60\textwidth}
{\lviol models -- Sneutrino relic abundance $\Omega h^{2}$ as a function of the sneutrino mass $m_{1}$. 
The [blue] band denotes the relic abundance for allowable models due to a variation of the sneutrino parameters in the intervals:
$80~\mbox{GeV} \leq m_{L} \leq 1000~\mbox{GeV}$ and 
$10^{-4}~\mbox{GeV} \leq m_{B} \leq 10^{2}~\mbox{GeV}$.
The other relevant supersymmetric parameters are: lightest neutralino 30\% heavier than the
sneutrino, higgs masses of 120 GeV for the lightest CP--even higgs $h$ and 400 GeV for the heaviest 
CP--even $H$ and the CP--odd $A$. The horizontal solid and dotted lines delimit the WMAP interval 
for cold dark matter.
\label{fig:cp-omega}}

\EPSFIGURE[t]{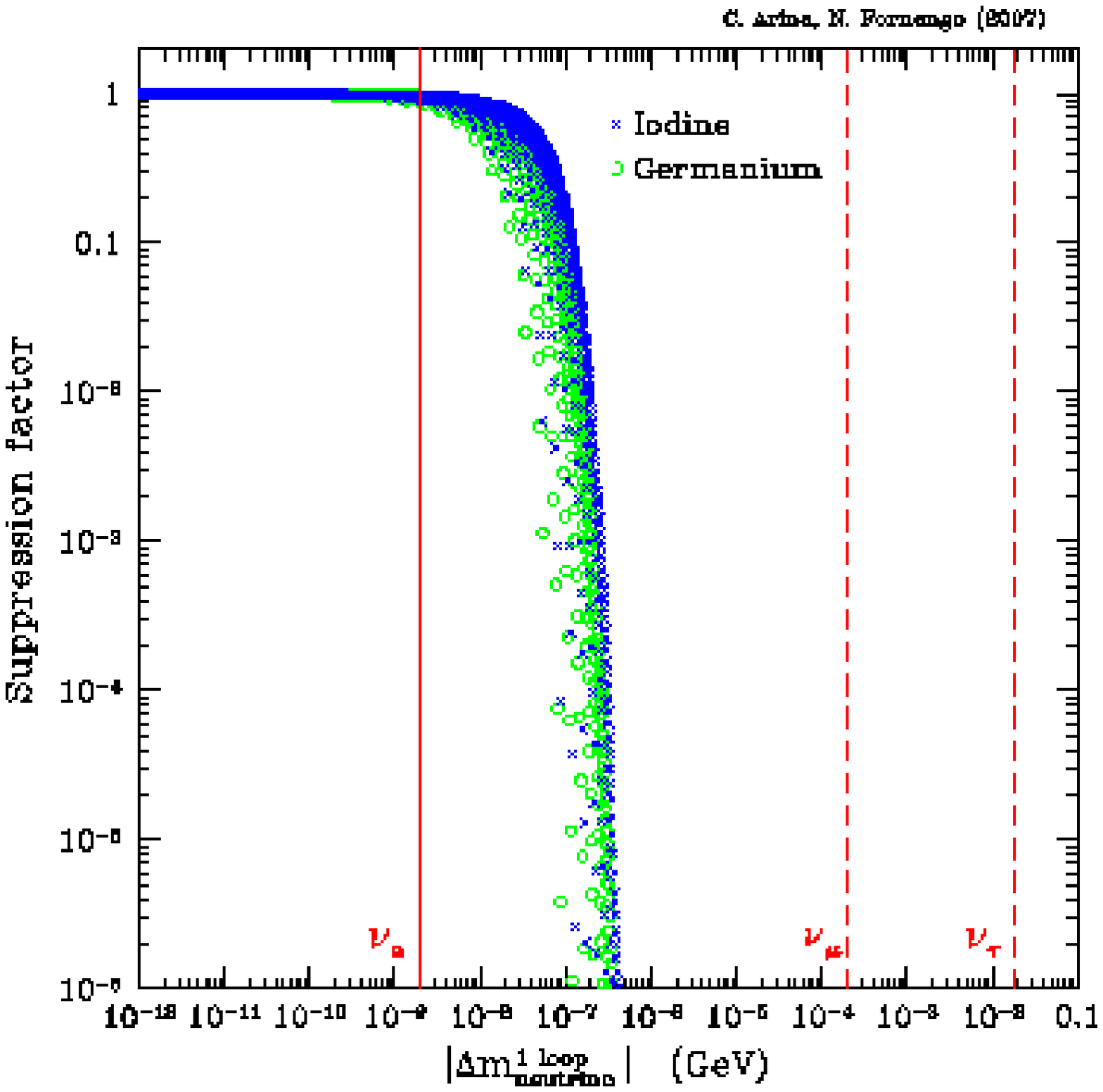,width=0.80\textwidth}
{\lviol models -- Suppression factor of the direct detection rate due to the off--diagonal $Z$--coupling,
plotted vs. the absolute value of the 1--loop contribution to the neutrino mass. The sneutrino mass parameters are varied as: $100~\mbox{GeV} \leq m_{L} \leq 1000~\mbox{GeV}$ and 
$10^{-4}~\mbox{GeV} \leq m_{B} \leq 10^{2}~\mbox{GeV}$. [Blue] crosses refer to the Iodine nucleus,
open [green] circles to the Germanium nucleus.
The vertical lines denote the
upper limits on the neutrino mass, as labelled.
\label{fig:cp-suppression}}

\EPSFIGURE[t]{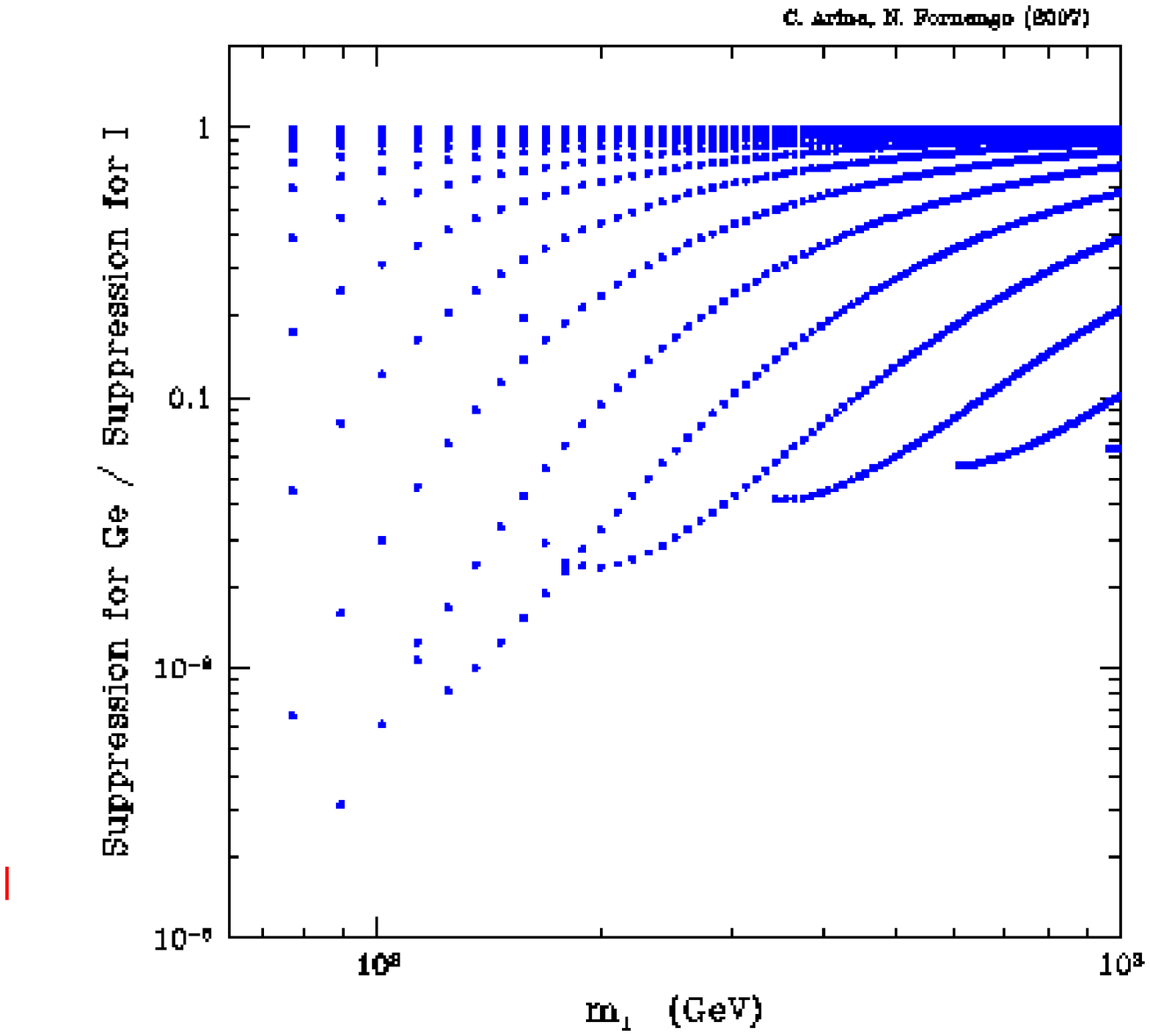,width=0.80\textwidth}
{\lviol models -- Relative suppression in the direct detection rate on Germanium and Iodine, as a function
of the sneutrino mass $m_{1}$. The sneutrino mass parameters are varied as: $100~\mbox{GeV} \leq m_{L} \leq 1000~\mbox{GeV}$ and $10^{-4}~\mbox{GeV} \leq m_{B} \leq 10^{2}~\mbox{GeV}$.
\label{fig:cp-relative}}

\EPSFIGURE[t]{./Figures/4_CP_Analysis/CP_direct_cdms.eps,width=0.60\textwidth}
{\lviol models -- Sneutrino--nucleon scattering cross section $\xi \sigma^{\rm (scalar)}_{\rm nucleon}$ 
as a function of the sneutrino mass $m_{1}$, for the same set of parameters of Fig. \ref{fig:cp-omega}.
The suppression factor in the direct detection rate, due to the off--diagonal $Z$--coupling, is included
in the theoretical predictions.
The dashed, dotted and dot--dashed curves denote the CDMS upper bounds  \cite{ArmelFunkhouser:2005zy,Akerib:2005kh}, as in 
Fig. \ref{fig:std-cdms}.
\label{fig:cp-direct}}

In Fig. \ref{fig:cp-mneutrino} we show the correlation between $|\mloop|$ and the sneutrino mass difference,
for a scan of the sneutrino mass parameters varied in the interval: $80~\mbox{GeV} \leq m_{L} \leq 1000~\mbox{GeV}$ and $10^{-4}~\mbox{GeV} \leq m_{B} \leq 10^{2}~\mbox{GeV}$. In this figure, the lightest neutralino has been
assumed to be a pure bino of 1 TeV mass. As for the neutrino mass bounds, we show in the plot
the limits of kinematical origin: 2 eV for electron--type neutrinos \cite{PDG}, 0.2 MeV for muon--type neutrinos \cite{PDG}
and 18 MeV for tau neutrinos \cite{PDG}. Solar and atmospheric neutrino results 
\cite{GonzalezGarcia:2000sq,GonzalezGarcia:2002dz,Maltoni:2004ei,Fogli:2005cq,Fogli:2004as}
as well as cosmological bounds on massive
light neutrinos \cite{Fogli:2004as,Seljak:2004xh,Lesgourgues:2006nd} 
are not compatible with the kinematical bounds on muon and tau neutrinos listed above, unless 
more than 3 families are present and/or additional sterile neutrinos are introduced with special mixing patterns
with active neutrinos. A mass bound of 2 eV is instead a more reliable upper bound on the neutrino mass \cite{Fogli:2004as,Seljak:2004xh,Lesgourgues:2006nd}, and 
our main conclusion will be based on this case.

Fig. \ref{fig:cp-mneutrino} shows that a mass bound of 2 eV on the neutrino mass implies a strong degeneracy
between the two sneutrino mass eigenstates. In this case, sneutrino phenomenology does not deviate significantly
from the minimal MSSM discussed in Section \ref{sec:std}. Only when a larger mass bound for neutrinos is allowed,
the split between the two mass eigenstates becomes sizeable, reaching the level of 100 MeV for the muon neutrino
kinematical mass bound, and even 30 GeV for the tau neutrino kinematical bound. These large corrections $\mloop$,
and the ensuing large sneutrino mass splittings, occur when one of the sneutrino mass eigenstate is light, opening up 
the possibility of light sneutrinos, which were precluded in the minimal MSSM. This is shown in Fig. 
\ref{fig:cp-light}, where the 1--loop correction $|\mloop|$ is plotted versus $m_1$. This shows that, under
this hypothesis, sneutrinos as light as 10 GeV are possible.

The off-diagonal $Z$ coupling affects also the relic abundance calculations \cite{Hall:1997ah}. In this case, the lightest
sneutrino is actually co--annihilating with the heavier eigenstate, when a $Z$--boson is exchanged (while this
is not the case, for instance, for higgs exchange: in that case the coupling is diagonal). Therefore,
when the mass splitting is large the $Z$--mediated (co)annihilation cross section gets reduced and the 
relic abundance may increase. The relic abundance for the \lviol models is shown in Fig. \ref{fig:cp-omega},
for a mass bound on the neutrino mass of 18 MeV. This would correspond to the case of a tau sneutrino
dark matter, since we are applying the tau neutrino kinematical mass bound. We see that the relic abundance
increases, and light sneutrinos down to 10 GeV are acceptable. If we reduce the mass bound on the neutrinos
down to to 2 eV, this figure reduces to the case of minimal MSSM discussed in Section \ref{sec:std}.

The off-diagonal $Z$ coupling has impact also on direct detection \cite{Hall:1997ah}. In this case, the sneutrino--nucleus
scattering for $Z$ exchange is no longer elastic, since we have to produce the heavier eigenstate. When
the mass difference between the two sneutrinos is large enough, elastic scattering in the detector is
suppressed. The lightest sneutrino can scatter on the nucleus only when:
\begin{equation}
\Delta m < \frac{\beta^2 m_1 m_{\cal N}}{2(m_1+m_{\cal N})}
\end{equation}
This is a nice realization of inelastic dark matter, which was discussed extensively in connection with direct
detection in Refs. \cite{Smith:2001hy,Smith:2002af,Tucker-Smith:2004jv}. 

We therefore have to include this effect in the calculation of the direct detection cross section. This inelasticity
effect produces a suppression in the direct detection rates, which depends on the energy of the recoil, the type
of nucleus and on the energy threshold of the detector. We define a suppression factor for direct detection as:
\begin{equation}
{\cal S} = \frac{{\cal R}(E_1,E_2;\Delta m)}{{\cal R}(E_1,E_2;0)} 
\end{equation}
where ${\cal R}(E_1,E_2;\Delta m)$ denotes the direct detection rate integrated in the energy range $(E_1,E_2)$
and calculated for a sneutrino mass difference $\Delta m$, while ${\cal R}(E_1,E_2;0)$ is the same quantity
calculated for $\Delta m=0$, {\em i.e.} in the standard case. The quenched energies $E^{\rm ee}_1$ $E^{\rm ee}_2$ have been chosen to be: 
$E^{\rm ee}_1 = 2$ KeV and $E^{\rm ee}_2 = 10$ KeV for the Iodine (representative for the DAMA/NAI
experiment) and 
$E^{\rm ee}_1 = 10$ KeV and $E^{\rm ee}_2 = 100$ KeV for the Germanium (representative for the CDMS experiment). The chosen values of $E^{\rm ee}$ correspond to the threshold energy of DAMA/NaI and CDMS. We apply this reduction factor
in the following way: instead of modifying the experimental result (which we cannot do separately for each configuration of the model parameter space), we compare the experimental results obtained for the standard 
case with a redefined scattering cross section:
\begin{equation}
\left [\xi \sigma^{\rm (scalar)}_{\rm nucleon} \right ]_{\rm eff} = 
{\cal S} (\xi \sigma^{\rm (scalar)}_{\rm nucleon})^{Z} +  (\xi \sigma^{\rm (scalar)}_{\rm nucleon})^{h,H} 
\label{eq:xisigmareduced}
\end{equation}
where we have applied the reduction factor only to the $Z$--mediated cross section. These suppression factors
(which apply to the $Z$--exchange case only)
are shown in Fig. \ref{fig:cp-suppression} for the case of a Iodine and a Germanium nucleus. 
The relative reduction between scattering on the Iodine and Germanium nuclei is shown in 
Fig. \ref{fig:cp-relative}, where it is manifest that for sneutrinos lighter than about 200 GeV the
detection rates in CDMS can be much more suppressed than in the DAMA/NAI experiment. This is a consequence
of the different responses of the two detectors to the dark matter velocity distribution function
and mass, as a consequence of the different quenches and threshold energies. This is a practical 
realization of the inelastic dark matter scenario able to explain why current CDMS sensitivity
could be insensitive to some cross sections which explain the DAMA/NAI effect \cite{Smith:2001hy,Smith:2002af,Tucker-Smith:2004jv}. We see that this 
is indeed a possibility for  sneutrinos in LR models.

The effective direct detection cross section of Eq. (\ref{eq:xisigmareduced}) is shown in Fig. \ref{fig:cp-direct}. The combination of the reduction effect due to the inelasticity of the $Z$--exchange contribution and the increase of the relic abundance, which affect \xisigma through the rescaling factor $\xi$, produces
a small effect for heavy sneutrinos. On the contrary, light sneutrinos possess a wide range of
variation of the direct detection cross section, most of which are not excluded by experimental constraints.
This is due to the large mass splitting of the two sneutrino states which effectively suppresses
direct detection through $Z$ exchange. By adopting the highest (dashed) line as a conservative limit from CDMS, we see that whenever $m_{1}<m_{Z}/2$
sneutrinos in \lviol models are viable dark matter candidates. This nevertheless occurs if we allow the
1--loop corrections to the neutrino mass to be as large as the tau--neutrino kinematical mass limit: when
a more reliable bound of 2 eV is assumed, we basically recover the minimal MSSM case (which also implies
a lower bound on the sneutrino mass of about 80 GeV).

In conclusion, once a neutrino mass bound of 2 eV is assumed, the \lviol models do not exhibit a phenomenology much richer than the minimal MSSM. Direct detection almost excludes this possibility, except for the occurrence of mass--matching condition between the sneutrino mass and the $Z$ or higgs masses (like in the case of minimal MSSM), and also in this case the compatibility with direct detection is marginal. 

For these reason we do not elaborate any further on the \lviol models, and we do not discuss 
neither the full supersymmetric parameter--space scan neither indirect detection signals for this case. 
                 
\section{Models with right--handed sneutrinos and lepton--number violating interactions: the case
for a see-saw neutrino mass}\label{sec:maj}

A supersymmetric model which can accommodate a Majorana mass--term for neutrinos and explain the 
observed neutrino mass pattern, and which relies on a renormalizable lagrangian, may be built by adding to the minimal MSSM right--handed fields $\tilde N^{I}$
and allowing for \lviol terms . The most general form of the superpotential which accomplishes this
conditions is \cite{Arkani-Hamed:2000bq,Grossman:1997is}:
\begin{equation}
W = \epsilon_{ij} (\mu \hat H^{1}_{i} \hat H^{2}_{j} - Y_{l}^{IJ} \hat H^{1}_{i} \hat L^{I}_{j} \hat R^{J}
+ Y_{\nu}^{IJ} \hat H^{2}_{i} \hat L^{I}_{j} \hat N^{J} ) + \frac{1}{2} M^{IJ} \hat N^I \hat N^J
\end{equation}
where $M^{IJ}$, $Y_{l}^{IJ}$ and $Y_{\nu}^{IJ}$ are matrices, which we again choose real and diagonal.
For the \lviol parameters we therefore assume: $M^{IJ}=M\;\delta^{IJ}$, in order to reduce the number of free parameters. The Dirac mass of the neutrinos is obtained as: $m_D^{I} = v_{2}Y_{\nu}^{II}$, while $M^{I}$ represent a Majorana mass--term for neutrinos.

The general form of the soft supersymmetry--breaking potential may be written as \cite{Dedes:2007ef}:
\begin{eqnarray}
V_{\rm soft} &=& (M_{L}^{2})^{IJ} \, \tilde L_{i}^{I \ast} \tilde L_{i}^{J} + 
(M_{N}^{2})^{IJ} \, \tilde N^{I \ast} \tilde N^{J} - \nonumber\\
& & 
 [(m^2_B)^{IJ}\tilde{N}^I\tilde{N}^J+\epsilon_{ij}(\Lambda_{l}^{IJ} H^{1}_{i} \tilde L^{I}_{j} \tilde R^{J} + 
\Lambda_{\nu}^{IJ} H^{2}_{i} \tilde L^{I}_{j} \tilde N^{J})  + \mbox{h.c.}]
\end{eqnarray}
where we again use the same assumptions of diagonality in flavour space for the matrices $(M_{L}^{2})^{IJ}$,
$(M_{N}^{2})^{IJ}$, $(m_{B}^{2})^{IJ}$, $\Lambda_{l}^{IJ}$ and $\Lambda_{\nu}^{IJ}$ as we already did before. The Dirac--mass parameter is not chosen as a free parameter: it is instead derived by the condition that the neutrino mass is determined by the see-saw mechanism. In this case: $m^{I}_{\nu} = m^{I}_{D}/M^{2}$. In our analyses we will fix, for definiteness, the neutrino mass
to be 2 eV, in order to saturate the bound which comes from the radiative contribution to the neutrino
mass discussed in the previous Section. The naming convention for this class of models is ``MAJ models``''.

Sneutrinos now are a superpositions of two complex fields: the left--handed field $\nu_{L}$ and the
right--handed field $\tilde N$. Since we introduced \lviol terms, it is convenient, as we did in the
previous Section, to work in a basis of CP eigenstates. By defining them accordingly to 
Eq. (\ref{eq:cpstates}), the mass matrix for the state vector 
$\Phi^{\dag}_{\rm MAJ} = (\snu_{+}^\ast \,\,\tilde{N}_{+}^\ast \,\, \snu_-^\ast \,\, \tilde{N}_-^\ast)$
has a the form:
\begin{eqnarray}
&& \mathcal{M}^2_{\rm MAJ}  = \\
&& \begin{pmatrix}
m^2_L+D^2+m^2_D & F^2 + m_D M & 0  & 0\cr
F^2+m_D M & m^2_N+M^2+m^2_D+m^2_B & 0 & 0 \cr
0 & 0 & m^2_L+D^2+m^2_D & F^2-m_D M \cr
0 & 0 & F^2- m_D M & m^2_N+M^2+m^2_D-m^2_B
\end{pmatrix}\nonumber
\label{eq:majmass}
\end{eqnarray}
where $D^2 = 0.5\, m_{Z}^{2} \cos(2\beta)$ and $F^2$ has already been defined in Eq. (\ref{eq:f2}).
The free parameters of the models for sneutrino sector are therefore: $m_L$, $m_N$, $M$, $m_B$ and $F^2$. 

Sneutrino mass eigenstates are obtained by diagonalizing Eq. (\ref{eq:majmass}). We define them as follows:
\begin{eqnarray}
\snu_i= Z_{i1}\snu_{+}+Z_{i2}\tilde{N}_+ +Z_{i3}\snu_{-}+Z_{i4}\tilde{N}_{-} \qquad i=1,2,3,4
\end{eqnarray}
The lightest state, which is our dark matter candidate, may now exhibit a mixing with the right--handed
field $\tilde N$ and the non--diagonal nature of the $Z$--coupling with respect of the CP eigenstates.
These models therefore share the properties of both LR models and \lviol models, but bring in
a more rich set of parameters. In general, the terms $F^{2}\pm m_{D} M$ induce left--right mixing on
the sneutrino eigenstates, while the $m_{B}^{2}$ term is responsible for CP splitting. The new Majorana--mass
parameter $M$ may lead to left--right mixing for its presence in the off-diagonal elements of the mass
matrix $\mathcal{M}^2_{\rm MAJ}$, and, if it is large, can drive the mass of the two heavier mass 
eigenstates, since it enters also in the diagonal elements. More specifically, if $F^{2}=0$ $M$
can lead only to a left--right mixing, while if $F\neq 0$ it can contribute also to the CP splitting.
A sizeable splitting however occurs when $m_{D} M \sim F^{2}$ are of the order of the diagonal
elements.

The actual phenomenology of the lightest sneutrino is
therefore due to the relative values of the various parameters. While $m_{L}$, which is in common
with the charged slepton sector, has to be necessarily larger than about 80--100 GeV to fulfill the
charged sleptons mass bounds, the other parameters are free in nature and not directly related to
the electroweak symmetry--breaking and its scale (at least at the tree level). The right--handed parameter $m_{N}$
and the CP--splitting parameter $m_{B}$ may be taken to be of the same order of magnitude as $m_{L}$,
but since they are in general independent from $m_{L}$ we assume them to vary freely, as we did
in the previous Sections. The same occurs for the parameter $F^{2}$. The
parameter $M$ is related to the Majorana mass of neutrinos: a natural scale for this parameter is therefore
a high--energy scale, larger than $10^{9}$ GeV, in order to obtain a see-saw neutrino mass below the eV scale with a Dirac--mass term of the order of the GeV scale. However, a value of $M=1$ TeV which produces
a eV neutrino mass with $m_{D}\sim 1$ MeV is an equally viable possibility, since $m_{D}$ is
a Dirac--type mass which originates from a neutrino Yukawa interaction of unknown strength. We therefore
distinguish two cases in our analysis: a low--scale Majorana mass, which we fix at the value of 1 TeV
(this class of models are named for convenience MAJ[A]); a large--scale Majorana mass, which we fix at
$10^{9}$ GeV (models MAJ[B]). We will see that a different phenomenology arises.
For a further discussion of these parameters and their ranges, see also Ref. \cite{Dedes:2007ef}.

The experimental constraints we adopt for the MAJ models (in addition to the one discussed in
Appendix \ref{app:mssm}) are the invisible $Z$--width for all the relevant mass eigenstates
at hand and the neutrino mass bound on the 1--loop corrections to the neutrino mass $|\mloop| \leq 2$ eV.
With our choice of the parameters, we do not have terms which mix different families
and therefore we do not generate contributions to the flavour--violating decays of charged leptons, as
in general may be the case \cite{Dedes:2007ef}. We therefore do not have additional bounds from limits on
processes like $\mu \rightarrow e+\gamma$ decays.

\subsection{Models with a TeV--scale Majorana mass--term}\label{sec:maja}

Let us first dicuss the case of TeV--scale Majorana mass term $M$. In this case, seizeable
mass splittings are possible and the mass bound from the radiative contribution to the
neutrino mass is relevant. Fig. \ref{fig:maja-mneutrino} shows $|\mloop|$ versus the mass
splitting of the two lightest sneutrino eigenstate, for $M=1$ TeV and a scan of the sneutrino relevant
parameters in the ranges: $10^{2}~\mbox{GeV} \leq m_{N} \leq 10^{3}~\mbox{GeV}$,
$1~\mbox{GeV} \leq m_{B} \leq 10^{3}~\mbox{GeV}$ and  $1~\mbox{GeV}^{2} \leq F^{2} \leq 10^{4}~\mbox{GeV}^{4}$. We see that, contrary to the case of the pure \lviol models discussed in the previous Section,
large sneutrino mass splittings are possible even with a neutrino mass bound of 2 eV.  This is related to the fact that the large mass splittings are here related to a large mixing with the right--handed field 
$\tilde N$: this therefore suppresses the sneutrino couplings and as a consequence also the 1-loop contribution to the neutrino mass. This is manifest in
Fig. \ref{fig:maja-sterile} where we show, for the same variation of parameters, the distribution
of the sterile component of the lightest sneutrino state versus the mass splitting. Typically,
when the sneutrino is mostly left--handed the mass splitting is also small, and the
sterile component increases with the increase of the mass separation.

\EPSFIGURE[t]{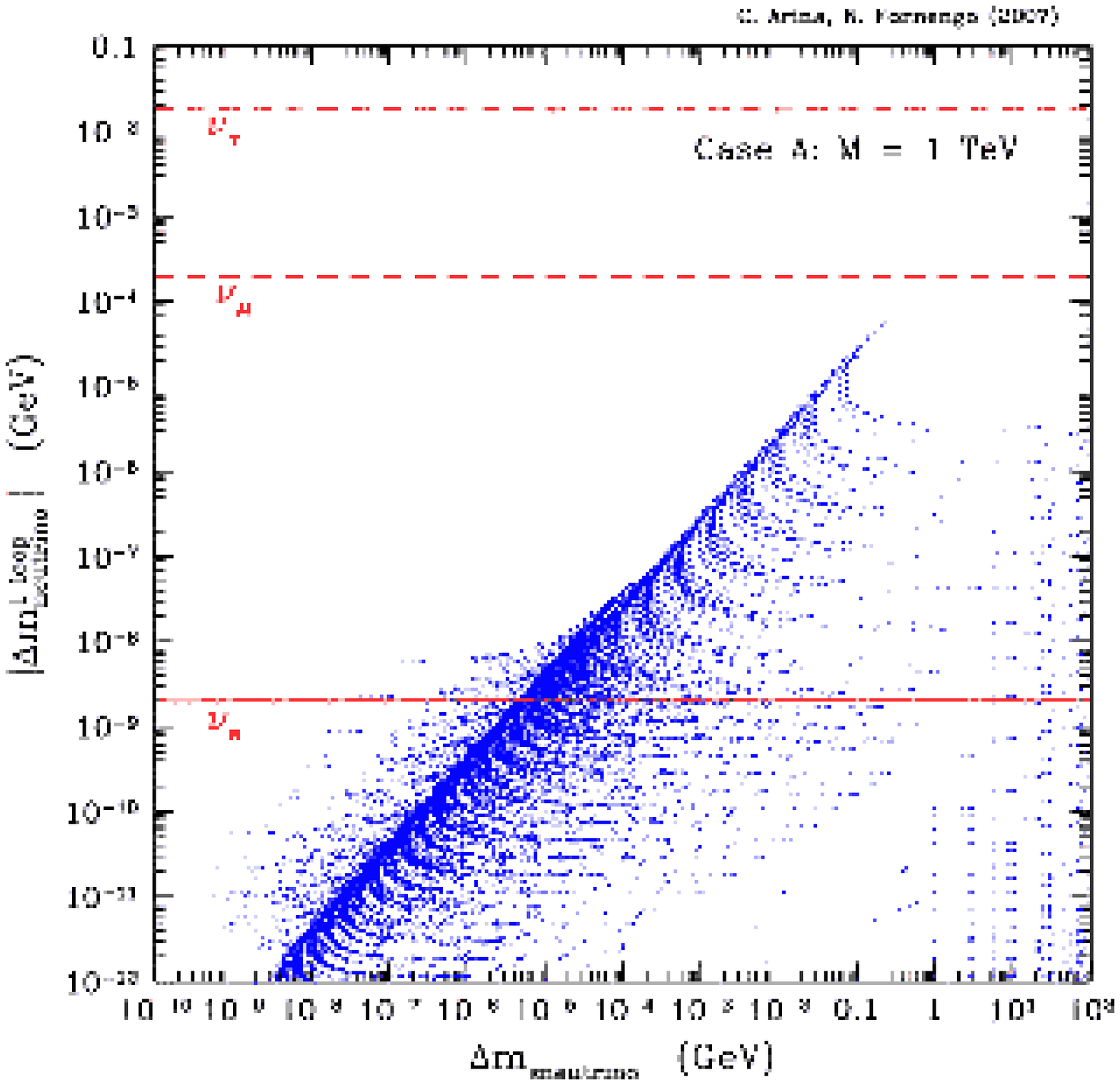,width=0.80\textwidth}
{MAJ[A] models -- Absolute value of the 1--loop contribution to the neutrino mass as a function of
the mass difference of the two lightest sneutrino CP eigenstates, for the case of a 
Majorana--mass parameter $M = 1$ TeV.
The other sneutrino mass parameters are varied as:
$10^{2}~\mbox{GeV} \leq m_{N} \leq 10^{3}~\mbox{GeV}$,
$1~\mbox{GeV} \leq m_{B} \leq 10^{3}~\mbox{GeV}$ and 
$1~\mbox{GeV}^{2} \leq F^{2} \leq 10^{4}~\mbox{GeV}^{4}$.
The horizontal lines denote the
upper limits on the neutrino mass, as labelled.
\label{fig:maja-mneutrino}}

\EPSFIGURE[t]{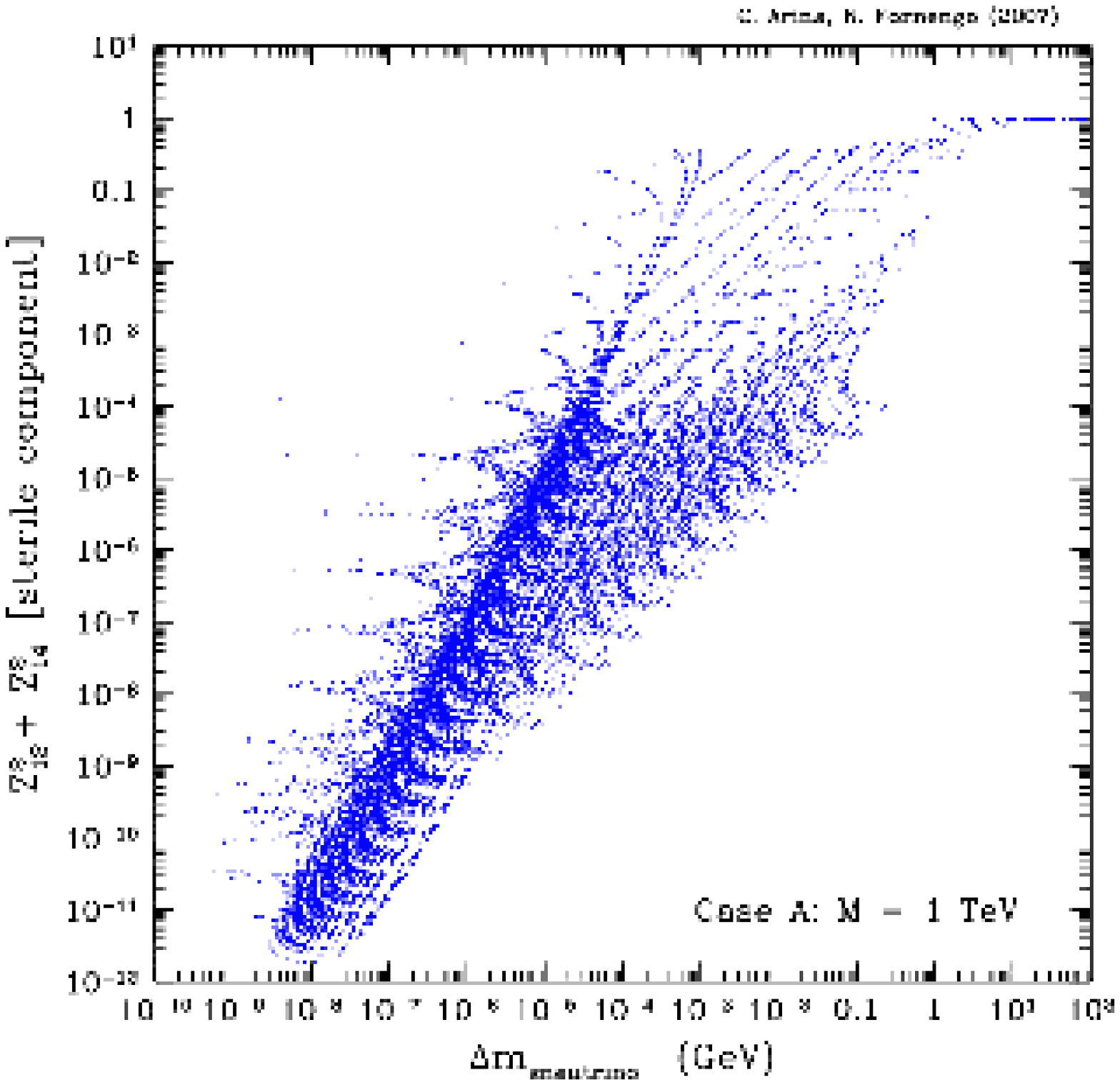,width=0.80\textwidth}
{MAJ[A] models -- Component of the lightest sneutrino into sterile fields as a function of
the mass splitting of the two lightest sneutrinos, for the case of a 
Majorana--mass parameter $M = 1$ TeV. The other sneutrino mass parameters are varied as
in Fig. \ref{fig:maja-mneutrino}.
\label{fig:maja-sterile}}

\EPSFIGURE[t]{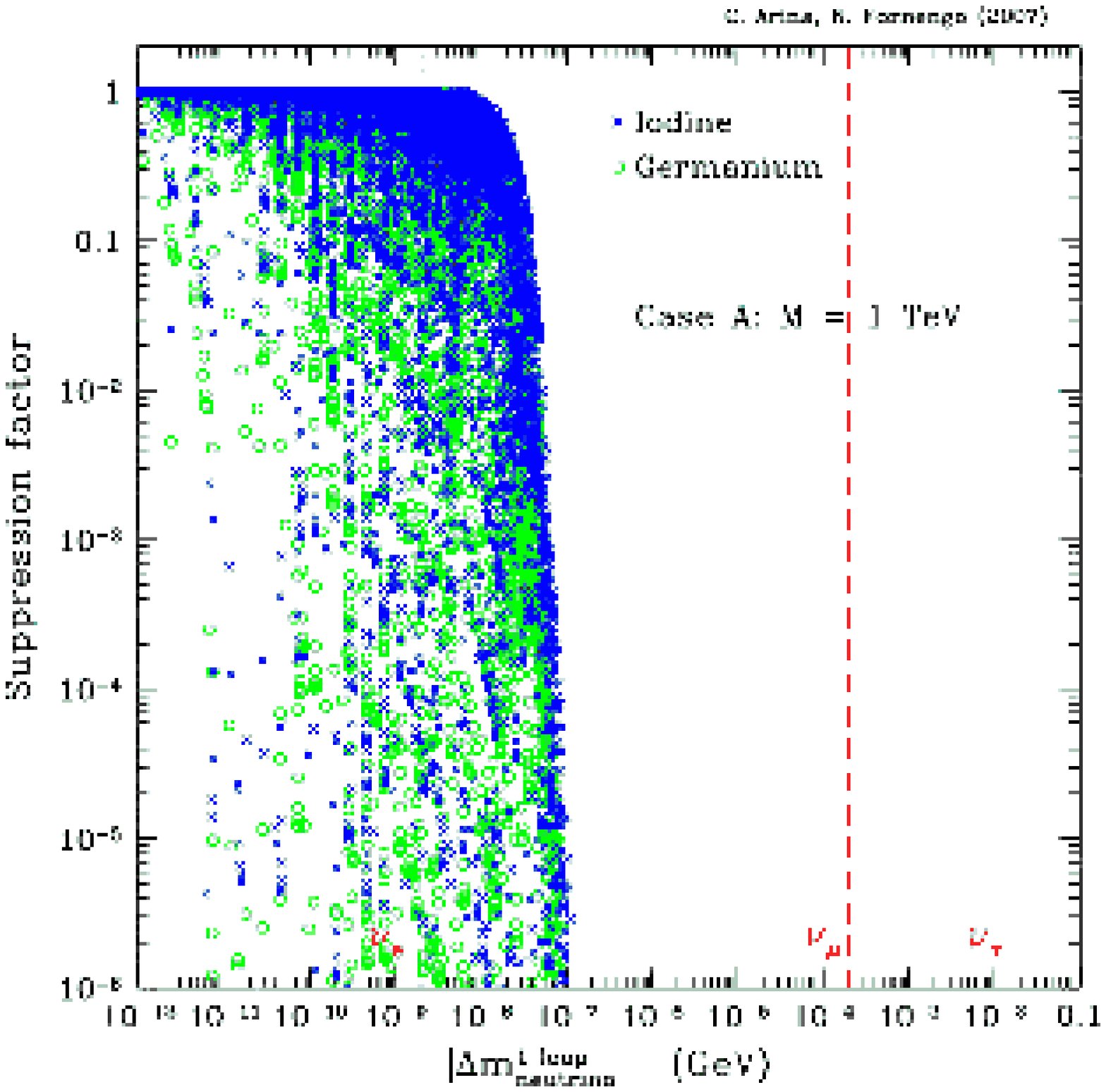,width=0.80\textwidth}
{MAJ[A] models -- Suppression factor of the direct detection rate due to the off--diagonal $Z$--coupling
plotted vs. the absolute value of the 1--loop contribution to the neutrino mass, for the case of a 
Majorana--mass parameter $M = 1$ TeV. The other sneutrino mass parameters are varied as
in Fig. \ref{fig:maja-mneutrino}. [Blue] crosses refer to the Iodine nucleus,
open [green] circles to the Germanium nucleus. The vertical lines denote the
upper limits on the neutrino mass, as labelled.
\label{fig:maja-suppression}}

\EPSFIGURE[t]{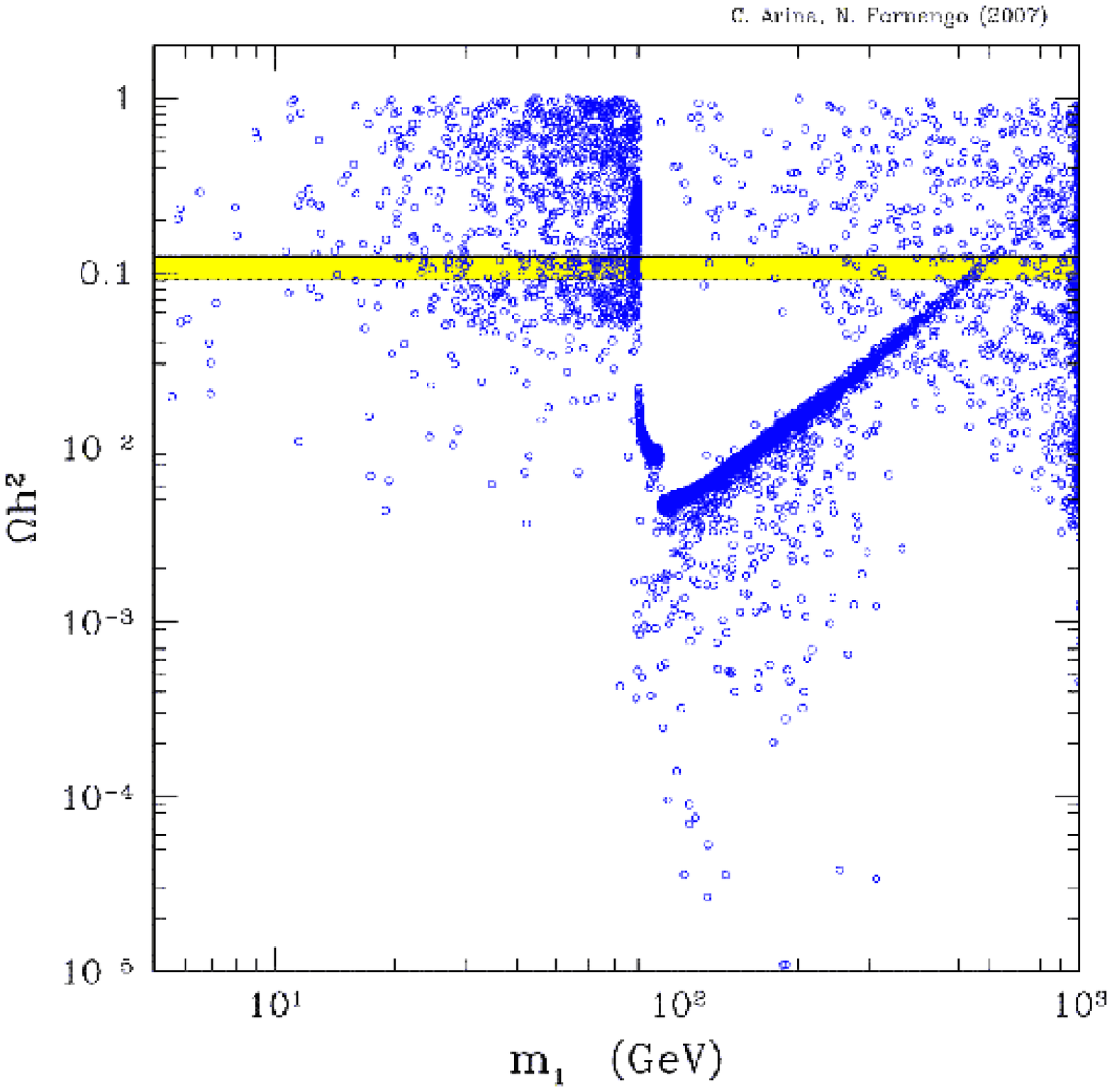,width=0.80\textwidth}
{MAJ[A] models -- Sneutrino relic abundance $\Omega h^{2}$ as a function of 
the sneutrino mass $m_{1}$, for the case of a Majorana--mass parameter $M = 1$ TeV and 
for a full scan of the supersymmetric parameter space. The sneutrino parameters (other than $M$) 
are varied as follows: 
$10^{2}~\mbox{GeV} \leq m_{N} \leq 10^{3}~\mbox{GeV}$,
$1~\mbox{GeV} \leq m_{B} \leq 10^{3}~\mbox{GeV}$ and 
$1~\mbox{GeV}^{2} \leq F^{2} \leq 10^{4}~\mbox{GeV}^{4}$.
All the models shown in the plot are acceptable from the point of view of all 
experimental constraints. The horizontal solid and 
dotted lines delimit the WMAP interval for cold dark matter.
\label{fig:maja-omegascan}}

\EPSFIGURE[t]{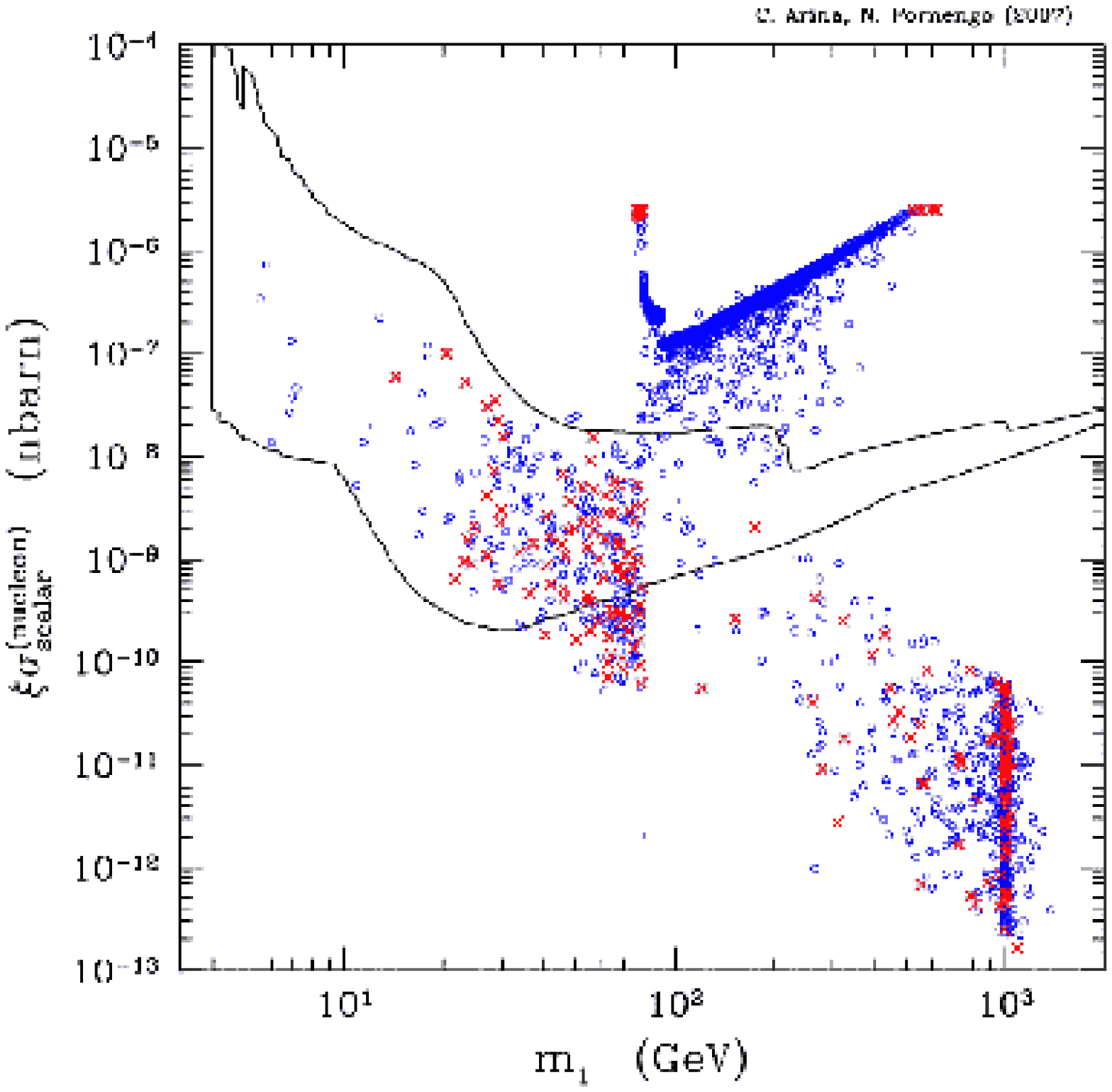,width=0.80\textwidth}
{MAJ[A] models -- Sneutrino--nucleon scattering cross section $\xi \sigma^{\rm (scalar)}_{\rm nucleon}$ 
as a function of the sneutrino mass $m_{1}$, for the case of a Majorana--mass parameter $M = 1$ TeV and 
for a full scan of the supersymmetric parameter space. Parameters are varied as in Fig. \ref{fig:maja-omegascan}.
[Red] crosses refer to models with sneutrino relic abundance
in the cosmologically relevant range; [blue] open circles refer to cosmologically subdominant 
sneutrinos. The solid curve shows the DAMA/NaI region, compatible with the annual 
modulation effect observed by the experiment \cite{Bernabei:2003za,Bernabei:2005hj,Bernabei:2005ca,Bernabei:2006ya,Bernabei:2007jz}.
\label{fig:maja-directscan}}

\EPSFIGURE[t]{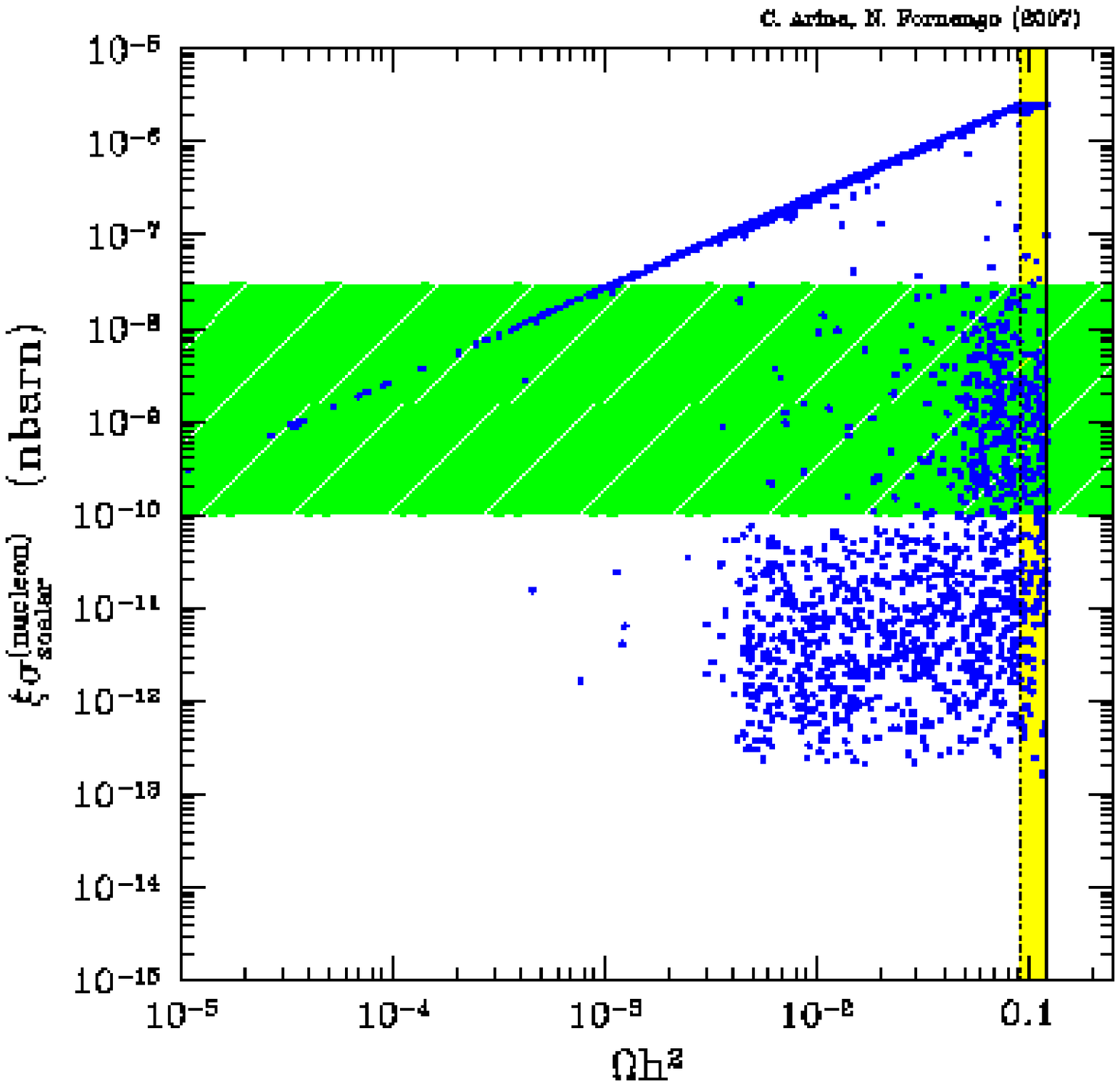,width=0.80\textwidth}
{MAJ[A] models -- Sneutrino--nucleon scattering cross section $\xi \sigma^{\rm (scalar)}_{\rm nucleon}$ 
vs. the sneutrino relic abundance $\Omega h^{2}$, for the case of a Majorana--mass parameter 
$M = 1$ TeV and for a full scan of the supersymmetric 
parameter space. Parameters are varied as in Fig. \ref{fig:maja-omegascan}. The horizontal [green] 
band denotes the current sensitivity of direct detection experiments; the vertical [yellow] band
delimits the WMAP interval for cold dark matter.
\label{fig:maja-sigmaomega}}

\EPSFIGURE[t]{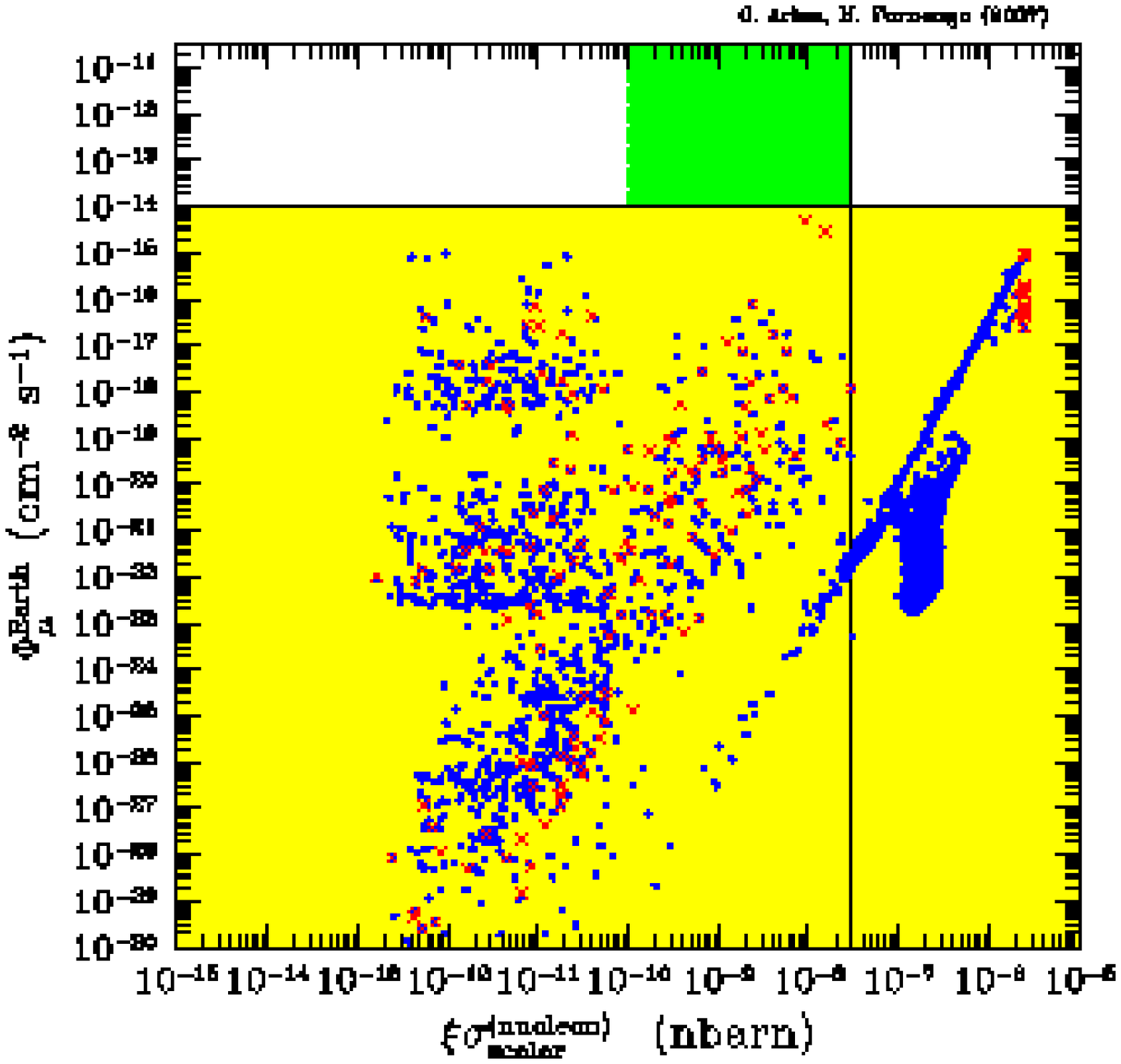,width=0.80\textwidth}
{MAJ[A] models -- Upgoing muon flux from sneutrino pair annihilation in the center of the Earth
$\Phi_{\mu}^{\rm Earth}$ vs. the sneutrino--nucleon scattering cross section 
$\xi \sigma^{\rm (scalar)}_{\rm nucleon}$, for the case of a Majorana--mass parameter $M = 1$ TeV
and a full scan of the supersymmetric parameter space.
Parameters are varied as in Fig. \ref{fig:maja-omegascan}.
[Red] crosses refer to models with sneutrino 
relic abundance in the cosmologically relevant range; [blue] dots refer to cosmologically 
subdominant sneutrinos. The horizontal solid line denotes the current upper limit from
neutrino telescopes. The vertical solid line denotes a conservative upper limit from direct detection 
searches and the [green] vertical shaded area refers to the current sensitivity in direct detection searches.
\label{fig:maja-fluxedirect}}

\EPSFIGURE[t]{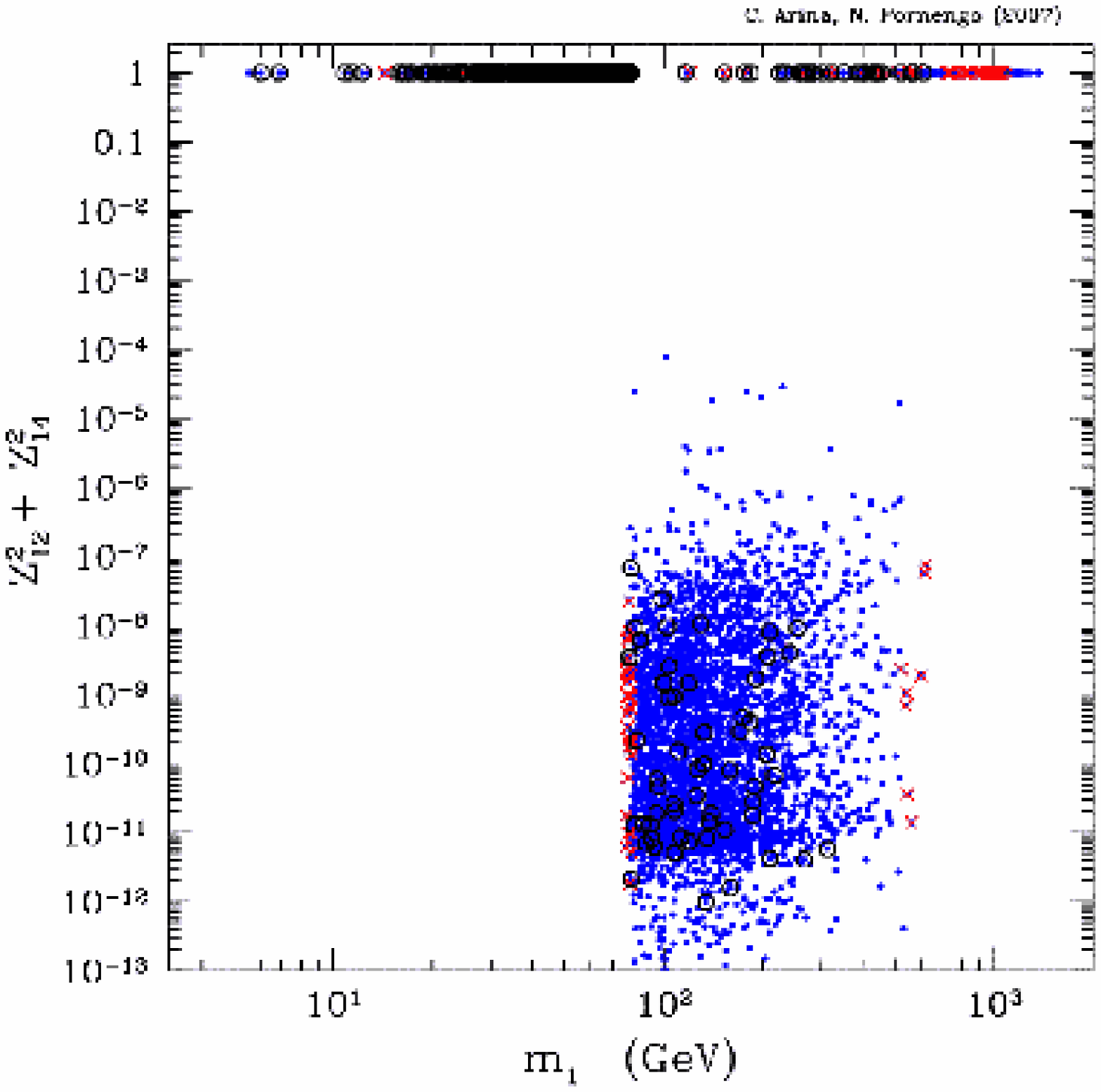,width=0.80\textwidth}
{MAJ[A] models -- Distribution of cosmologically acceptable models in the 
$m_{1}$ vs. sterile--component plane, for the case of a Majorana--mass parameter $M = 1$ TeV
and a full scan of the supersymmetric parameter space.
Parameters are varied as in Fig. \ref{fig:maja-omegascan}. [Red] crosses refer to models with sneutrino relic abundance
in the cosmologically relevant range; [blue] dots refer to cosmologically subdominant 
sneutrinos; open dots mark the configurations which have a direct--detection cross section
in the current sensitivity range.
\label{fig:maja-m1sterile}}

\EPSFIGURE[t]{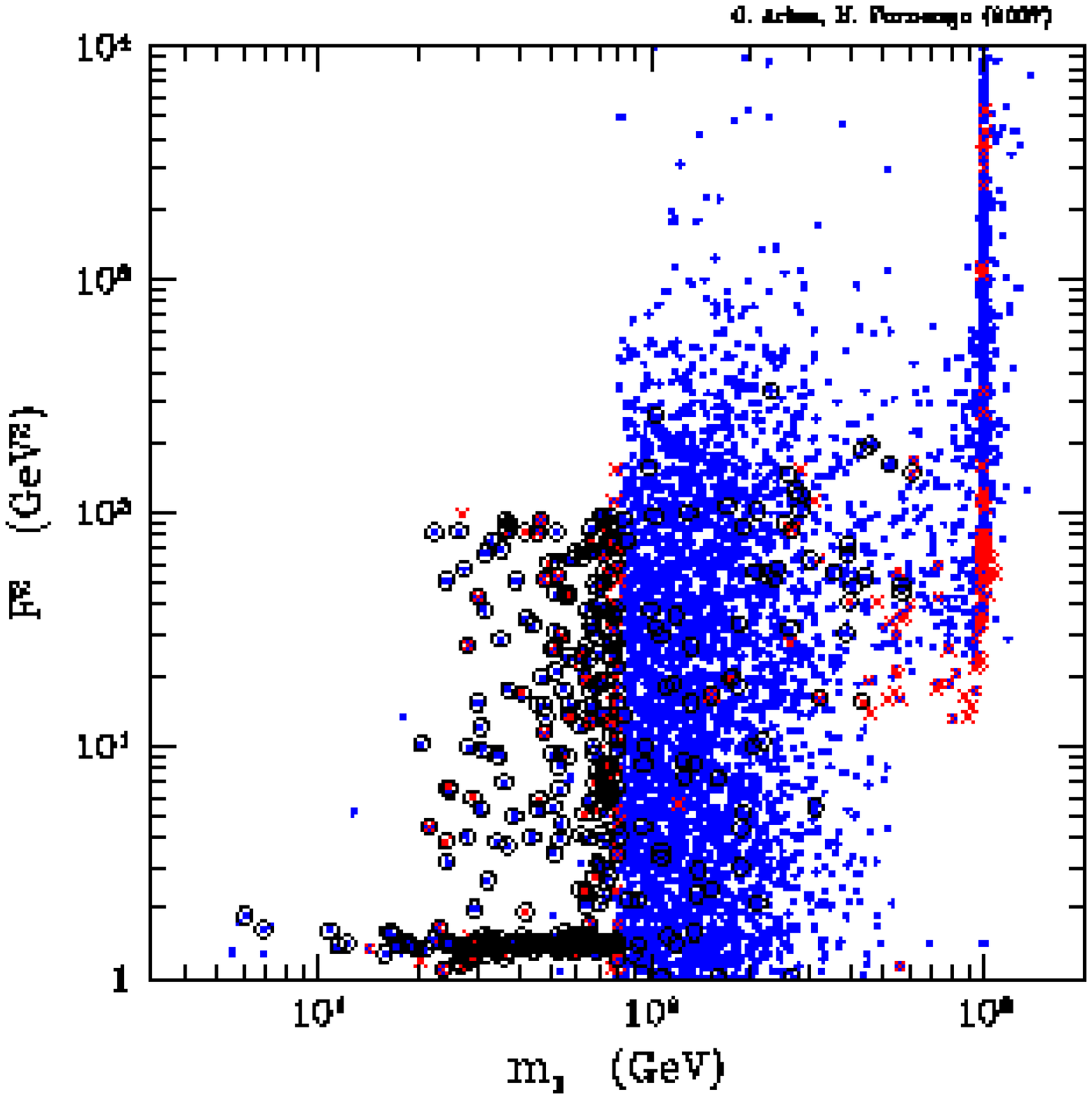,width=0.80\textwidth}
{MAJ[A] models -- Distribution of cosmologically acceptable models in the 
$m_{1}$ -- $F^{2}$ plane, for the case of a Majorana--mass parameter $M = 1$ TeV
and a full scan of the supersymmetric parameter space.
Parameters are varied as in Fig. \ref{fig:maja-omegascan}. [Red] crosses refer to models with sneutrino relic abundance
in the cosmologically relevant range; [blue] dots refer to cosmologically subdominant 
sneutrinos; open dots mark the configurations which have a direct--detection cross section
in the current sensitivity range.
\label{fig:maja-m1f2}}

\EPSFIGURE[t]{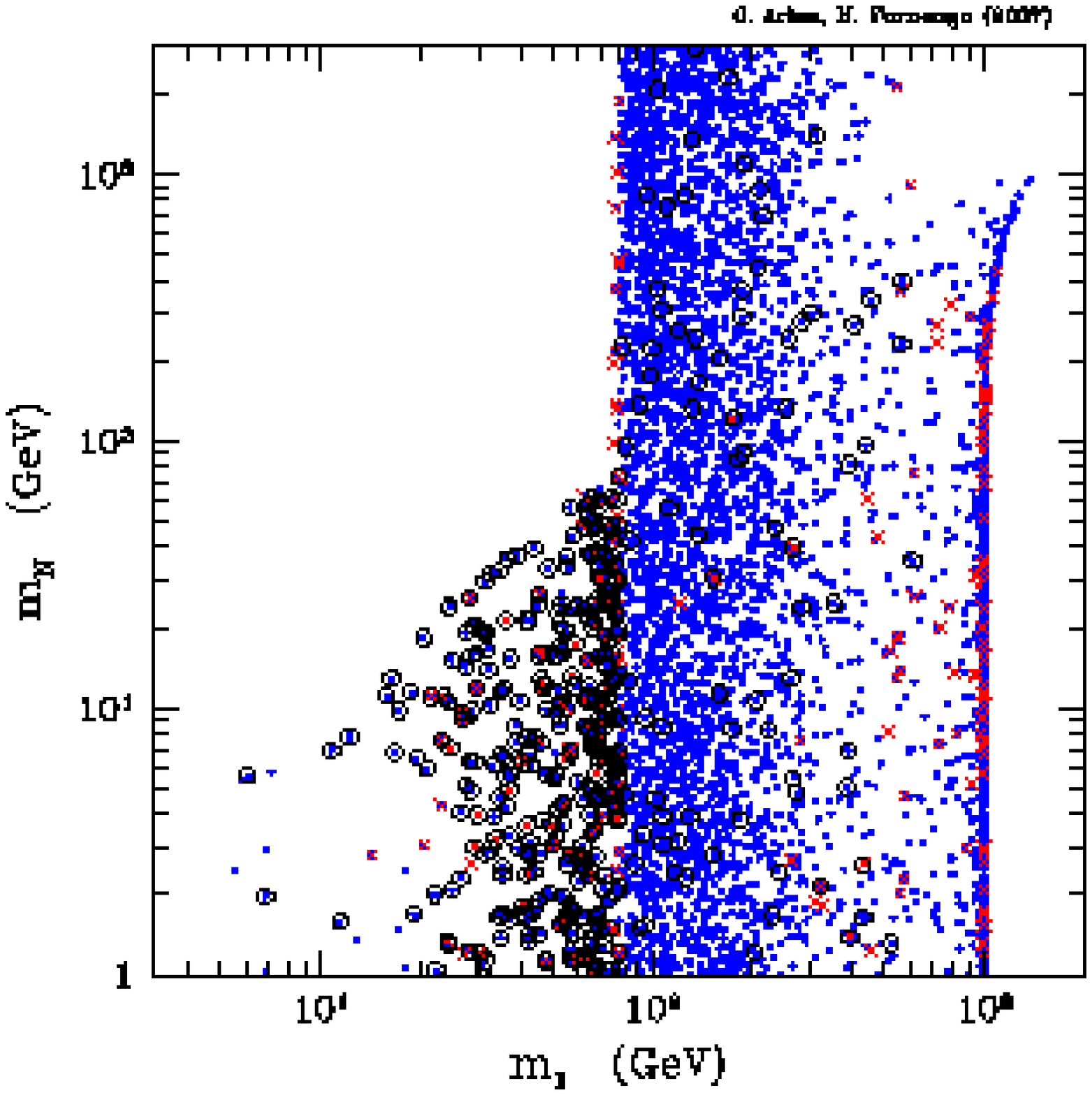,width=0.80\textwidth}
{MAJ[A] models -- Distribution of cosmologically acceptable models in the 
$m_{1}$ -- $m_{N}$ plane, for the case of a Majorana--mass parameter $M = 1$ TeV
and a full scan of the supersymmetric parameter space.
Parameters are varied as in Fig. \ref{fig:maja-omegascan}. [Red] crosses refer to models with sneutrino relic abundance
in the cosmologically relevant range; [blue] dots refer to cosmologically subdominant 
sneutrinos; open dots mark the configurations which have a direct--detection cross section
in the current sensitivity range.
\label{fig:maja-m1mn}}

\EPSFIGURE[t]{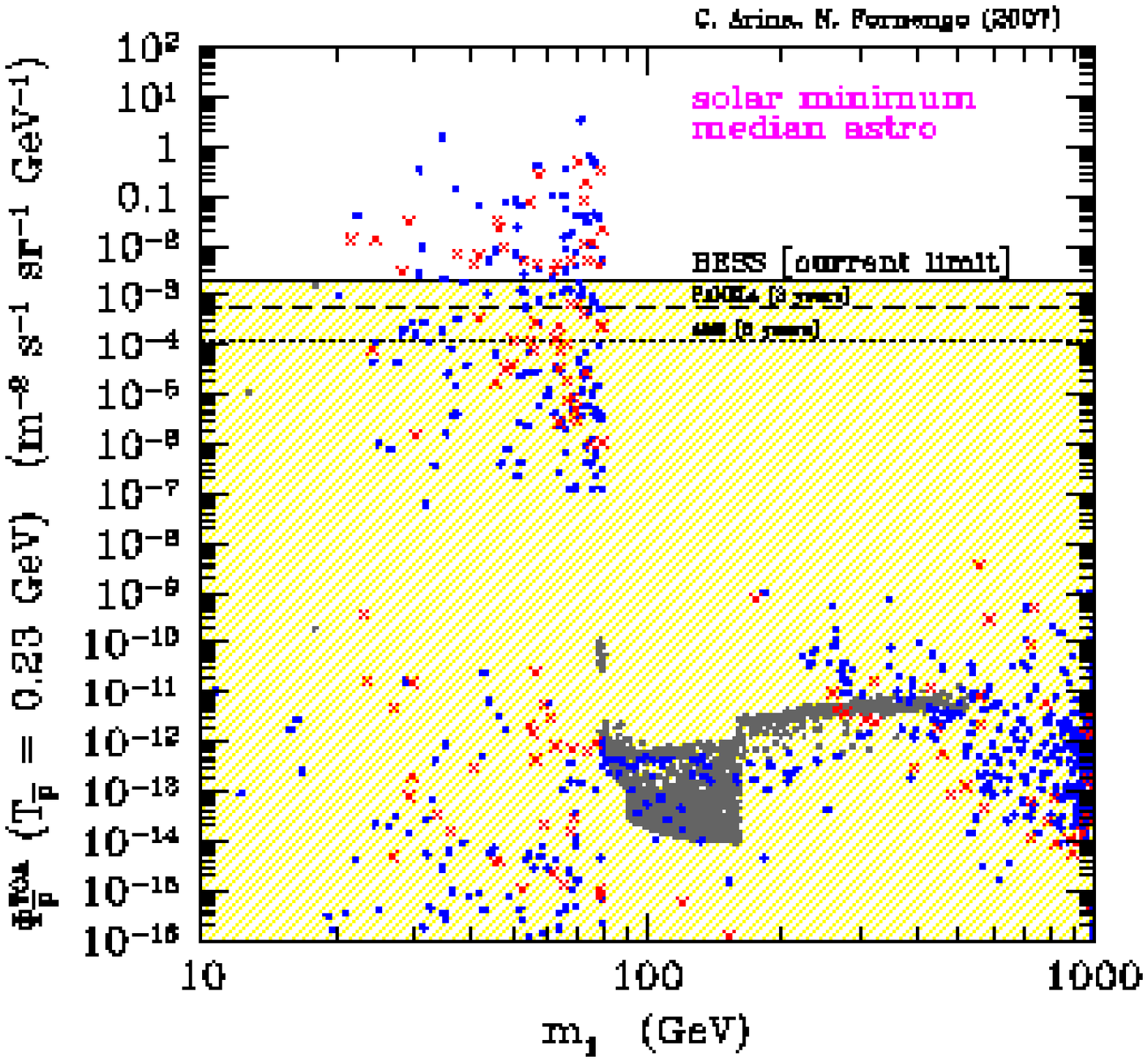,width=0.80\textwidth}
{MAJ[A] models -- Antiproton flux at the antiproton kinetic energy $T_{\bar p} = 0.23$ GeV as a 
function of the sneutrino mass $m_{1}$, for the galactic propagation parameters which provide the
median value of antiproton flux and for a solar activity at its minimum. 
The plot refers to the case of a Majorana--mass parameter $M = 1$ TeV
and a full scan of the supersymmetric parameter space.
Parameters are varied as in Fig. \ref{fig:maja-omegascan}.
[Red] crosses refer to models with sneutrino relic  abundance in the cosmologically relevant range; [blue] dots refer to cosmologically 
subdominant sneutrinos; light gray points denote configurations which are excluded by direct
detection searches. The [yellow] shaded area denotes the amount of exotic antiprotons which can be accommodated in the BESS data \cite{Orito:1999re,Maeno:2000qx} . The dashed and dotted lines show the
PAMELA \cite{pam} and AMS \cite{ams} sensitivities to exotic antiprotons for 3 years missions, respectively. 
\label{fig:maja-pbar023}}

\EPSFIGURE[t]{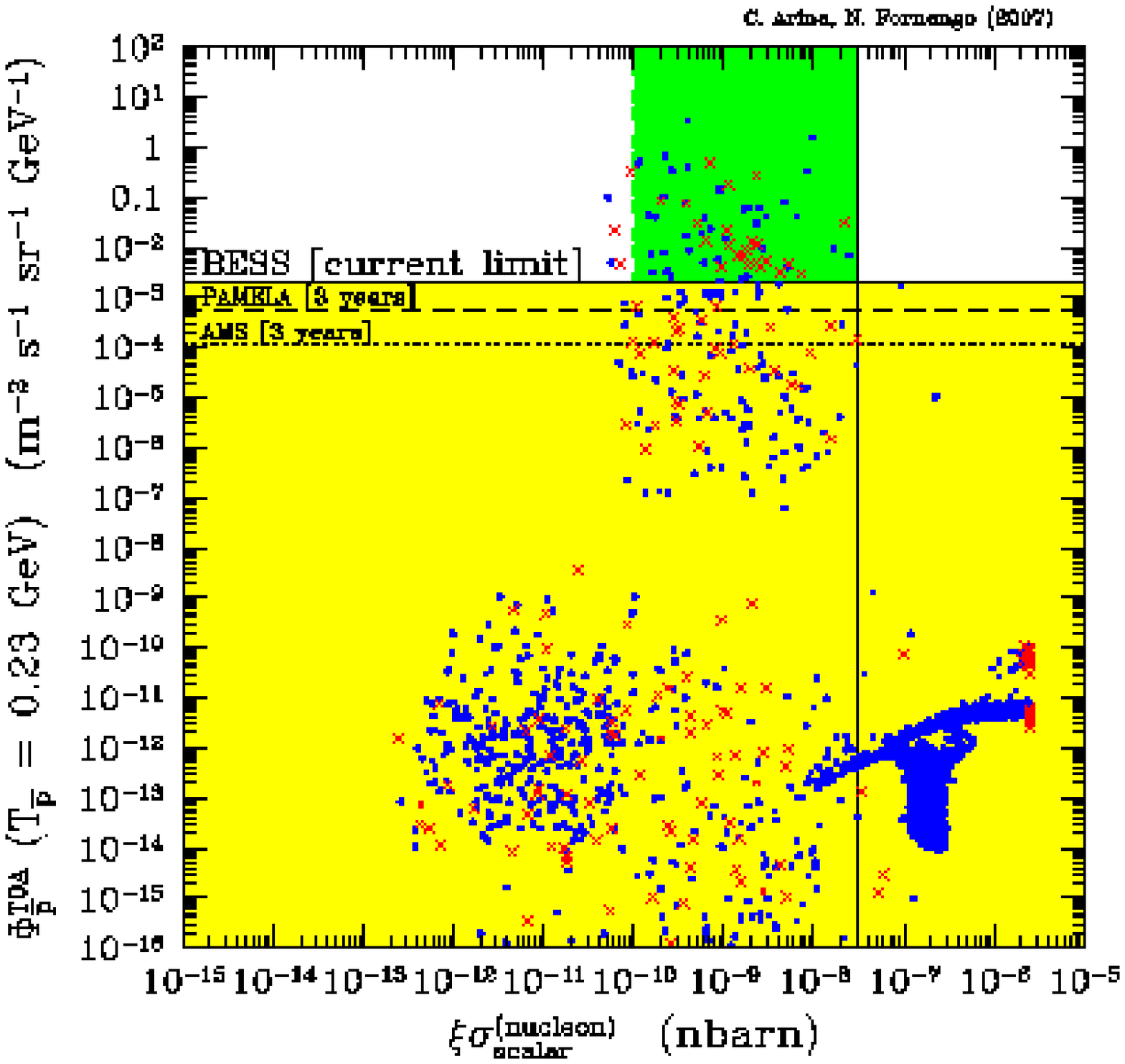,width=0.80\textwidth}
{MAJ[A] models -- Antiproton flux at the antiproton kinetic energy $T_{\bar p} = 0.23$ GeV vs. the
sneutrino--nucleon scattering cross section $\xi \sigma^{\rm (scalar)}_{\rm nucleon}$,
for the case of a Majorana--mass parameter $M = 1$ TeV
and a full scan of the supersymmetric parameter space.
Parameters are varied as in Fig. \ref{fig:maja-omegascan}.
[Red] crosses refer to models with sneutrino relic abundance in the cosmologically relevant range; [blue] dots refer to cosmologically subdominant sneutrinos. The horizontal solid line denotes the upper 
limit from BESS \cite{Orito:1999re,Maeno:2000qx} and the [yellow] shaded area shows the amount of exotic antiprotons which 
can be accommodated in the BESS data. The dashed and dotted horizontal lines show the
PAMELA \cite{pam} and AMS \cite{ams} sensitivities to exotic antiprotons for 3 years missions, respectively. The vertical solid line denotes a conservative upper limit from direct detection searches
and the [green] vertical shaded area refers to the current sensitivity in direct detection searches.
\label{fig:maja-pbardirect}}

\EPSFIGURE[t]{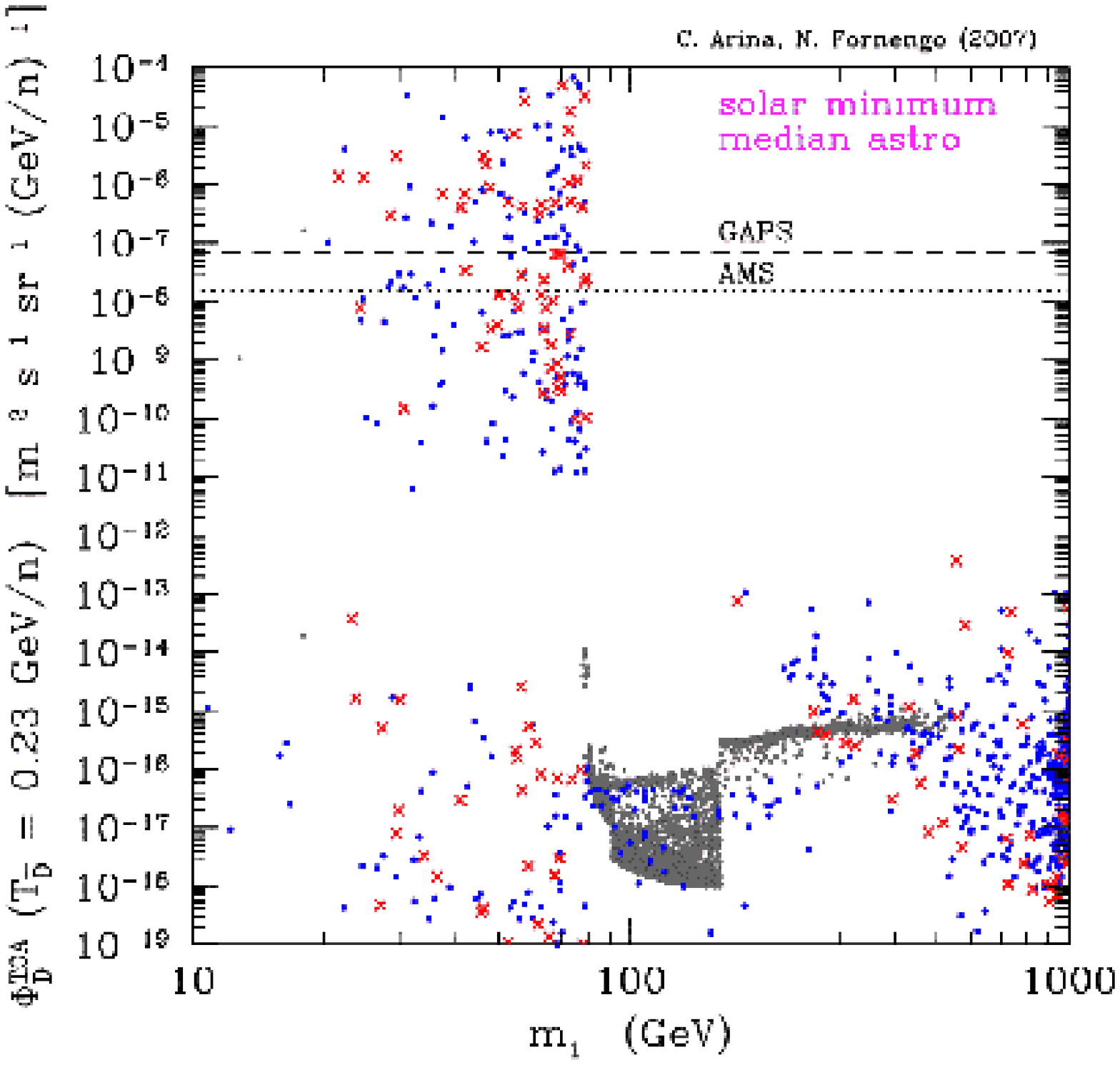,width=0.80\textwidth}
{MAJ[A] models -- Antideuteron flux at the antideuteron kinetic energy (per nucleon) $T_{\bar p} = 0.23$ GeV/n, as a function of the sneutrino mass $m_{1}$,
for the case of a Majorana--mass parameter $M = 1$ TeV
and a full scan of the supersymmetric parameter space.
Parameters are varied as in Fig. \ref{fig:maja-omegascan}. Notations are as in Fig. \ref{fig:lr-pbar023}.
The dashed and dotted lines show the GAPS \cite{Mori:2001dv,koglin} and AMS \cite{ams} sensitivities.
\label{fig:maja-dbar023}}

\EPSFIGURE[t]{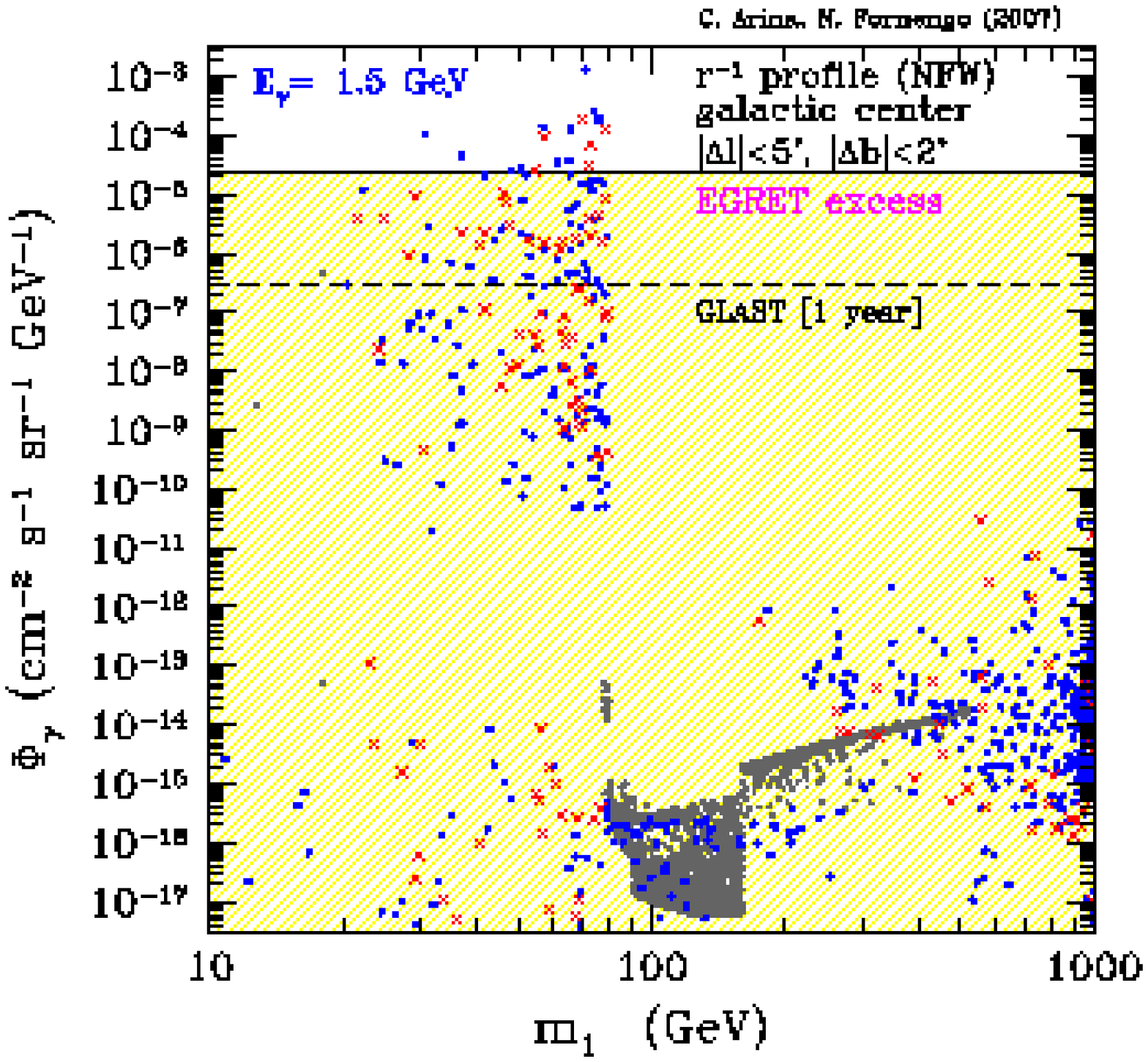,width=0.80\textwidth}
{MAJ[A] models -- Gamma--ray flux from the galactic center at the photon energy $E_{\gamma} = 1.5$ GeV,
as a function of the sneutrino mass $m_{1}$, for a galactic profile of Moore's type \cite{Moore:1999nt,Moore:1999gc} and for the angular resolution of EGRET \cite{Egret,MayerHasselwander:1998hg}.
The plot refers to the case of a Majorana--mass parameter $M = 1$ TeV
and a full scan of the supersymmetric parameter space.
Parameters are varied as in Fig. \ref{fig:maja-omegascan}. [Red] crosses refer to models with sneutrino relic  abundance in the cosmologically relevant range; [blue] dots refer to cosmologically 
subdominant sneutrinos; light gray points denote configurations which are excluded by direct
detection searches. The [yellow] shaded area denotes the amount of exotic gamma--rays compatible
with the EGRET excess \cite{Egret,MayerHasselwander:1998hg}. The dashed line shows the GLAST \cite{glast} sensitivity for a 1 year data-taking and for the same EGRET angular bin.
\label{fig:maja-gamma}}

The large sneutrino mass splitting which is now allowed also
for a neutrino mass bound of 2 eV is relevant for the direct detection suppression effect: this
is at variance with the \lviol models, where this bound implied a strong restriction of the
sneutrino phenomenology, bringing it to the minimal MSSM case. In addition to the mass splitting,
a sizeable sterile component now also appears. The suppression factor is shown if Fig. \ref{fig:maja-suppression}, which can be compared to the corresponding case of \lviol models of Fig. \ref{fig:cp-suppression}.

The relic abundance of the MAJ[A] models is shown in Fig. \ref{fig:maja-omegascan}. It is remarkable
that in the whole mass range from 5 GeV to 1 TeV sneutrinos can explain the required amount of dark
matter in the Universe. For the same mass range, sneutrinos may as well be a subdominant component.

Direct detection is shown in Fig. \ref{fig:maja-directscan}. We see that three different populations
arise: configurations on the upper right are clearly excluded by direct detection searches.
Most of them refer to subdominant sneutrinos. Configurations on the lower right part of the
plot are allowed but well below current direct detection sensitivity. Configurations
on the center and left part of the plot all fall inside the current sensitivity region, and
in this specific plot show that could explain the annual modulation effect observed by  the DAMA/NaI
experiment. We notice that a large fraction of the configurations which fall inside the DAMA/NaI region are
also cosmologically dominant. The correlation of the direct detection cross section \xisigma and the
relic abundance are plotted in Fig. \ref{fig:maja-sigmaomega}, which shows how almost all the
cosmologically relevant configurations are under investigation or under reach of direct detection
studies. This can be confronted with the case of pure LR models shown in Fig.  \ref{fig:lr-sigmaomega},
where a large fraction of the cosmologically relevant configurations were either already excluded
by direct detection or with very low values of \xisigma. We remind that now \xisigma contains the suppression factor as discussed in Eq. (\ref{eq:xisigmareduced}).

Contrary to the case of LR models, indirect detection in not currently sensitive to the MAJ[A] models. 
Fig. \ref{fig:maja-fluxedirect} shows the correlation between the upgoing muon flux from the Earth
as a function of \xisigma. We notice that all the configurations are below the current limits from
neutrino telescopes. However, there is a large fraction of configurations which could be accessible by
an increase of sensitivity, although the increase should be sizeable.

Figs. \ref{fig:maja-m1sterile} shows the distribution of the sterile components with respect to the
sneutrino mass. We see that, when sneutrinos are light, they are typically right--handed, and therefore
they do not couple to the $Z$ boson (although they still have couplings with higgses). For
masses above 80 GeV they may exhibit both right-- or left--handed behaviour.

The distribution of the cosmologically allowed configurations in the sneutrino parameter space is 
shown in Figs. \ref{fig:maja-m1f2} and \ref{fig:maja-m1mn}. Light states typically have small values
of both $m_{N}$ and $F^{2}$, and as it was already shown if Fig. \ref{fig:maja-directscan} they
are almost all acceptable and currently explored by direct detection experiments. The distribution
of configurations in these sectors of the parameter space are somewhat different from the analogous
results for the LR models, shown in Fig. \ref{fig:lr-mnml} and \ref{fig:lr-mnf}: in that case,
larger values of $F^{2}$ were required also for light sneutrinos, a difference which is traced into
the new structure of the sneutrino mass matrix and neutrino couplings which arise here from the
introduction of \lviol terms in the lagrangian.

The antiproton flux at $T_{\bar p} = 0.23$ GeV is shown in Fig. \ref{fig:maja-pbar023}. We notice that
a fraction of the configurations with low mass are currently excluded by the BESS data. This is at variance
with the case of LR models, where instead the antiproton flux for light sneutrinos were all well below
the BESS sensitivity. Another remarkable difference from the LR models is that now the configurations
which are excluded by direct detection refer to very low antiproton fluxes, practically undetectable. These
two features show that for the MAJ[A] models antiproton searches and direct detection studies exhibit a high
degree of complementarity, both in the ability to exclude model configurations and the prospects of
detection. This is clearly summarized in Fig. \ref{fig:maja-pbardirect}, where the high level of
complementarity is manifest. It is also remarkable that a fraction of configurations which are
currently under study by direct detection experiments (either just inside the CDMS sensitivity range
or, even more interestingly, inside the DAMA/NaI annual modulation region) have a chance of
detection by the future PAMELA and AMS flights, or by some slightely more sensitive and future experiments.

The antideuteron searches are as well appealing, as Fig. \ref{fig:maja-dbar023} shows: the configurations
for light sneutrinos above the GAPS sensitivity line are almost all excluded already by the BESS data on
antiprotons, but those under the GAPS and AMS sensitivity line are under reach. Figs.  \ref{fig:maja-pbar023} and \ref{fig:maja-dbar023} also show that for MAJ[A] models, antimatter searches are
not sensitive to heavy sneutrinos.

Also the gamma--ray signal is interesting, since EGRET and GLAST are sensitive to sneutrino
dark matter also for dark matter profiles of the NFW type \cite{Navarro:1995iw,Navarro:1996gj}. This is at variance with the LR models, where
a steeper profile was needed in order to obtain sizeable fluxes. Fig. \ref{fig:maja-gamma} shows
the gamma--ray flux at $E_{\gamma} = 1.5$ GeV from the center of the Galaxy, for a EGRET--like angular resolution. Like for the antimatter signals, only sneutrinos with masses below 80 GeV produce sizeable signals. There are configuration which are at the level of explaining the EGRET excess. GLAST will
be sensitive to a large fraction of the sneutino configurations.

\subsection{Models with a large--scale Majorana mass--parameter}\label{sec:majb}

As a second example of this class of models we consider the case of a large--scale Majorana mass--parameter,
which we fix at $M=10^{9}$ GeV. We call these models MAJ[B]. In this case, the large value of $M$ drives the two highest sneutrino mass eigenstates to be very heavy and decoupled from the rest. The model for the light sneutrino sector is somehow less rich phenomenologically as compared to the case of a TeV--scale Majorana mass--parameter. The 1--loop neutrino mass contribution $|\mloop|$ is shown in Fig. \ref{fig:majb-mneutrino} as a function of the sneutrino mass difference. Contrary to the case of \lviol models, larger
sneutrino mass splittings are allowed, up to almost 10 MeV. This is an interesting range for the
inelasticity properties in direct detection dark matter \cite{Smith:2001hy,Smith:2002af,Tucker-Smith:2004jv}. Directly from the MAJ[A] case, instead, larger
mass splittings are not possible now. This is related to the fact that light sneutrinos are not
obtained in this class of models. This is shown in Fig. \ref{fig:majb-omegascan} where the relic
abundance is plotted versus $m_{1}$. We see that now cosmologically relevant sneutrinosare present only for very restricted mass ranges, one around  80--90 GeV and the second at 500--600 GeV. For masses
above 600 GeV sneutrinos are cosmologically excluded. In the whole interval 90--500 GeV, sneutrinos are subdominant.

Direct detection comes back to be a strong constraint. Fig. \ref{fig:majb-directscan} shows that direct
detection bounds exclude most of the configurations, and in particular all the cosmologically relevant ones.
Nevertheless, sneutrinos in the mass range 90--300 GeV are allowed, although subdominant.

Indirect detection for MAJ[B] models is not very appealing: the antiproton flux at $T_{\bar p}= 0.23$ GeV
is shown in Fig. \ref{fig:majb-pbar023}: configurations which pass the direct detection bound all refer to
very low antiproton fluxes, practically undetectable. A similar situation occurs for antideuterons and
gamma--rays.

\EPSFIGURE[t]{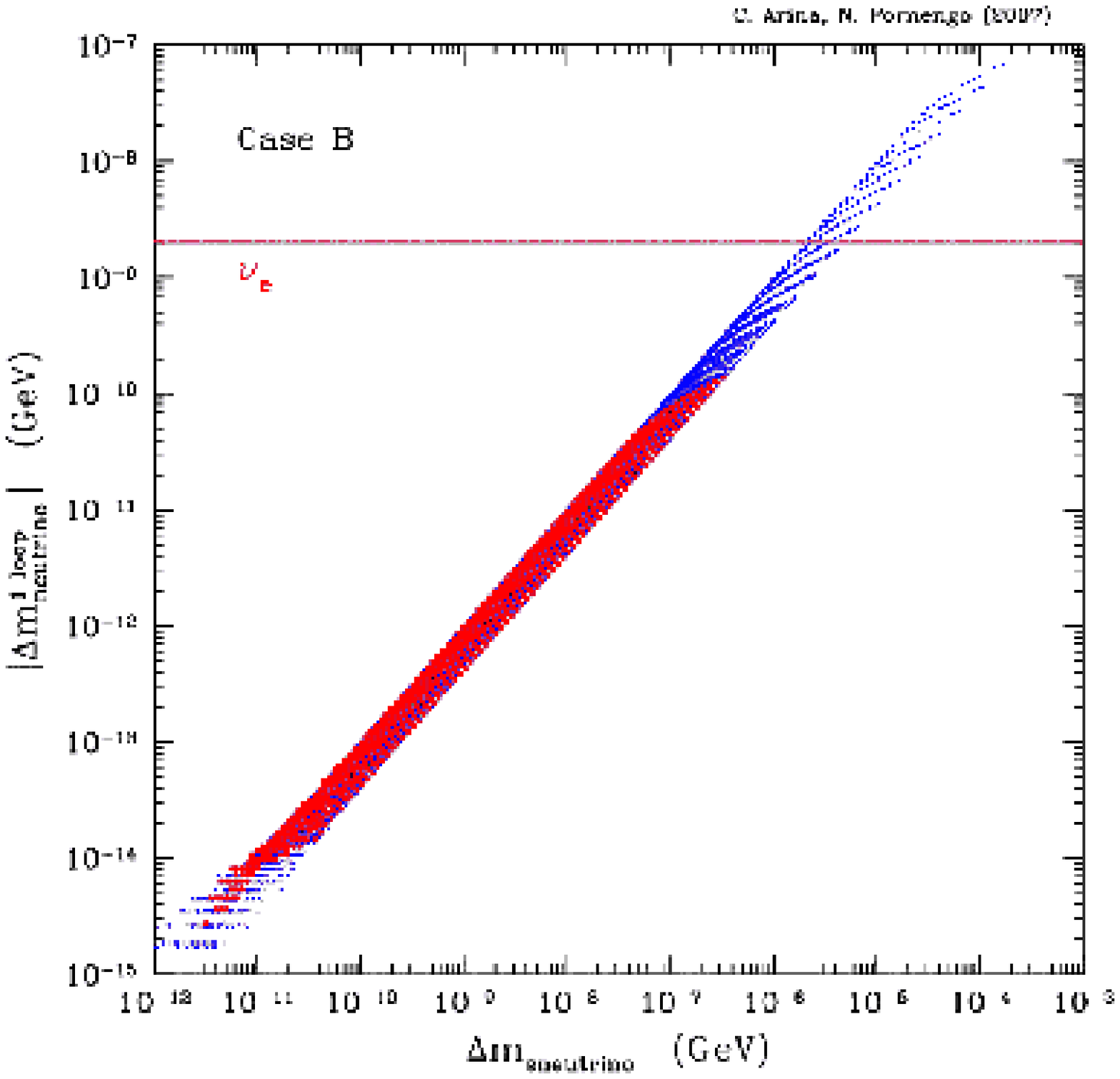,width=0.80\textwidth}
{MAJ[B] models -- Absolute value of the 1--loop contribution to the neutrino mass as a function of
the mass difference of the two lightest sneutrino CP eigenstates, for the case of a 
Majorana--mass parameter $M = 10^{9}$ GeV.
The other sneutrino mass parameters are varied as:
$m_{N} = 0$,
$10^{3}~\mbox{GeV} \leq m_{B} \leq 10^{8}~\mbox{GeV}$ and 
$1~\mbox{GeV}^{2} \leq F^{2} \leq 10^{4}~\mbox{GeV}^{4}$.
The horizontal lines denote the
upper limits on the electron neutrino mass.
\label{fig:majb-mneutrino}}

\EPSFIGURE[t]{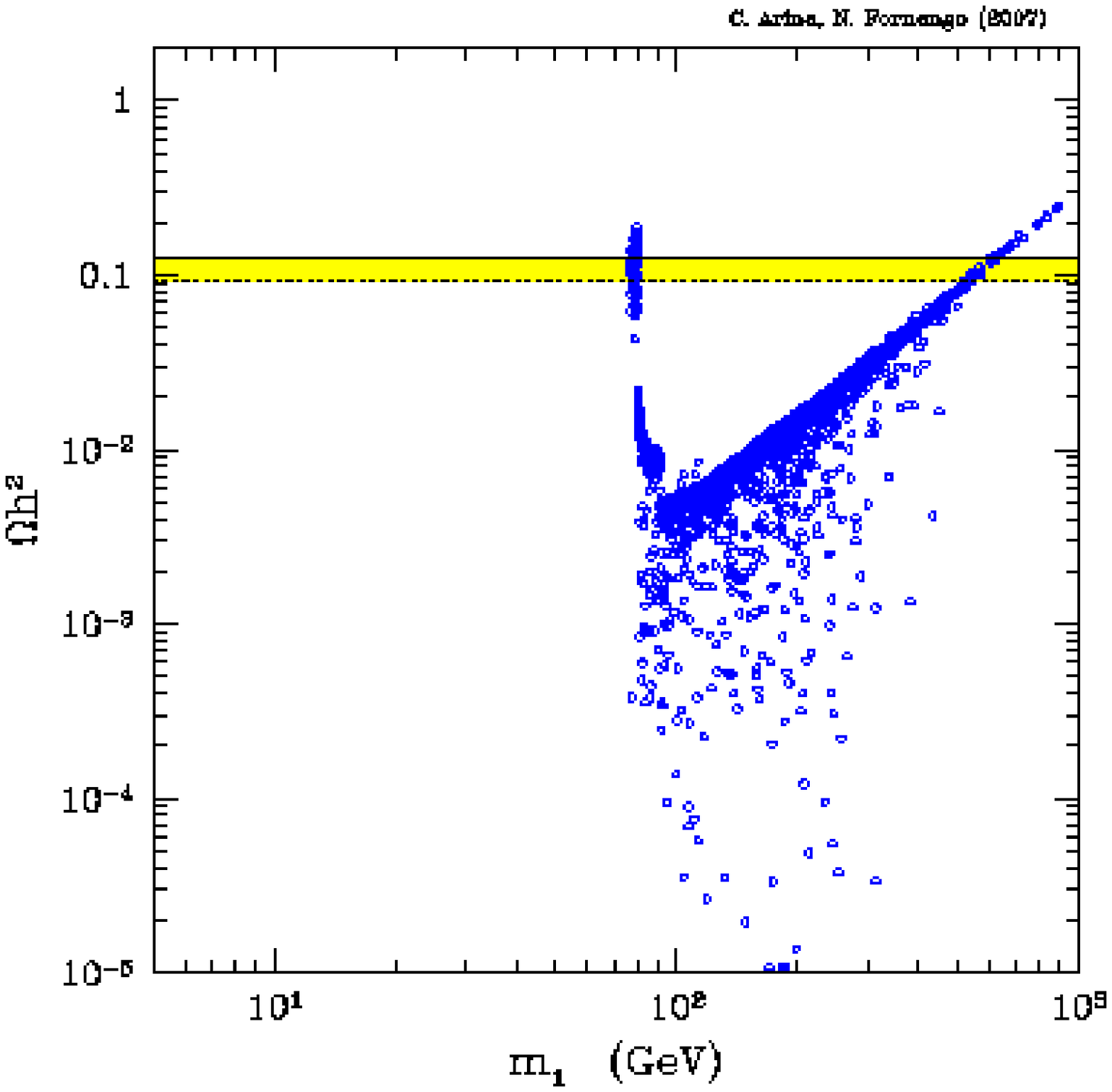,width=0.80\textwidth}
{MAJ[B] models -- Sneutrino relic abundance $\Omega h^{2}$ as a function of 
the sneutrino mass $m_{1}$, for the case of a Majorana--mass parameter $M = 10^{9}$ GeV and 
for a full scan of the supersymmetric parameter space. The sneutrino parameters (other than $M$) 
are varied as follows: 
$m_{N} = 0$,
$10^{3}~\mbox{GeV} \leq m_{B} \leq 10^{8}~\mbox{GeV}$ and 
$1~\mbox{GeV}^{2} \leq F^{2} \leq 10^{4}~\mbox{GeV}^{4}$.
All the models shown in the plot are acceptable from the point of view of all 
experimental constraints. The horizontal solid and 
dotted lines delimit the WMAP interval for cold dark matter.
\label{fig:majb-omegascan}}

\EPSFIGURE[t]{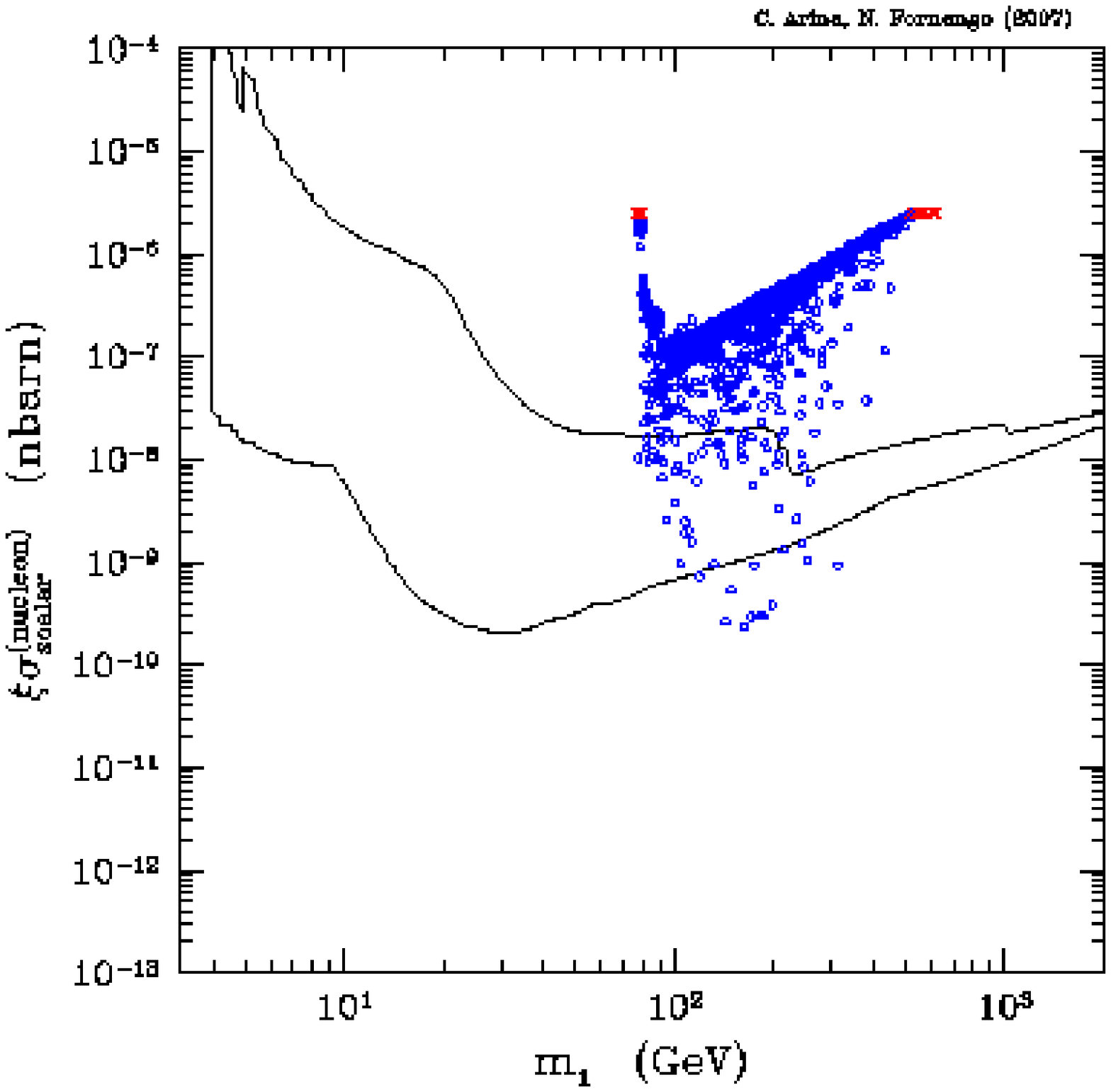,width=0.80\textwidth}
{MAJ[B] models -- Sneutrino--nucleon scattering cross section $\xi \sigma^{\rm (scalar)}_{\rm nucleon}$ 
as a function of the sneutrino mass $m_{1}$, for the case of a Majorana--mass parameter 
$M = 10^{9}$ GeV  and for a full scan of the supersymmetric parameter space. Parameters are 
varied as in Fig. \ref{fig:majb-omegascan}.
[Red] crosses refer to models with sneutrino relic abundance
in the cosmologically relevant range; [blue] open circles refer to cosmologically subdominant 
sneutrinos. The solid curve shows the DAMA/NaI region, compatible with the annual 
modulation effect observed by the experiment \cite{Bernabei:2003za,Bernabei:2005hj,Bernabei:2005ca,Bernabei:2006ya,Bernabei:2007jz}.
\label{fig:majb-directscan}}

\EPSFIGURE[t]{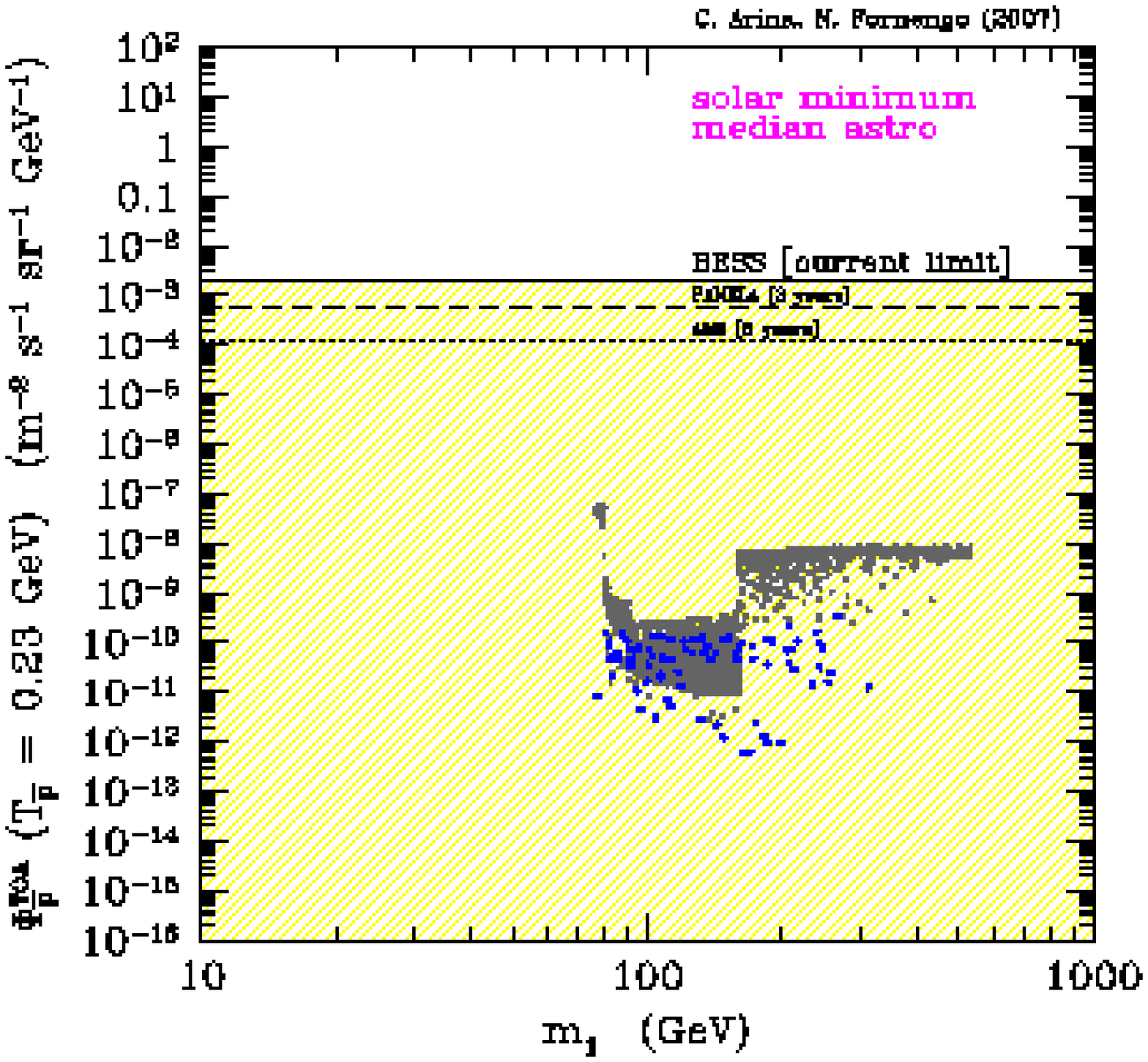,width=0.80\textwidth}
{MAJ[B] models -- Antiproton flux at the antiproton kinetic energy $T_{\bar p} = 0.23$ GeV as a 
function of the sneutrino mass $m_{1}$, for the galactic propagation parameters which provide the
median value of antiproton flux and for a solar activity at its minimum. 
The plot refers to the case of a Majorana--mass parameter $M = 10^{9}$ GeV 
and a full scan of the supersymmetric parameter space.
Parameters are varied as in Fig. \ref{fig:majb-omegascan}.
[Red] crosses refer to models with sneutrino relic  abundance in the cosmologically relevant range; [blue] dots refer to cosmologically 
subdominant sneutrinos; light gray points denote configurations which are excluded by direct
detection searches. The [yellow] shaded area denotes the amount of exotic antiprotons which can be accommodated in the BESS data \cite{Orito:1999re,Maeno:2000qx} . The dashed and dotted lines show the
PAMELA \cite{pam} and AMS \cite{ams} sensitivities to exotic antiprotons for 3 years missions, respectively. 
\label{fig:majb-pbar023}}

\section{Conclusions}\label{sec:conclusions}

In this paper we have analyzed the possibility to have sneutrinos as dark matter candidates. 
We have discussed various supersymmetric models, and specifically extensions of the Minimal
Supersymmetric Standard Model which could lead to explanations of the neutrino mass problem.

We have first re--analyzed the minimal version of the MSSM. Sneutrinos are here typically subdominant
dark matter components, with low values of the relic abundance in all the mass range from 50-70 (their
lower mass bound in the MSSM) up to 700 GeV. They may represent the dominant dark matter component
for masses in the 600-700 GeV range. This possibility is actually excluded by direct detection searches,
which allow sneutrinos to be a subdominant dark matter component only very marginally, and for mass--
matching conditions that pose the sneutrino annihilation cross--section on one of the higgs poles or
the $Z$ pole.

The inclusion of right--handed superfields to the supersymmetric lagrangian allows for a much
richer phenomenology. The mixing with right--handed fields suppresses the $Z$--coupling and
leads to some increase of the relic abundance and to some decrease of the direct detection rate.
The reduced $Z$ coupling is also instrumental in allowing light sneutrinos, by circumventing
the invisible $Z$--width bound. From a full scan of the supersymmetric parameter space, we find that 
cosmologically dominant relic sneutrinos are present in the mass range from 15 GeV up to 1 TeV
(where we stop our scan). When the full supersymmetric parameter space is considered, we find that
15 GeV is actually the mass lower bound, induced by the cosmological limit on the relic abundance.
Direct detection is acceptable for all the allowed mass range. We also find that cosmologically
dominant sneutrinos are accepted by direct detection, and a large fraction of the supersymmetric
configurations predict direct detection rates at the level of the current experimental sensitivities
(including the possibility to explain the DAMA/NaI annual modulation effect). Indirect detection rates
offer good possibilities: antiproton fluxes are under reach of the PAMELA and AMS detectors in the
50--200 GeV mass range. The configurations accessible to indirect searches are typically cosmologically
subdominant. The same occurs also for the antideuteron signal, which is accessible by GAPS and AMS
to the same configurations to which PAMELA and AMS are sensitive for antiprotons. This offers 
a great opportunity for dark matter searches: a signal detectable in one antimatter channel 
by two different detectors, will be detectable also in the other channels, again by two different detectors.
Gamma-rays from the center of the galaxy do not provide very large signals: we predict fluxes
not too far from the EGRET excess in the 50--200 GeV mass range, but this requires a very steep
dark matter density profile toward the galactic center (of the $r^{-1.5}$ type). GLAST will be sensitive
to configurations in this same mass range, again for a very steep profile. 

Models with a lepton--number violating term, introduced as non--renormalizable 5--dimensional operator,
do not lead to a phenomenology very different from the standard MSSM, once a mass bound of 2 eV on the 1-loop correction to the neutrino mass (which is induced by the \lviol terms) is considered. Only for
a neutrino mass bound of 18 MeV, which corresponds to the kinematical mass bound for the tau neutrino,
some increase of the relic abundance is possible. However, the direct detection limit strongly bounds
these models, making them almost marginal.

A renormalizable lagrangian with both right--handed fields and \lviol terms, which offers the possibility
to include neutrino masses via a see-saw mechanism, again offers very rich sneutrino phenomenology. In the case of a TeV--scale Majorana mass--parameter, sneutrinos may be the dominant dark matter component for masses in the range 5 GeV up to 1 TeV (upper bound in our scan). The direct detection is
nicely evaded in all the mass range, and most of the configurations fall inside the current experimental sensitivity range (including the possibility to explain the DAMA/NaI annual modulation effect). Antiproton
fluxes are a stringent bound for light sneutrinos, complementary to the bound imposed by direct detection
which instead is more severe for heavier particles. Many configurations for masses below 80--90 GeV will
be explored by PAMELA and AMS, while for masses above 90 GeV antiproton searches loose sensitivity.
Also for antideuterons, AMS and GAPS will have sensitivity to probe a fraction of the configurations
for masses below 80--90 GeV. For these light sneutrinos, also gamma--rays provide a significant
probe, also for NFW profiles. GLAST will have sensitivity to a fraction of those configurations
with mass below 80--90 GeV. Finally, models with a large Majorana mass-parameter are strongly bounded
by direct detection: configurations with masses in the range 90--300 GeV are not excluded by
direct detection, but they all refer to cosmologically subdominant sneutrinos. Indirect detection rates
are typically very suppressed.
  
We therefore conclude that sneutrinos offer a rich phenomenology as dark matter candidates,
and they are a viable alternative to relic neutralinos in a wide class of supersymmetric models.
Their phenomenology is also linked and constrained by neutrino physics through the problem
of the origin of neutrino masses.

\acknowledgments 
We acknowledge Research Grants funded jointly by the Italian Ministero
dell'Istruzione, dell'Universit\`a e della Ricerca (MIUR), by the
University of Torino and by the Istituto Nazionale di Fisica Nucleare
(INFN) within the {\sl Astroparticle Physics Project}.

\section{Appendix: The supersymmetric model}
\label{app:mssm}

The supersymmetric model we adopt in this paper is an effective MSSM scheme
at the electroweak scale. The free independent parameters of the model, which are not
directly related to the sneutrino sector, are the following:
the SU(2) gaugino--mass parameter $M_2$; the ratio between the
U(1) and SU(2) gaugino--mass parameters $R \equiv M_1/M_2$ 
(in the GUT--induced case $R=5/3 \tan^2\theta_W\simeq 0.5$,
where $\theta_W$ is the Weinberg angle); the Higgs--mixing parameter $\mu$; 
$\tan\beta = v_2/v_1$; the mass of the pseudoscalar higgs $m_A$;
a common soft--mass for all the squarks $m_{Q}$ (both
right-- and left--handed); a common dimensionless trilinear parameter
for the third family $A$ ($A_{\tilde b} = A_{\tilde t} \equiv A
m_{\tilde q}$ and $A_{\tilde \tau} \equiv A m_{\tilde l}$; the
trilinear parameters for the other families being set equal to
zero). The masses of the CP--even and charged higgses are calculated
from $m_A$ (and the other relevant supersymmetric parameters) by employing 2--loop corrections.

The full scans of the parameter space are performed over the following
ranges of the MSSM parameters: $1 \leq \tan \beta \leq 50$, $100 \, {\rm GeV}
\leq |\mu| \leq 3000 \, {\rm GeV}, 100 \, {\rm GeV}
\leq M_2 \leq 3000 \, {\rm GeV},
100 \, {\rm GeV} \leq m_{Q} \leq 3000 \, {\rm GeV }$,
$90\, {\rm GeV }\leq m_A \leq 1000 \, {\rm GeV }$, $-3
\leq A \leq 3$. As for the $R$ parameter, we use either its mSUGRA value $R=0.5$
or we scan over the interval $0.005 \leq R \leq 0.5$, depending on the case at study.
In order to have the sneutrino as a dark matter candidate, we accept only parameter configurations
for which the lightest sneutrino is also the lightest among all the supersymmetric particles.

We impose the following experimental constraints: accelerators data on
supersymmetric particles and Higgs--boson searches (CERN $e^+ e^-$ collider LEP2
\cite{Colaleo:2001tc,delphi,:2001xwa,lep2} and Collider Detectors D0 and CDF at Fermilab \cite{Affolder:2000rg,Abazov:2006fe});
measurements of the $b \rightarrow s + \gamma$ decay process \cite{bsgamma}:
we adopt the interval 2.89 $\leq B(b \rightarrow s + \gamma) \cdot 10^{-4} \leq$ 4.21,
which is larger by 25\% with respect to the experimental
determination  \cite{bsgamma} in order to take into account theoretical
uncertainties in the SUSY contributions  \cite{bsgamma_theorySUSY} to the
branching ratio of the process (for the Standard Model calculation, we employ
the recent NNLO results from Ref.  \cite{bsgamma_theorySM}); the upper bound on
the branching ratio $BR(B_s^{0} \rightarrow \mu^{-} + \mu^{+})$ \cite{bsmumu}: we
take $BR(B_s^{0} \rightarrow \mu^{-} + \mu^{+}) < 1.2 \cdot 10^{-7}$;
measurements of the muon anomalous magnetic moment $a_\mu \equiv (g_{\mu} -
2)/2$: for the deviation $\Delta a_{\mu}$ of the  experimental world average
from the theoretical evaluation within the Standard Model we use here the range
$-98 \leq \Delta a_{\mu} \cdot 10^{11} \leq 565 $, derived from the latest
experimental  \cite{bennet} and theoretical \cite{bijnens} data. The invisible $Z$--width
constraints is also imposed on neutralinos lighter than $m_Z/2$ which occur in the gaugino
non--universal models.


\end{document}